\documentclass[fleqn,usenatbib]{mnras}

\usepackage{newtxtext,newtxmath}

\usepackage[T1]{fontenc}

\DeclareRobustCommand{\VAN}[3]{#2}
\let\VANthebibliography\thebibliography
\def\thebibliography{\DeclareRobustCommand{\VAN}[3]{##3}\VANthebibliography}

\usepackage{graphicx}	
\usepackage{amsmath}	
\usepackage{caption}
\usepackage{subcaption}

\title[Spectral and temporal properties of Mrk~530]{Accretion dynamics and coronal geometry in Mrk~530: Insights from 24 years of X-ray monitoring}

\author[Dash et al.]{
Priyadarshee P. Dash,$^{1,2}$\thanks{ankitppdash@gmail.com}
Prantik Nandi$^{3}$\thanks{prantiknandi007@gmail.com},
Sachindra Naik$^{1}$\thanks{snaik@prl.res.in},
Narendranath Layek$^{1,2}$,
Sandip K. Chakrabarti$^{3}$
\\
$^1$Astronomy and Astrophysics Division, Physical Research Laboratory, Navrangpura, Ahmedabad - 380009, Gujarat, India\\
$^2$Indian Institute of Technology Gandhinagar, Palaj, Gandhinagar - 382055, Gujarat, India\\
$^3$Indian Centre for Space Physics, Netaji Nagar, Kolkata, 700099, India\\
}

\date{Accepted XXX. Received YYY; in original form ZZZ}

\pubyear{\the\year{}}

\begin{document}
\label{firstpage}
\pagerange{\pageref{firstpage}--\pageref{lastpage}}
\maketitle

\begin{abstract}
We present a long-term broadband study of the Seyfert galaxy Mrk~530 spanning $\sim$24 yr (2001--2024). The source remains largely stable across epochs, except in 2018, when a possible quasi-periodic oscillation is observed simultaneously in the UV and X-ray bands, with characteristic timescales of $\sim$90 and $\sim$60 days, characterized by low coherence. Time-resolved spectral analysis shows that this epoch is characterized by comparable coronal cooling and compressional heating timescales, a condition conducive to oscillatory behavior in the inner accretion flow. Other epochs exhibit a clear mismatch between these timescales, and no such variability is observed. The X-ray spectral properties display significant long-term evolution. The photon index and luminosity vary systematically, while a soft excess is present only in early epochs (2001--2006) and weakens thereafter, consistent with an evolving warm corona. Physically motivated models indicate that changes in the accretion rate regulate both the spectral slope and coronal geometry, with higher disc accretion rates producing enhanced cooling, a more compact corona, and softer spectra, and lower rates yielding an expanded hot flow and harder emission. These results suggest that accretion-driven coupling between the disc and corona governs both the long-term spectral evolution and transient short-timescale variability in Mrk~530.

\end{abstract}




\section{Introduction}

Active Galactic Nuclei (AGN) are among the most luminous ($L_{bol} \sim 10^{47} {\rm erg\ s^{-1}}$; \cite{2015MNRAS.447.3368B, Duras_2020}) sources in the Universe, powered by accretion onto supermassive black holes (SMBHs; \citealt{lynden1969galactic}). As matter falls towards the central SMBH, it may settle into a geometrically thin but optically thick accretion disc around the SMBH where gravitational potential energy is efficiently converted into radiation \citep{1973A&A....24..337S}. The luminosity and the peak of this thermal radiation depend on the mass accretion rate and the mass of the black hole. For Seyfert AGN, this thermal peak typically lies in the optical/UV band \citep{1989ApJ...346...68S}.

The X-ray emission in AGNs typically originates in the innermost regions, where disc photons undergo inverse Compton scattering in a hot ($T \sim 10^{8-9}$ K) electron cloud \citep{1980A&A....86..121S,1991ApJ...380L..51H,1993ApJ...413..507H} often referred to as the hot corona or Compton cloud or CENBOL \citep{1980A&A....86..121S,1985A&A...143..374S, 1995ApJ...455..623C,nandi2019spectral, Nandi2026}. This process produces a power-law continuum with high-energy cutoffs ($\sim$100–300 keV) \citep{1980A&A....86..121S,1985A&A...143..374S,1991ApJ...380L..51H,1993ApJ...413..507H,1994ApJ...428L..13N,1995ApJ...455..623C}. Alongside the primary power-law continuum, several additional components are commonly observed in the X-ray spectra of AGN. In the hard X-ray domain (10–50 keV range), a broad hump-like feature known as the reflection hump \citep{1999agnc.book.....K} is often visible, which arises from the Compton scattering of high-energy photons off the accretion disc \citep{1990ApJ...363L...1Z,1999ASPC..161..178R,2013MNRAS.430.1694D}. The most prominent discrete feature in the X-ray spectra of AGN is the Fe~K$\alpha$ fluorescence line at 6.4 keV \citep{1991MNRAS.249..352G,1993MNRAS.262..179M}, which provides a sensitive probe of the inner accretion disc and circumnuclear material. At lower energies ($<2$ keV), Seyfert~1 galaxies frequently exhibit an additional component above the extrapolated power law termed as the soft excess \citep{1984ApJ...281...90H,1985ApJ...297..633S,2021MNRAS.507..687J,2021PASA...38...42K}. The physical origin of this feature remains debated, although different perspectives are given by different theories. In particular, the warm Comptonization model with $kT \sim 0.1$–0.2 keV \citep{2009MNRAS.394..250M,Done2012,Petrucci2020, Nandi2026} and the blurred reflection from an ionized accretion disc model \citep{2002MNRAS.331L..35F,2005MNRAS.358..211R,2010ApJ...718..695G} are used to explain this excess emission. More recent developments include high-density relativistic reflection models \citep{2019ApJ...871...88G} and hybrid frameworks combining warm Comptonization and ionized reflection components \citep{2021ApJ...913...13X, 2022MNRAS.515..353X}. Alternative explanations include relativistically smeared absorption in partially ionized material \citep{2004MNRAS.349L...7G} or pure Comptonization of disc photons in a vertically and/or radially stratified corona \citep{2021MNRAS.506.3111N, nandi2023survey, nandi2024accretion, 2025ApJ...981...74L, Nandi2026}. A systematic observational comparison of these scenarios has been presented by \citet{waddell2024erosita}.

At higher energies, the observed spectral cut-off provides constraints on the temperature of the hot corona, while reflection features probe both the geometry of the X-ray–emitting region and the ionization state of the accretion disc.
Together, these components provide complementary constraints on the disc–corona geometry and their interaction with the central black hole, demonstrating the diagnostic capability of broadband X-ray spectroscopy for AGN accretion studies.

In addition to their spectral complexity, AGN also exhibit temporal variability over a broad range of timescales. In particular interest, the quasi-periodic oscillations (QPOs), widely detected in Galactic X-ray binaries \citep{1997ApJ...482L.167Z, Ingram_2019}, are generally classified into two main categories based on their characteristic frequencies. Low-frequency QPOs (LFQPOs; $\leq$ 30 Hz) are typically associated with geometric effects or instabilities in the accretion flow, whereas high-frequency QPOs (HFQPOs; 40-450 Hz) are thought to originate in the innermost regions of the accretion flow, close to the black hole \citep{2006ARA&A..44...49R}. In AGN, however, the detection of QPOs remains rare \citep{1996ApJ...464..760H, 2018ApJ...853..193Z}, primarily because characteristic variability timescales are expected to scale inversely with black hole mass $(\nu \propto M_{BH}^{-1})$. As a consequence, QPO frequencies in AGN are shifted to much lower values than those observed in X-ray binaries, often extending beyond the temporal coverage or sensitivity of typical X-ray monitoring campaigns. In particular, LFQPO analogues in AGN are predicted to occur on timescales of days to years, making their detection especially challenging with conventional observations.

Despite these observational limitations, the identification of QPO-like signals in AGN is of considerable significance, as such features provide a direct link between characteristic variability timescales and accretion dynamics across the black hole mass spectrum. To date, significant X-ray QPO detections in AGN have been reported in only a small number of sources, including the Narrow-Line Seyfert 1 (NLS1) galaxies RE J1034+396 \citep{2008Natur.455..369G}, 1H 0707-495 \citep{2016ApJ...819L..19P}, and Mrk 766 \citep{2017ApJ...849....9Z}, as well as the Seyfert 2 galaxy 1ES 1927+654 \citep{2025Natur.638..370M}. These rare detections offer valuable constraints on the physical mechanisms governing accretion and variability in supermassive black hole systems.

Physically, AGN QPOs are often interpreted as signatures of oscillatory behaviour in the innermost accretion flow and/or corona \citep{2001A&A...374L..19A,2003ApJ...584L..83M}. Another theory proposes that the Lense–Thirring precession of a misaligned inner flow could generate QPOs in AGN \citep{1999ApJ...513..827M,2011MNRAS.415.2323I}. From an advective flow point of view, QPOs could be generated by oscillations of a shock front or Comptonizing region \citep{1996ApJ...457..805M,chakrabarti2004effect,Chakrabarti2015}. Sometimes, instabilities in the accretion disc–corona system that couple thermal and dynamical timescales could produce QPOs in AGNs \citep{1999A&A...349.1003T,2003ApJ...593..980L}. The detection of simultaneous QPOs in multiple energy bands combined with phase-lag measurements can be used as a powerful tool to constrain the geometry of the innermost regions of AGN.

Mrk~530 (NGC~7603) is a nearby Seyfert~1.5 galaxy at a redshift of $z=0.029$ (comoving distance $=118.3$ Mpc; \cite{2011ApJ...731...68L}), hosting a central black hole of mass $M_{\rm BH}=1.15\times10^8~M_{\odot }$ \citep{2012AJ....143...49W}. This source has been investigated extensively across multiple wavebands, including the optical/UV \citep{2016ApJ...822...45T,2017ApJ...846..102M} and radio \citep{2004A&A...425...99L,2013A&A...554A..85S}. Remarkably, Mrk~530 was the first reported case of a Changing-Look AGN in the optical/UV regime \citep{ricci2023changing}, characterized by the appearance or disappearance of broad emission lines on timescales of years to decades. In 1974, the source was classified as Seyfert~1, with broad H$\alpha$ and Balmer lines detected; by 1975 it had transitioned to Seyfert~1.9, with the broad H$\beta$ line vanishing while H$\alpha$ persisted \citep{1976ApJ...210L.117T}. The broad H$\beta$ component reappeared in 1987 and 1993, confirming a reversal of this transition \citep{1995ApJ...440..141G}. Long-term monitoring further revealed strong continuum and line variability over $\sim20$ yr \citep{2000A&A...361..901K}. In X-rays, only one detailed study exists, based on a {\it Suzaku} observation in 2012 \citep{2018MNRAS.478.4214E}. The analysis reported sinusoidal-like variability in the 0.3–10 keV and 1–3 keV bands, with spectral evolution from a soft to a harder state during the observation. A prominent soft excess was also detected, consistent with the source’s complex and variable nature. Despite these findings, a systematic, multi-epoch X-ray investigation of Mrk~530 has not yet been undertaken, leaving its long-term spectral and temporal evolution poorly constrained.

In this work, we present the results from our analysis of long-term (~24 years, 2001 - 2024) observations of Mrk~530 with {\it XMM-Newton} and {\it Swift} observatories. The paper is structured in the following way. Section~\ref{obs_data} provides an overview of the observational data and describes the procedures followed for data reduction. The results are presented in detail in Section~\ref{timing} and Section~\ref{spectral}. Then, we discuss our key findings in Section~\ref{discussion}, and finally, our conclusions are summarized in Section~\ref{conclusions}.   

\section{Observations and Data Reduction}
\label{obs_data}

MRK~530 was observed with \textit{Swift} and \textit{XMM-Newton} observatories in different epochs between 2001 and 2024. The details of the observations are listed in Table~\ref{tab:log}. These archival data were accessed through HEASARC\footnote{\url{http://heasarc.gsfc.nasa.gov/}}, and were reduced and analyzed using \texttt{HEAsoft} v6.34.

\subsection{Swift}
\label{sec_swift}
The {\it Swift} satellite \citep{2004ApJ...611.1005G} provides simultaneous coverage from optical/UV to hard X-rays. We used data from the Ultraviolet/Optical Telescope (UVOT; \citealt{2005SSRv..120...95R}) and the X-ray Telescope (XRT; \citealt{2005SSRv..120..165B}) covering 0.3–10 keV. For XRT, spectra and light curves were extracted with the `XRT product builder'\footnote{\url{http://swift.ac.uk/user_objects/}} \citep{2009MNRAS.397.1177E}, which processes observations in both photon-counting (PC) and window-timing (WT) modes. To ensure adequate signal-to-noise, we grouped the XRT data into seven sets (XRT1–XRT7; Table~\ref{tab:log}). The 2006 epochs (XRT1, XRT2) were analyzed individually due to long exposures, while XRT3 combines two consecutive observations from 2016. The 2018 dataset (XRT4) contains 45 observations, analyzed separately despite low SNR owing to the unusual variability of the source in that year. Later epochs were combined by year: XRT5 (2022), XRT6 (2023), and XRT7 (2024).

The \textit{Swift}/UVOT provides data in three optical filters (V, B, and U) and three UV filters (UVW1, UVM2, and UVW2). In the year 2018, UVOT had a total of 45 exposures with 3 observations of the UVW2 filter and 42 observations from the UVM2 filter. We primarily utilize the UVM2 ($\lambda_{\rm eff}=2246$ \AA) filter data in this work. Starting from level~II images, we performed aperture photometry using {\tt UVOTSOURCE}, adopting a 5 arcsec circular source region centered on Mrk~530 and an annular background region with inner and outer radii of 6 and 18 arcsec. We identified six epochs (MJD 58238, 58272, 58288, 58298, 58301, and 58316) for which the UV observations produced negative count rates of the source during photometry. This occurs when the source happens to fall on localised regions of the UVOT detector with very little throughput. Consequently, observations of these epochs were excluded from further scientific analysis.

\begin{center} 
\begin{table}
\caption{Log of observations of MRK~530}
\resizebox{\columnwidth}{!}{\begin{tabular}{c|c|c|c|c}
\hline
ID & Date & Obs.ID & Observatory/ & Total Exposure\\
   & (yyyy-mm-dd) &  & Instrument & (ks) \\
\hline
XMM1 & 2001-06-03 & 0066950301 & XMM-Newton/MOS & 12.20\\\\
XMM2 & 2001-11-28 & 0066950401 & XMM-Newton/PN & 13.18\\\\
XRT1 & 2006-04-28 & 00035365002 & Swift/XRT & 11.19 \\ \\
XRT2 & 2006-05-08 & 00035365003 & Swift/XRT & 10.35\\\\
XMM3 & 2006-06-14 & 0305600601 & XMM-Newton/PN & 16.82\\\\
XRT3 & 2016-01-19 - & 00049538008 - & Swift/XRT & 5.08\\
 & 2016-01-20 & 00049538009 & &  \\\\
XRT4 & 2018-01 - & 00049538010 - & Swift/XRT \& & 38.50\\ 
 & 2018-12 & 00049538058 & Swift/UVOT & \\\\
XRT5 & 2022-01- & 00015007001- & Swift/XRT & 10.95\\
 & 2022-12 & 00015007007 & &  \\\\
XRT6 & 2023-01 - & 00015007008 - & Swift/XRT & 25.90\\
 & 2023-12 & 00015007032 & &  \\\\
XRT7 & 2024-01- & 00015007033 - & Swift/XRT & 24.91\\
 & 2024-12 & 00015007049 & &  \\
\hline
\end{tabular}}
\label{tab:log}
\end{table}
\end{center}

\subsection{XMM-Newton}
Mrk~530 was observed with {\it XMM-Newton} \citep{2001A&A...365L...1J} on three occasions between 2001 and 2006 (Table~\ref{tab:log}). We used archival data from the {\it European Photon Imaging Camera} (EPIC), which consists of three CCD detectors covering 0.3–10 keV: MOS1 and MOS2 \citep{2001A&A...365L..27T} and PN \citep{2001A&A...365L..18S}. For XMM2 and XMM3, we analyzed the PN data, while for XMM1, only MOS data were available. The raw data were processed with the {\tt epchain} and {\tt emchain} tasks of the {\tt SAS} v18.0.0 package. We retained only unflagged events with {\tt PATTERN $\leq 4$} (PN) and {\tt PATTERN $\leq 12$} (MOS). To account for the background flaring events, we extracted a light curve in the high-energy band ($E \ge 10$ keV). Good-time-interval (GTI) files were then generated from this light curve using the task {\tt TABGTIGEN}, applying an appropriate count rate threshold (cts$ = 0.5$) to identify and include events not affected by background flaring.
Source spectra and light curves were extracted from a 15 arcsec circular region centered on the source, while background events were obtained from a concentric annulus with radii 25–75 arcsec. The response files were generated with {\tt ARFGEN} and {\tt RMFGEN}, and background-corrected light curves were produced using {\tt epiclccorr}.

\section{Timing Analysis}
\label{timing}
We begin our analysis of Mrk~530 using the light curves extracted from both \textit{XMM-Newton} and \textit{Swift} observations. For its high signal-to-noise ratio and continuous coverage for long intervals in each observation, the \textit{XMM-Newton} data are used to investigate short-term variability of the source. In contrast, the \textit{Swift} monitoring of Mrk~530 spans nearly two decades and is used to probe the long-term temporal behavior.

\subsection{Short-term Variability}
\label{timing_xmm}
Timing analysis of Mrk~530 was performed on the X-ray light curves from the {\it XMM-Newton} observations. We extracted light curves in the soft (0.3--2 keV) and hard (2--10 keV) X-ray energy bands, binned at 500~s, and further divided them into narrow energy ranges to probe inter-band correlations. The 2--6 keV continuum, likely produced in a hot Compton cloud \citep{1980A&A....86..121S,1985A&A...143..374S}, was split into 2--3, 3--4, 4--5, and 5--6 keV sub-bands to investigate its origin. The 6--7 keV band is generally dominated by Fe~K$\alpha$ emission \citep{1991MNRAS.249..352G,1993MNRAS.262..179M}. The 7--10 keV range approaches the reflection-dominated regime and was further divided into 7--8 and 8--10 keV sub-bands. Emission below $\sim$2 keV may be affected by the presence of soft excess and absorption features. To study them, we divided this energy range into 0.3--0.5, 0.5--1.0, and 1.0--2.0 keV bands for this work (see Figure~\ref{fig:lightcurves}). This fine energy binning enables disentangling contributions from the disc, corona, reflection, and soft excess. 

\subsubsection{Fractional Variability}
\label{sec:F_var}

To quantify the temporal variability of Mrk~530, we calculated the fractional variability $F_{\rm var}$ \citep{1996ApJ...470..364E,1997ApJ...476...70N,2003MNRAS.345.1271V,2023MNRAS.521.5440K,2024MNRAS.528.5269L} across different energy bands. For a light curve with $N$ data points of count rate $x_i$, error $\sigma_i$, mean $\mu$, and variance $\sigma^2$, the intrinsic excess variance is

\begin{equation}
\sigma^2_{\rm XS}=\sigma^2-\frac{1}{N}\sum_{i=1}^N\sigma_i^2, 
\end{equation}
and the fractional variability is
\begin{equation}
F_{\rm var}=\sqrt{\frac{\sigma^2_{\rm XS}}{\mu^2}}.
\end{equation}
The normalized excess variance is $\sigma^2_{\rm NXS}=\sigma^2_{\rm XS}/\mu^2$, with uncertainties in both $\sigma^2_{\rm NXS}$ and $F_{\rm var}$ estimated following \cite{2003MNRAS.345.1271V}. In addition, the peak-to-peak amplitude is defined as $R=x_{\rm max}/x_{\rm min}$, where $x_{\rm max}$ and $x_{\rm min}$ are the maximum and minimum count rates, respectively. The derived parameters from the variability analysis of the {\it XMM-Newton} observations are listed in Table~\ref{fvar}.

The XMM1 and XMM2 data show large uncertainties featuring negative $\sigma^2_{\rm NXS}$ across most bands, indicating that the variability was indistinguishable from statistical noise. The overall $F_{\rm var}$ is generally low, suggesting minimal flux fluctuations over the observation timescale. Exceptions include the 5.0--6.0 keV band in XMM1 ($F_{\rm var}\sim37.1\pm9.0\%$) and the 8.0--10.0 keV band in XMM2 ($F_{\rm var}\sim14.5\pm21.4\%$), where $\sigma^2_{\rm NXS}$ is positive, although the uncertainties remain high. In the soft X-ray band ($<2$ keV), XMM2 shows modest variability ($F_{\rm var}\sim2.2\pm0.6\%$) well below 10\%, suggesting a relatively stable soft excess plausibly arising from a warm Comptonizing corona \citep{2009MNRAS.394..250M, Done2012, Petrucci2020}, ionized reflection from the accretion disc \citep{2002MNRAS.331L..35F, 2005MNRAS.358..211R, 2010ApJ...718..695G} or due to the presence of a distant neutral absorber.

\begin{figure}
    \centering
    \includegraphics[width=\columnwidth]{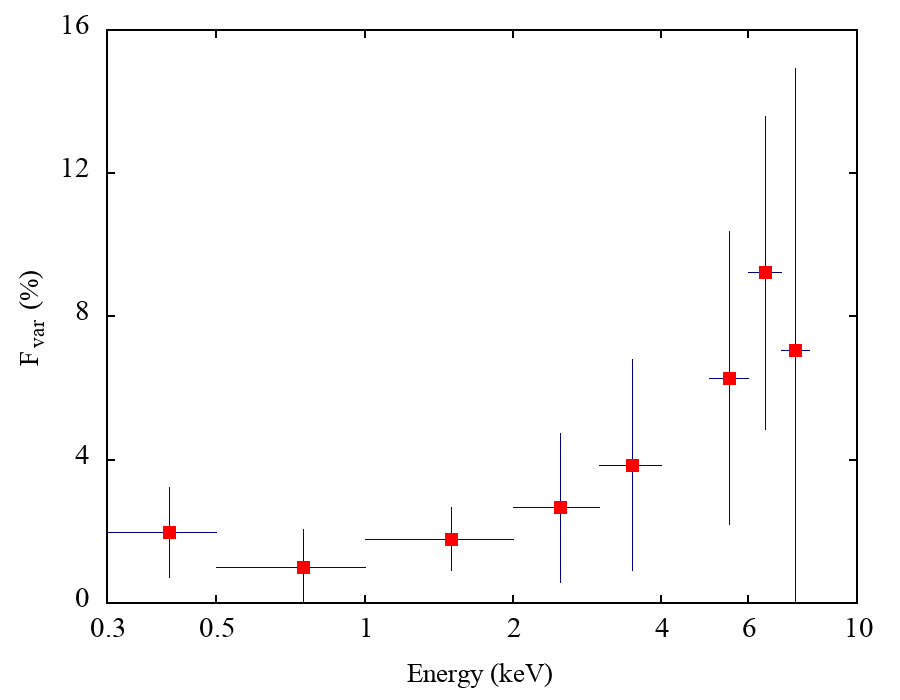}
    \caption{Energy-dependent fractional variability of Mrk~530 in the 2006 {\it XMM-Newton} (XMM3) observation.}
    \label{fig:fvar2006}
\end{figure}

In the XMM3 observation, the variability across most of the energy bands is better constrained and shows a clear increase in value towards the hard X-ray regime (Figure~\ref{fig:fvar2006}). The 6.0--7.0 keV band is found to be the most variable ($F_{\rm var}\sim9.2\pm4.4\%$), likely due to changes in the Fe~K$\alpha$ line. The primary continuum (2.0--10.0 keV) shows moderate variability ($F_{\rm var}\sim4.2\pm1.0\%$) consistent with intrinsic fluctuations of the hot corona, while the soft band (0.3--2.0 keV) shows lower variability ($F_{\rm var}\sim1.5\pm0.4\%$). This supports a relatively stable soft excess component, consistent with the earlier 2001 results. We also note that the results from this variability analysis are insensitive to the time bin (500s), with both higher and lower binning of the data yielding results that are consistent within uncertainties.

The variability characteristics observed in the {\it XMM-Newton} observations likely depend on the geometry and the activities around the central engine \citep{Ponti2012, Middei2017}. The absence of variability in the XMM1 and XMM2 observations suggests a stable corona and soft excess, or that intrinsic variations were hidden by large statistical uncertainties. Such low variability is typical of low-flux states or reduced accretion activity \citep{Uttley2003, Paolillo2004}. In contrast, XMM3 captures the source in a more dynamic state (Table~\ref{tab:Fvar}). Moderate variability in the 2--10 keV continuum, with the amplitude increasing towards higher energies, is consistent with enhanced Comptonisation producing more energetic photons \citep{2000ApJ...531L..41C}. These variations originate in the hot corona, while the soft band ($<2$ keV) shows weaker variability, consistent with a relatively stable soft excess \citep{Kammoun2015, Gliozzi2020, Ding2022}. This component is often linked to a warm, optically thick corona \citep{Done2012, noda2018explaining, Petrucci2020} or ionized reflection \citep{2005MNRAS.358..211R, 2006MNRAS.365.1067C, 2010ApJ...718..695G}. Both scenarios are expected to show smoother or slower variability compared to the rapidly changing hot corona. The shift of nearly non-variable (XMM1 \& XMM2) to a low-variable (XMM3) state may therefore reflect changes in accretion rate, coronal heating efficiency, or reflector geometry. To better understand, we present a detailed spectral analysis in the following sections (Section~\ref{spectral}). 

 \begin{table*}
	\centering
	\caption{Variability statistics in different energy ranges for the {\it XMM-Newton} observations, calculated using light curves with 500~s time bin.}
	\label{tab:Fvar}
	\begin{tabular}{c c c c c c c c c c} 
	\hline
ID & Energy band & N & $x_{\rm max}$ &  $x_{\rm min}$ & $\mu$ & $R=\frac{x_{\rm max}}{x_{\rm min}}$ & $\sigma^{2}_{\rm NXS}$ &  $F_{\rm var}$ \\
 & (keV)  &  & (count $s^{-1})$ & (count $s^{-1}$) & (count $s^{-1}$)    &    & ($10^{-2}$)     &($\%$) \\
 \hline
      & 0.3 - 0.5 & 24 & 0.83   & 0.56   & 0.71 & 1.47   & -0.08 & - \\
      & 0.5 - 1.0 & 24 & 1.56 & 1.20 & 1.41 & 1.30 & -0.19 & - \\
      & 1.0 - 2.0 & 24 & 1.93 & 1.32 & 1.59 & 1.46 & $0.31 \pm 0.19$ & $5.56 \pm 1.90$ \\
      & 2.0 - 3.0 & 24 & 0.54 & 0.35 & 0.41 & 1.55 & -0.03 & -\\
      & 3.0 - 4.0 & 24 & 0.26 & 0.15 & 0.21 & 1.70 & -1.41 & - \\
XMM1 & 4.0 - 5.0 & 24 & 0.18 & 0.09 & 0.13 & 1.94 & -2.59 & - \\
      & 5.0 - 6.0 & 24 & 0.15 & 0.02 & 0.08 & 8.74 & $13.77\pm5.34$ & $37.11\pm8.97$ \\
      & 6.0 - 7.0 & 24 & 0.09 & 0.02 & 0.05 & 4.15 & -5.75 & - \\
      & 7.0 - 8.0 & 24 & 0.07 & 0.01 & 0.02 & 9.42 & -1.84 & - \\
      & 8.0 - 10.0 & 24 & 0.05 & 0.01 & 0.02 & 10.16 & -25.40 & - \\
      & 0.3 - 2.0 & 24 & 3.91 & 3.39 & 3.69 & 1.16 & -0.02 & - \\
      & 2.0 - 10.0 & 24 & 1.05 & 0.80 & 0.93 & 1.32 & -0.07 & - \\
                       \hline
      & 0.3 - 0.5 & 20 & 4.47 & 3.78 & 3.98 & 1.18  & $0.01\pm0.06$  & $0.74\pm3.76$ \\
      & 0.5 - 1.0 & 20 & 6.79 & 5.77 & 6.13 & 1.18 & $0.09 \pm 0.05$ & $3.01 \pm 1.00$ \\
      & 1.0 - 2.0 & 20 & 4.57 & 3.90 & 4.16 & 1.17 & $0.02 \pm 0.05$ & $1.42 \pm 1.86$ \\
      & 2.0 - 3.0 & 20 & 1.13 & 0.83 & 0.99 & 1.38 & -0.01 & -\\
      & 3.0 - 4.0 & 20 & 0.48 & 0.32 & 0.41 & 1.49 & -0.06 & - \\
XMM2 & 4.0 - 5.0 & 20 & 0.32 & 0.21 & 0.25 & 1.53 & -0.94 & - \\
      & 5.0 - 6.0 & 20 & 0.26 & 0.13 & 0.19 & 2.06 & -0.04 & - \\
      & 6.0 - 7.0 & 20 & 0.18 & 0.09 & 0.13 & 1.94 & -0.04 & - \\
      & 7.0 - 8.0 & 20 & 0.13 & 0.01 & 0.08 & 13.90 & $0.37 \pm 3.94$ & $6.12 \pm 32.16$ \\
      & 8.0 - 10.0 & 20 & 0.17 & 0.03 & 0.08 & 6.00 & $2.10 \pm 6.18$ & $14.50 \pm 21.42$ \\
      & 0.3 - 2.0 & 20 & 15.51 & 13.89 & 14.47 & 1.11 & $0.05 \pm 0.02$ & $2.21 \pm 0.65$ \\
      & 2.0 - 10.0 & 20 & 2.32 & 2.05 & 2.17 & 1.13 & -0.20 & - \\
     \hline
      & 0.3 - 0.5  & 33 & 2.83   & 2.40   & 2.61 & 1.18   & $0.04 \pm 0.05$ & $1.98 \pm 1.26$ \\
      & 0.5 - 1.0 & 33 & 5.24 & 4.62 & 4.98 & 1.14 & $0.01 \pm 0.02$ & $1.01  \pm 1.06$ \\
      & 1.0 - 2.0 & 33 & 4.27 & 3.74 & 4.01 & 1.14 & $0.04 \pm 0.03$ & $1.80 \pm 0.87$ \\
      & 2.0 - 3.0 & 33 & 1.14 & 0.90 & 1.01 & 1.28 & $0.04 \pm 1.0$ & $2.67 \pm 2.07$\\
      & 3.0 - 4.0 & 33 & 0.60 & 0.42 & 0.50 & 1.46 & $0.15 \pm 0.22$ & $3.86 \pm 2.94$ \\
XMM3 & 4.0 - 5.0 & 33 & 0.40 & 0.27 & 0.33 & 1.51 & -0.09 & - \\
      & 5.0 - 6.0 & 33 & 0.30 & 0.17 & 0.23 & 1.79 & $0.40 \pm 0.51$ & $6.29 \pm 4.09$ \\
      & 6.0 - 7.0 & 33 & 0.24 & 0.11 & 0.16 & 2.13 & $0.85 \pm 0.78$ & $9.23 \pm 4.38$ \\
      & 7.0 - 8.0 & 33 & 0.14 & 0.05 & 0.10 & 2.91 & $0.50 \pm 1.10$ & $7.07 \pm 7.88$ \\
      & 8.0 - 10.0 & 33 & 0.12 & 0.05 & 0.08 & 2.12 & -2.67 & - \\
      & 0.3 - 2.0 & 33 & 12.14 & 11.13 & 11.62 & 1.09 & $0.02 \pm 0.01 $ & $1.54 \pm 0.45$ \\
      & 2.0 - 10.0 & 33 & 2.77 & 2.18 & 2.44 & 1.27 & $0.18 \pm 0.06$ & $4.19 \pm 0.98$ \\
     \hline
     \label{fvar}
       \end{tabular}
\end{table*}

\subsubsection{Correlation Analysis}
\label{correlation}

We perform cross-correlation analysis between the different X-ray energy bands mentioned in Section~\ref{timing_xmm} using the {\it XMM-Newton} observations. Such a study can help us to understand the connection between different components of the X-ray spectrum. This analysis employed the $\zeta$-transformed discrete correlation function (ZDCF)\footnote{\url{https://www.weizmann.ac.il/particle/tal/research-activities/software}} \citep{1997ASSL..218..163A, alexander2013improved}, an improved form of the discrete correlation function (DCF; \cite{edelson1988discrete}) that is well suited to sparse, unevenly sampled data. To estimate the uncertainties in correlation coefficients and lag, we carried out 108,000 Monte Carlo simulations of the light curves.

For the 2001 and 2006 observations, {\it XMM-Newton} data suffered from low exposure times (a few kiloseconds) for this AGN (see Table~\ref{tab:log}). Consequently, the resulting lightcurves contain a few observational data points and are dominated by high uncertainties (see Figure~\ref{fig:lightcurves}). For correlation analysis, we adopted the 3–4 keV band as the reference, which is dominated by Comptonized photons from the hot corona. Using the ZDCF method, we cross-correlated this band with other energy ranges across XMM1, XMM2, and XMM3, but no significant correlation was detected. The typical timescale for the variability of Compton cloud in Seyfert AGNs is in the order of $10^4 - 10^5$ seconds, consistent with thermal timescales of hot corona \citep{Hu2022}. The short exposures in the {\it XMM-Newton} observations are insufficient to sample such fluctuations robustly. In such low-exposure regimes, the variability signature can be dominated by stochastic fluctuations rather than physical connections between energy bands \citep{Beckmann2007}. The presence of large statistical uncertainties further complicates the detection of inter-band lags or reverberation effects. As a result, any reverberation or other effects that may be present in the X-ray spectra of Mrk~530 remain undetectable in these observations. 

\subsection{Long-term Variability}
\label{sec:2018_lc}
We examined the long-term variability of Mrk~530 during the 2018 campaign with the \textit{Swift} observatory, which provided 45 intervals of XRT and UVOT coverage. For this work, we used the one-year light curves from both instruments. The XRT light curve was extracted in the 0.3–10 keV band with a bin size of 1 day (86.4 ks). For the UVOT data, only the UVM2 filter was available with a total of 42 observations. We constructed the UVM2 light curve with the same bin size. The resulting UV and X-ray light curves are presented in Figure~\ref{fig:wwz}.

\subsubsection{Periodicity Analysis}
We performed temporal analysis on the X-ray and UV light curves of Mrk~530 to probe long-term variability. The weighted average count rates were $26.77\pm0.53$ (UV) and $0.84\pm0.05$ (X-ray). A constant model fit using these averages yielded poor fits, confirming variability in both bands. The fractional variabilities were estimated at $19.56\pm2.51\%$ (UV) and $32.88\pm3.99\%$ (X-ray), with the stronger X-ray variability consistent with emission from a compact hot corona close to the black hole \citep{Panagiotou2022}. During the analysis, we observed that the UV/X-ray light curves display a quasi-sinusoidal behaviour. The UV light curve shows two unequal peaks, while the X-ray light curve exhibits two peaks of equal amplitude. We modelled them with a sine function, $y=A\sin(\omega t+\phi)+c$, with best-fit parameters listed in Table~\ref{tab:sine} and fits shown in Figure~\ref{fig:wwz}. We find that the UV and X-ray variations do not occur on the same timescale. The periodicity estimated from the UV light curve is $87.15\pm2.44$ days, whereas the X-ray light curve yields a shorter period of $59.34\pm2.44$ days. Furthermore, we found that the UV and X-ray lightcurves have a clear phase offset \citep{2018MNRAS.474.5351P}. The light curves were found to have a phase difference of $0.60$, corresponding to a time delay of $6.75\pm2.44$ days. To examine and quantify the possibility of the presence of a possible QPO in the UV and X-ray light curves, we use the Lomb-Scargle (LS) Periodogram \citep{lomb1976least,scargle1982studies} and the weighted wavelet Z-transform (WWZ; \cite{1996AJ....112.1709F}) analysis techniques. 

\subsubsection{Lomb-Scargle Periodogram}
The Lomb–Scargle (LS) periodogram \citep{lomb1976least, scargle1982studies} is widely used to search for periodicities in irregularly sampled data by fitting sine waves with a $\chi^2$ statistic. This method reduces the effects of uneven sampling and can reveal periodic or quasi-periodic signals present in the data. For a trial frequency $\omega$, the LS power from time series values $y_i$ at $t_i$ is given by, 

\begin{equation}
    P(\omega)=\frac{1}{2}\left[\frac{\left[\Sigma^{N}_{i=1}y_i~cos~\omega(t_i-\tau)\right]^2}{\Sigma^{N}_{i=1}cos^2~\omega(t_i-\tau)}+\frac{\left[\Sigma^{N}_{i=1}y_i~sin~\omega(t_i-\tau)\right]^2}{\Sigma^{N}_{i=1}sin^2~\omega(t_i-\tau)}\right], 
\end{equation} \\
where $\tau$ is the phase correction,
\begin{equation}
    tan~(2\omega \tau)=\frac{\Sigma_{i=1}^{N}sin~(2\omega t_i)}{\Sigma_{i=1}^{N}cos~(2\omega t_i)}.
\end{equation}
The period corresponding to the highest power is the best-fit period.

The periodogram power spectral density (PSD) for Mrk~530 is shown in Figure~\ref{fig:wwz}. To test the significance of the result, we generated $10^4$ artificial light curves with the same PSD using the method of \cite{1995A&A...300..707T}, where the original PSD was fit with a power law. The resulting distribution of powers provided confidence intervals, with 68\%, 95\%, and 99\% levels, which are overplotted in Figure \ref{fig:wwz}. The observed peaks lie well above the 99\% global confidence threshold, confirming their statistical significance. 

The findings from the LS-periodogram analysis of the 2018 X-ray and UV light curves remain consistent with the periods obtained from the sine function fitting of the data. The periods are estimated at $\sim$91 days ($\omega=0.69\pm0.42\times10^{-1}$ rad/day) for the UV light curve and $\sim$61 days ($\omega=1.03\pm0.36\times10^{-1}$ rad/day) for the X-ray light curve (Table~\ref{tab:sine}). The uncertainties are calculated by fitting a Gaussian function around the respective maxima. The LS periodograms feature a broad peak well above the 99\% confidence level for both the UV and the X-ray light curves.

\begin{table}
    \centering
    \caption{Parameters obtained from the periodicity analysis of X-ray and UV light curves of Mrk~530 in 2018. When the estimated error in the period was smaller than the average light curve sampling interval ($\Delta t = 2.44$ days), we adopted $\Delta t$ as the uncertainty.}
    \resizebox{\columnwidth}{!}{\begin{tabular}{c|c|c|c|c}
    \hline\\
        & Parameters & Sine Fit & LS-Periodogram & WWZ \\
         \hline\\\vspace{0.2cm}
         & Amplitude (A) & $7.47\pm0.14$ & - & $109.00\pm7.73$\\\vspace{0.2cm}
        & Frequency ($\omega;~\times10^{-1}$) & $0.72\pm0.01$ & $0.69\pm0.42$ & $0.70\pm0.06$\\\vspace{0.2cm}
    UV & Period (days) & $87.15\pm2.44$ & $91.06\pm55.43$ & $89.76\pm7.69$\\\vspace{0.2cm}
        & Offset (c) &$28.13\pm0.13$& - & -\\\vspace{0.2cm}
        & Phase ($\phi$) &$-2.58\pm0.02$ & - & -\\
        \hline\\\vspace{0.2cm}
        & Amplitude (A) & $0.33\pm0.01$& - & $30.17\pm0.36$\\\vspace{0.2cm}
        & Frequency ($\omega;~\times10^{-1}$) &$1.06\pm0.01$ & $1.03\pm0.36$ & $1.06\pm0.06$\\\vspace{0.2cm}
    X-ray & Period (days) &$59.34\pm2.44$ & $61.06\pm21.32$ & $59.28\pm3.40$\\\vspace{0.2cm}
        & Offset (c) & $0.74\pm0.01$ &- & -\\\vspace{0.2cm}
        & Phase ($\phi$) & $-1.85\pm0.03$ & -& -\\
    \hline    
    \end{tabular}}

    \label{tab:sine}
\end{table}

\subsubsection{Weighted Wavelet Z-transform}
\begin{figure*}
    \centering

    \begin{minipage}[b]{0.48\linewidth}
        \centering
        \includegraphics[width=\linewidth]{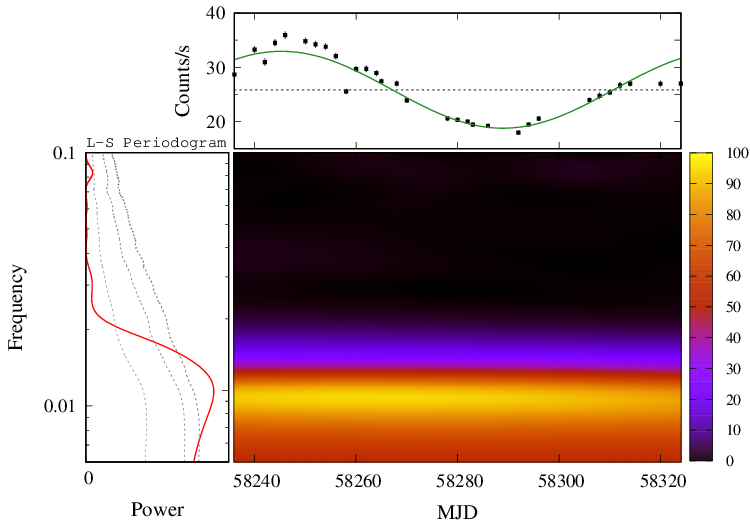}
    \end{minipage}
    \hspace{0.1cm}
    \begin{minipage}[b]{0.48\linewidth}
        \centering
        \includegraphics[width=\linewidth]{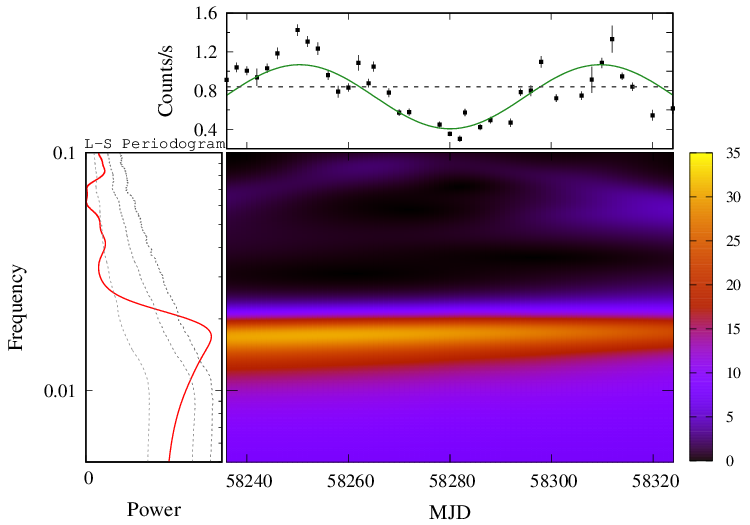}
    \end{minipage}
    \hspace{0.1cm}
    \caption{Plots obtained from the timing analysis of Mrk~530 in the 2018 \textit{Swift/}XRT \& UVOT observations. The upper panel shows the lightcurves in the UV (left) and X-ray (right) energy bands. The grey line depicts the average counts/s, and the green curve is the sine function fit. The lower left panel shows the results from the LS-periodogram along with $1\sigma$, $2\sigma$, and $3\sigma$ confidence levels estimated from simulating surrogate light curves. The contour plot represents the corresponding WWZ response.} 
    \label{fig:wwz}
\end{figure*}

Sometimes, a quasi-periodic signal may not remain constant throughout the observation. This can occur if the QPO evolves with time in its amplitude and/or frequency. The wavelet technique characterizes such non-stationary periodicity in a time series by decomposing it simultaneously in both frequency and time domains \citep{Torrence1998}. In this context, the weighted wavelet $z$-transform (WWZ) is a commonly used method for quantifying periodicities in time series by employing the $z$-statistic \citep{1996AJ....112.1709F}. The color-scaled density plot of WWZ power (Figure~\ref{fig:wwz}) shows a period of $\sim$90 days ($\omega=0.70\pm0.06\times10^{-1}$ rad/day) for the UV light curve and a period of $\sim$59 days ($\omega=1.06\pm0.06\times10^{-1}$ rad/day) for the X-ray light curve (Table~\ref{tab:sine}).

The distribution of the QPO power in the WWZ analysis of the UV light curve is uniform across the observing window, indicating a temporally coherent modulation. In contrast, the WWZ analysis of the X-ray light curve exhibits an asymmetric and time-localized structure. This behaviour suggests that the UV variability likely originates from a more extended and stable emitting region, while the X-ray variability is associated with a compact corona or inner accretion flow \citep{Poutanen1997, Ingram2009}. In such regions, rapid structural or radiative changes can naturally distort the X-ray QPO signal, whereas oscillations arising at larger radii, responsible for the UV emission, are expected to remain comparatively stable \citep{1996ApJ...457..805M}. Although the periodic modulations are detected in both the UV and X-ray bands, their coherence is found to be weak. The coherence of a QPO quantifies how sharply defined the oscillation is in the frequency domain and is expressed by the quality factor $Q=\nu_0/\Delta \nu_{FWHM}$, where $\nu_0$ is the central frequency at which the QPO signal peaks and $\Delta \nu_{FWHM}$ is the corresponding width. Physically, a high value of $Q$ indicates a narrow, well-defined oscillation that persists coherently over many cycles, whereas a low $Q$ value implies a broad peak corresponding to a weakly coherent or short-lived modulation. In our case, the quality factors of the periodic variations found in the UV and X-ray light curves are calculated to be $Q_{UV}\approx1.0~\text{and}~Q_{X}\approx1.6$, which classify them as low-coherence oscillations. Furthermore, fewer than two full cycles are observed over the available light curves. As a result, we consider these features as QPO candidates rather than established quasi-periodic oscillations.

\subsubsection{Correlation in Long-term Study}
The 2018 observations revealed distinct timescales for UV and X-ray photons, with periodicities of $\sim90$ days and $\sim59$ days, respectively. A clear phase offset was observed between the light curves, corresponding to a delay of $6.75\pm2.44$ days. While these results suggest differences in the physical origin or propagation of variability in the two bands, they do not address whether these variations are physically connected. To investigate this, we applied the discrete correlation function (DCF; \citealt{edelson1988discrete}) and the $\zeta$-transformed discrete correlation function (ZDCF; \citealt{1997ASSL..218..163A, alexander2013improved}) methods to the UV and X-ray light curves. 

\begin{figure}
    \centering
    \includegraphics[width=\columnwidth]{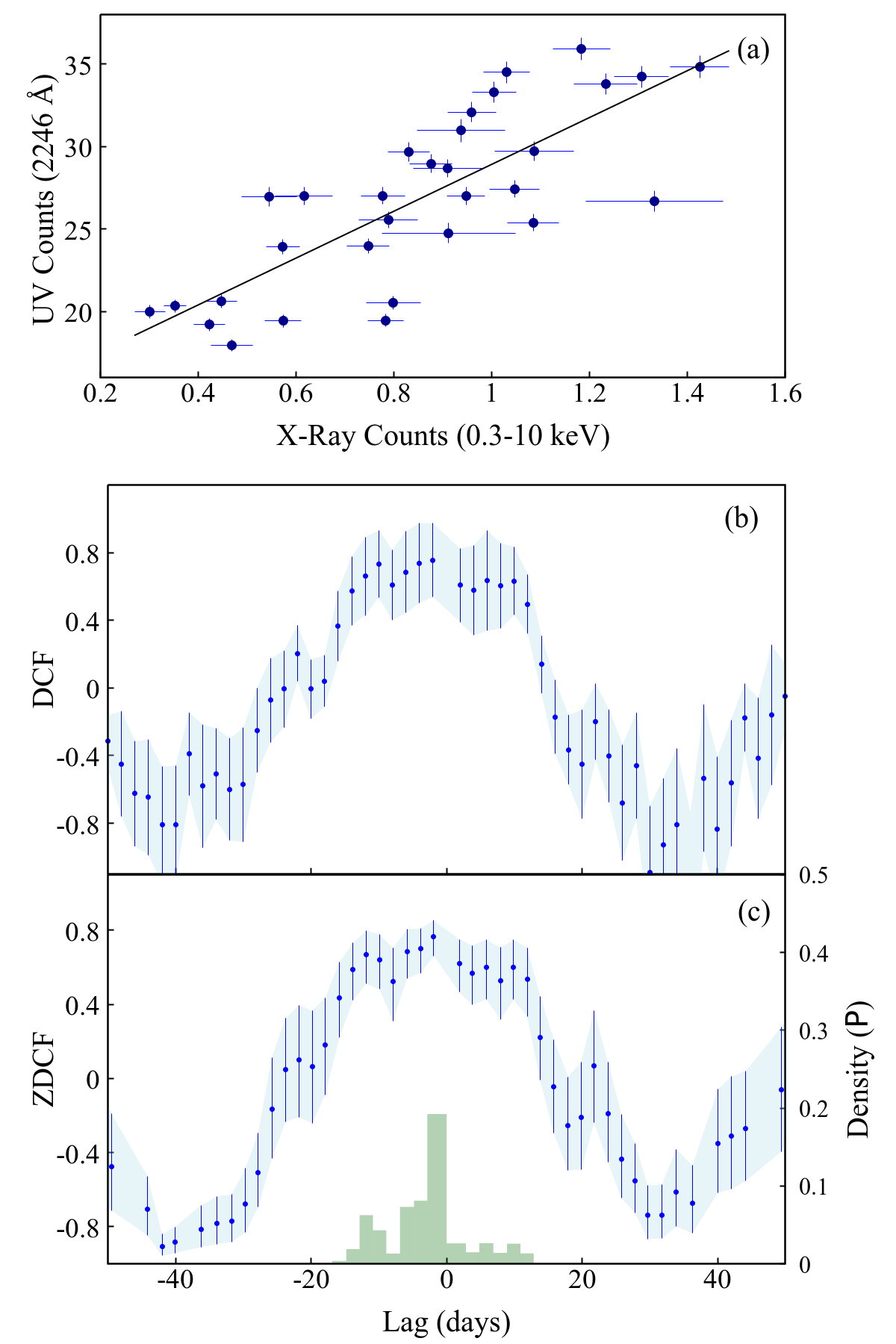}
    \caption{Correlation plots between the X-ray and the UV light curves of Mrk~530 in 2018. The count-count plot is presented in the top panel (a). The corresponding DCF and ZDCF plots are shown in the bottom panels (b) \& (c), respectively. The likelihood of each ZDCF peak is shown in light green.}
    \label{fig:zdcf}
\end{figure}

The strong Pearson correlation coefficient\footnote{\url{https://www.socscistatistics.com/tests/pearson/default2.aspx}} ($PCC\approx0.79$, $p-$value < 0.01) between the UV and X-ray light curves (Figure~\ref{fig:zdcf}a) indicates a close connection between variability in the two bands. The DCF shows a peak correlation of $0.75\pm0.24$ at a time delay of $-4.0\pm2.0$ days (Figure~\ref{fig:zdcf}b), while the ZDCF analysis yields a peak correlation of $0.77^{+0.09}_{-0.10}$ with a time delay of $-1.91^{+0.90}_{-0.10}$ days (Figure~\ref{fig:zdcf}c). In both cases, the negative lag indicates that variations in the UV band lead those in the X-rays. This behaviour implies that changes in the lower-energy emission precede the response at higher energies, consistent with scenarios in which perturbations propagate through the emitting regions. Alternatively, differences in radiative cooling and Compton scattering timescales may also introduce similar inter-band delays. However, a strong inter-band correlation and a measurable time lag, by themselves, do not uniquely constrain the geometry or spatial co-location of the emitting regions. This is because multiple physical scenarios, such as the propagation of accretion-rate fluctuations or complex transfer functions, can reproduce similar correlation signatures \citep{AU2006, 2014xru..confE..25U}. Finally, we caution the fact that although the peak correlations are comparable within uncertainties, the inferred time lags from the two methods are not mutually consistent, likely due to the broadness of the correlation peak. Hence, we note that this lag estimate should be interpreted with care.

The {\it Swift}/XRT monitoring obtained in 2023 and 2024 consists of only $\sim16$ observations per year with a cadence of approximately one week. Given the characteristic variability timescales inferred from the 2018 campaign ($\sim$60--90 days), these datasets are insufficient to robustly assess the persistence of the modulation. Consistently, Lomb–Scargle periodogram analysis of the 2023 and 2024 XRT light curves does not reveal any statistically significant periodic signal at the previously inferred timescales (Figure~\ref{fig:periodogram}).

\section{Spectral Analysis}
\label{spectral}
To investigate spectral properties in the 0.3--10 keV energy range of Mrk~530 over ~24 years (2001--2024), we use \textit{XMM-Newton} and \textit{Swift}/XRT data. Spectral analysis is performed in \textsc{XSPEC} v12.14.1 \citep{1996ASPC..101...17A} using $\chi^{2}$ statistics with a minimum of 20 counts per bin. To estimate the uncertainties in each spectral parameter, we use the \texttt{error} command and the Monte Carlo Markov Chain (MCMC) method in \textsc{XSPEC} software at $90\%$ confidence level. The unabsorbed X-ray luminosities are calculated using the \texttt{clumin} command in \textsc{XSPEC} applied to the \texttt{powerlaw} model. We adopt cosmological parameters $H_0=75~\rm km~s^{-1}~Mpc^{-1}$ and $q_0=0$ following \cite{2011ApJ...731...68L}.   

\subsection{Characterization of X-ray Spectrum}

To date, no comprehensive long-term X-ray spectral study of Mrk~530 has been performed. The only dedicated X-ray spectral analysis of this source was performed by \citet{2018MNRAS.478.4214E}, in which a single, long $(\sim101~\text{ks})$ \textit{Suzaku} observation in 2012 was analysed, and a prominent soft excess component was reported. To investigate the long-term spectral behaviour of Mrk~530, we analyze all available X-ray observations acquired between 2001 and 2024. As a first step, we apply phenomenological models to each epoch in order to characterize the gross spectral properties, including the continuum slope. We subsequently replace these empirical descriptions with physically motivated models to explore the accretion-related processes responsible for shaping the observed spectral characteristics.

\begin{figure}
    \centering
    \includegraphics[width=\columnwidth]{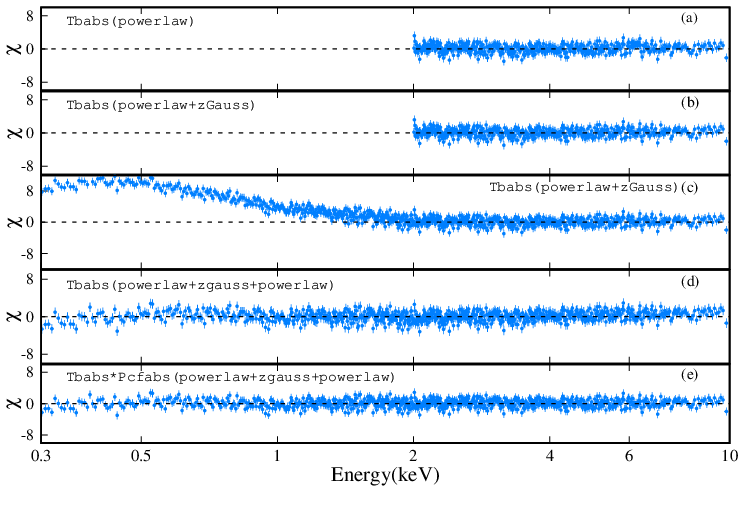}
    \caption{The residuals obtained during the phenomenological spectral fitting of the 2006 XMM3 observation of Mrk~530.}
    \label{fig:chi_plot}
\end{figure}

\subsection{Phenomenological Models}
\label{sec:pheno}
Initially, the primary continuum of Mrk~530 in the 2--10 keV energy range was taken into account to fit the data using a phenomenological model. According to theory, photons in this range arise from inverse Compton scattering of seed photons by a hot electron cloud \citep{1980A&A....86..121S}, producing non-thermal emission. Thus, we adopted a {\tt Powerlaw} model for the continuum, including Galactic absorption via {\tt TBabs} with a fixed $N_H=3.84\times10^{20}~\rm cm^{-2}$, obtained from the HEASARC $N_H$ calculator\footnote{\url{https://heasarc.gsfc.nasa.gov/cgi-bin/Tools/w3nh/w3nh.pl}}. Spectral fitting of {\it XMM-Newton} data featured positive residuals around the 6--7 keV energy range (Figure~\ref{fig:chi_plot}a), indicative of an Fe~K$\alpha$ line, which was modeled with a Gaussian component ({\tt zGauss}) added to the {\tt Powerlaw}. The baseline model for the {\it XMM-Newton} spectra is therefore {\tt TBabs $\times$ (Powerlaw + zGauss)} (Figure~\ref{fig:chi_plot}b). In contrast, the {\it Swift}/XRT spectra do not show any evidence of an Fe~K$\alpha$ line. This absence could be attributed to the limited spectral resolution and shorter exposures of {\it Swift}/XRT, or it may indicate that the Fe line was intrinsically weak or absent during these observations. Consequently, the {\tt zGauss} component was excluded when modeling the {\it Swift}/XRT spectra.

\begin{figure*}
\centering
\includegraphics[trim={0 0cm 0cm 0},scale=1.0]{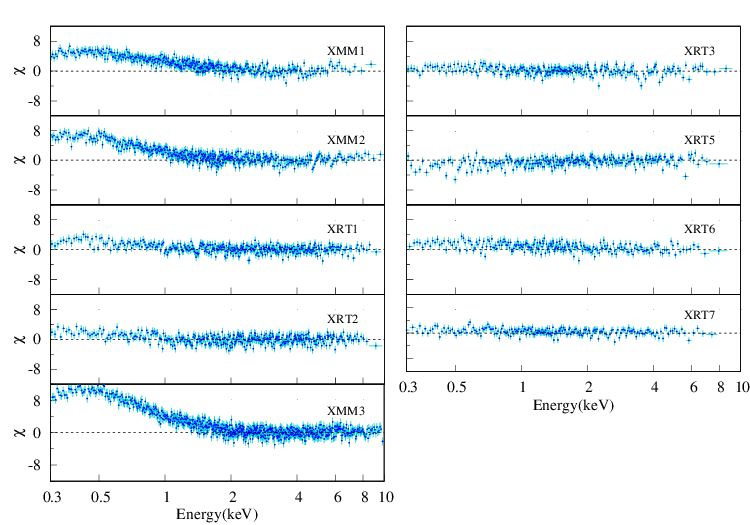}
    \caption{Residuals showing the variation in soft excess in Mrk~530 over the duration of 24 years of X-ray observations. All spectra are fitted with a Galactic absorbed power-law in the 2--10 keV energy range and then extended to lower energies.}
\label{fig:softexcess}
\end{figure*}

After modeling the primary continuum, we extended the fits to the full 0.3–10.0 keV range (Figure~\ref{fig:chi_plot}c), revealing distinct features across observations (Figure~\ref{fig:softexcess}). The soft excess is prominent in the XMM1 observation, while its strength gradually decreases from XMM1 to XMM2, XRT1, XRT2, before increasing in XMM3. To model this component, an additional {\tt powerlaw} was added to the baseline broadband model (Figure~\ref{fig:chi_plot}d). As low-energy photons (<2 keV) are affected by line-of-sight absorption, we included a partial covering absorber ({\tt Pcfabs}) along with Galactic absorption (Figure~\ref{fig:chi_plot}e). Therefore, the final phenomenological model used to fit the entire X-ray spectrum of Mrk~530 over the 0.3–10.0 keV range is: {\tt TBabs} $\times$ {\tt Pcfabs} $\times$ ({\tt powerlaw + zGauss + powerlaw}). As in the case of the Fe K$\alpha$ line, the presence of an intrinsic absorber is clearly detected in {\it XMM-Newton} observations with parameters that are marginally consistent within uncertainties across epochs (2001-2006). A simultaneous fit to the three {\it XMM-Newton} spectra gives a hydrogen column density of $N_H=26.00^{+9.31}_{-6.75}\times10^{22}~cm^{-2}$ and a covering fraction of $C_f=0.34\pm0.05$. These best-fitting absorber parameters were therefore fixed for all observations obtained between 2001 and 2006. Once the spectra for the entire 0.3--10 keV energy range were modeled, we used the {\tt clumin} convolution model to compute the luminosities of the primary continuum ($L_{PC}$) and the soft excess $(L_{SE})$ component, both evaluated over the 0.3–10.0 keV band. The resulting phenomenological model parameters for all observations are reported in Table~\ref{tab_pheno}.

\begin{table*}
\caption{The best fit parameters for the baseline phenomenological model \texttt{TBabs}$\times$\texttt{Pcfabs}$\times$\texttt{(powerlaw+zGauss+powerlaw)} for all the observations of Mrk~530 used in this work. The soft excess (SE) and primary continuum (PC) luminosities are calculated in the energy range 0.3--10 keV.}
\resizebox{\textwidth}{!}{\begin{tabular}{c c c c c c c c c c c c}
\hline\\
ID & $N_H$ & $C_f$ & $\Gamma_{PC}$ & $Norm_{PC}^{\dagger}$ & $log ~L_{PC}$ & $\Gamma_{SE}$ & $Norm_{SE}^{\dagger}$ & $log~ L_{SE}$ & Fe k$\alpha$ & EW & $\chi^2/dof$ \\
               & ($10^{22}~cm^{-2}$) & & & $(10^{-3})$ & $log ~(erg s^{-1})$ & & $(10^{-3})$ & $log~ (erg s^{-1})$ & (keV) & (eV) &\\
\hline \\
\vspace{0.2cm}
XMM1 &  $26.00^{+9.31}_{-6.75}$ & $0.34^{+0.05}_{-0.05}$ & $1.73^{+0.10}_{-0.11}$ & $4.50^{+0.86}_{-1.07}$ & $43.71^{+0.08}_{-0.11}$ & $2.47^{+0.10}_{-0.10}$ & $8.37^{+1.18}_{-0.98}$ & $43.84^{+0.06}_{-0.05}$ & $6.38^{+0.11}_{-0.11}$ & $168^{+106}_{-107}$ & $442.01/418$\\
\vspace{0.2cm}
XMM2 & $26^f$ & $0.34^f$ & $2.06^{+0.03}_{-0.03}$ & $5.44^{+1.01}_{-1.44}$ & $43.70^{+0.07}_{-0.13}$ & $2.59^{+0.11}_{-0.11}$ & $6.14^{+1.52}_{-1.08}$ & $43.68^{+0.10}_{-0.07}$ & $6.43^{+0.07}_{-0.07}$ & $65^{+52}_{-51}$ & $591.13/502$\\
\vspace{0.2cm}
XRT1 & $26^f$ & $0.34^f$ & $1.84^{+0.09}_{-0.08}$ & $7.76^{+1.34}_{-2.99}$ & $43.91^{+0.07}_{-0.21}$ & $2.58^{+0.34}_{-0.32}$ & $4.65^{+3.22}_{-1.56}$ & $43.58^{+0.24}_{-0.16}$ & - & - & $200.35/212$\\
\vspace{0.2cm}
XRT2 & $26^f$ & $0.34^f$ & $1.87^{+0.07}_{-0.06}$ & $11.21^{+0.95}_{-1.52}$ & $44.06^{+0.04}_{-0.06}$ & $2.98^{+0.40}_{-0.36}$ & $3.57^{+1.79}_{-1.19}$ & $43.50^{+0.15}_{-0.11}$ & - & - & $233.07/215$\\
\vspace{0.2cm}
XMM3 & $26^f$ & $0.34^f$ & $1.91^{+0.02}_{-0.02}$ & $6.12^{+0.40}_{-0.45}$ & $43.78^{+0.03}_{-0.03}$ & $2.66^{+0.06}_{-0.06}$ & $6.73^{+0.50}_{-0.45}$ & $43.75^{+0.03}_{-0.03}$ & $6.39^{+0.06}_{-0.07}$ & $44^{+27}_{-23}$ & $844.24/806$\\
\vspace{0.2cm}
XRT3 & - & - & $1.92^{+0.10}_{-0.10}$ & $2.53^{+0.96}_{-0.96}$ & $43.40^{+0.14}_{-0.21}$ & $2.20^{+1.05}_{-0.21}$ & $2.23^{+2.74}_{-1.88}$ & $43.29^{+0.45}_{-0.76}$ & - & - & $91.86/82$\\ 
\vspace{0.2cm}
XRT5 & - & - & $1.63^{+0.05}_{-0.05}$ & $2.95^{+0.12}_{-0.12}$ & $43.56^{+0.02}_{-0.02}$ & - & - & - & - & - & $90.26/93$\\ 
\vspace{0.2cm}
XRT6 & - & - & $1.71^{+0.03}_{-0.03}$ & $3.88^{+0.28}_{-0.28}$ & $43.64^{+0.03}_{-0.03}$ & $2.22^{+0.56}_{-0.39}$ & $1.35^{+3.14}_{-0.74}$ & $43.06^{+0.68}_{-0.36}$ & - & - & $247.17/214$\\
\vspace{0.2cm}
XRT7 & - & - & $1.67^{+0.09}_{-0.10}$ & $3.72^{+0.58}_{-0.60}$ & $43.64^{+0.04}_{-0.04}$ & $2.42^{+1.20}_{-0.80}$ & $0.44^{+3.83}_{-0.34}$ & $42.58^{+1.09}_{-0.45}$ & - & - & $196.15/210$\\ 
\hline
\end{tabular}}
\leftline{\textdagger in the unit of photons/keV/$cm^2$/s}
\label{tab_pheno}
\end{table*}

\par

We begin our analysis with the 2001 {\it XMM-Newton} observation (XMM1). Spectral fitting results a primary continuum of photon index $\Gamma_{PC}=1.73^{+0.10}_{-0.11}$ and a corresponding luminosity of $log~L_{PC}=43.71^{+0.08}_{-0.11}$. A Fe~K$\alpha$ line is detected at $6.38\pm0.11$ keV with equivalent width $168\pm107$ eV. The soft excess is prominent, exhibiting the highest luminosity among all observations, $log~L_{SE}=43.84^{+0.06}_{-0.05}$, and a photon index of $\Gamma_{SE}=2.47\pm0.10$.

Five months later, the XMM2 observation shows a steeper primary continuum with $\Gamma_{PC}=2.06\pm0.03$, while the luminosity remains consistent (within uncertainties) at $log~L_{PC}=43.70^{+0.07}_{-0.13}$. The Fe~K$\alpha$ line is weaker at $6.43\pm0.07$ keV with a reduced equivalent width of $65\pm52$ eV. The soft excess in this epoch exhibits a marginally steeper photon index of $\Gamma_{SE}=2.59\pm0.11$ and a lower luminosity of $log~L_{SE}=43.68^{+0.10}_{-0.07}$. All parameter values obtained from the phenomenological model fitting are listed in Table~\ref{tab_pheno}, and their long-term variations are shown in Figure~\ref{fig:longterm}.

In 2006, Mrk~530 was observed in three epochs with \textit{Swift}/XRT (XRT1, XRT2) and \textit{XMM-Newton} (XMM3). The primary continuum remains relatively stable across these observations, with photon indices of $\Gamma_{PC}=1.84\pm0.09$ (XRT1), $1.87\pm0.07$ (XRT2), and $1.91\pm0.02$ (XMM3). The corresponding luminosities are $log~L_{PC}=43.91^{+0.07}_{-0.21}$, $44.06^{+0.04}_{-0.06}$, and $43.78^{+0.03}_{-0.03}$, respectively. The soft excess photon indices are also consistent (within uncertainties) with $\Gamma_{SE}=2.58^{+0.34}_{-0.32}$ (XRT1), $2.98^{+0.40}_{-0.36}$ (XRT2), and $2.66\pm0.06$ (XMM3). The corresponding soft excess luminosities show modest variations ranging from $log~L_{SE}=43.58^{+0.24}_{-0.16}$ (XRT1) to $43.50^{+0.15}_{-0.11}$ (XRT2) and $43.75\pm0.03$ (XMM3). An Fe~K$\alpha$ emission line is detected only in the XMM3 spectrum, at $6.39^{+0.06}_{-0.07}$ keV with equivalent width of $44^{+27}_{-23}$ eV. Overall, the 2006 observations indicate stable photon indices and soft excess slopes and modest luminosity variations.

 After a 10-year gap, Mrk~530 was observed again in 2016 with \textit{Swift}/XRT, although the exposure time for this observation was relatively low. The primary continuum remained consistent with earlier epochs ($\Gamma_{PC}=1.92\pm0.10$) but showed reduced luminosity ($log~L_{PC}=43.40^{+0.14}_{-0.21}$). The soft excess weakened significantly (Figure~\ref{fig:softexcess}), with a poorly constrained slope $\Gamma_{SE}=2.20^{+1.05}_{-0.21}$ and lower luminosity $log~L_{SE}=43.29^{+0.45}_{-0.76}$, marking a notable decline from the observations in 2006. No Fe~K$\alpha$ line or intrinsic absorber was detected, likely due to the limited exposure and resolution. Both the continuum and soft excess thus indicate a low-flux state, where weaker spectral features could not be reliably constrained.

In 2022, the combined \textit{Swift}/XRT dataset (XRT5) showed distinct results compared to all previous epochs. The spectrum was well fitted by a single power law, as adding a second component reduced the primary normalization to negligible values. Consequently, no soft excess was detected in this observation (Figure~\ref{fig:softexcess}). The primary continuum was well constrained with $\Gamma_{PC}=1.63\pm0.05$ and $log~L_{PC}=43.56\pm0.02$. No evidence of an intrinsic absorber was found.

The subsequent combined observations from 2023 and 2024 (XRT6 and XRT7) exhibited broadly consistent spectral properties. The primary continuum was estimated at $\Gamma_{PC}=1.71\pm0.03$ (XRT6) and $1.67\pm0.10$ (XRT7), with identical luminosities of $log~L_{PC}=43.64\pm0.04$. The soft excess was below the detection limit (Figure~\ref{fig:softexcess}), with poorly constrained slopes of $\Gamma_{SE}$ estimated at $2.22^{+0.56}_{-0.39}$ (XRT6) and $2.42^{+1.20}_{-0.80}$ (XRT7). The soft excess luminosity shows a decline from $log~L_{SE}=43.06^{+0.68}_{-0.36}$ to $42.58^{+1.09}_{-0.45}$. Similar to the XRT3 and XRT5 observations, no intrinsic absorber was detected. The values of the parameters estimated through the phenomenological model fitting of the spectra across all observations are listed in Table~\ref{tab_pheno}, and the variation of relevant parameters is plotted in Figure~\ref{fig:longterm}.

Overall, these recent observations indicate that Mrk~530 has transitioned into a state characterized by a comparatively harder primary continuum, accompanied by a progressive weakening of the soft excess. In addition, the intrinsic absorber is not statistically required by the Swift/XRT spectra obtained during these epochs. This behaviour is consistent with a long-term reduction in the relative contribution of the warm Comptonizing component associated with the soft-excess emission.

\subsection{Physical Models}
\label{sec:physical}

While the phenomenological models provide a useful first-order characterization of the spectral components and their long-term variability, they do not directly probe the physical processes responsible for these changes. In particular, the origin of the primary continuum, the mechanism driving the soft excess, and the role of absorption require a more physically motivated spectral fitting. Therefore, in this section, we replace the empirical models with self-consistent physical models. These models constrain the properties of the Comptonizing corona, the accretion disk, and the intervening absorbers, thereby offering deeper insights into the accretion physics of Mrk~530. To investigate the origin of the observed variability and build a clearer physical picture of the source, we apply two physical models generally used in AGN studies: \textsc{AGNSED} and \textsc{TCAF}, to all available X-ray spectra.

\subsubsection{AGNSED}

\texttt{AGNSED} is a physical model developed by \cite{2018MNRAS.480.1247K} to describe the broadband continuum emission in AGNs. The model incorporates a standard accretion disk that produces thermal emission like a blackbody \citep{1973A&A....24..337S,1973blho.conf..343N}. The hard X-ray emission is accounted for by a hot corona of relativistic electrons \citep{1980A&A....86..121S,2015MNRAS.451.4375F} that generate X-rays via inverse Comptonization up to $\sim 300$ keV \citep{1991ApJ...380L..51H}. The soft excess generally observed below 2 keV is explained by incorporating a warm, optically thick Comptonizing region \citep{1998MNRAS.301..179M,Done2012}. Each component is parameterized to quantify its physical contribution to the AGN spectrum. The hot corona is described by its electron temperature ($kT_{e,hot}$) and radial extent ($R_{hot}$) in units of the gravitational radius ($R_{g}$), with the emerging spectrum characterized by a photon index ($\Gamma_{hot}$). The warm Comptonizing region is similarly defined by its temperature ($kT_{e,warm}$), size ($R_{warm}$), and photon index ($\Gamma_{warm}$). The mass accretion rate is represented by $ \log \dot{m}$, the logarithmic ratio of the true mass accretion rate $\dot{m}$ to the Eddington accretion rate $\dot{m}_{\rm Edd}$. The model further quantifies the central engine through parameters such as black hole mass ($M_{\rm BH}$), dimensionless spin parameter ($a$), inclination angle ($\cos i$), and logarithm of the outer disc radius ($\log r_{\rm out}$). In addition, user-specified fixed parameters include the comoving distance (in Mpc) and the \texttt{reprocess} flag, which accounts for reflection from the accretion disc.

\begin{figure*}
    \centering
    \includegraphics[width=\textwidth]{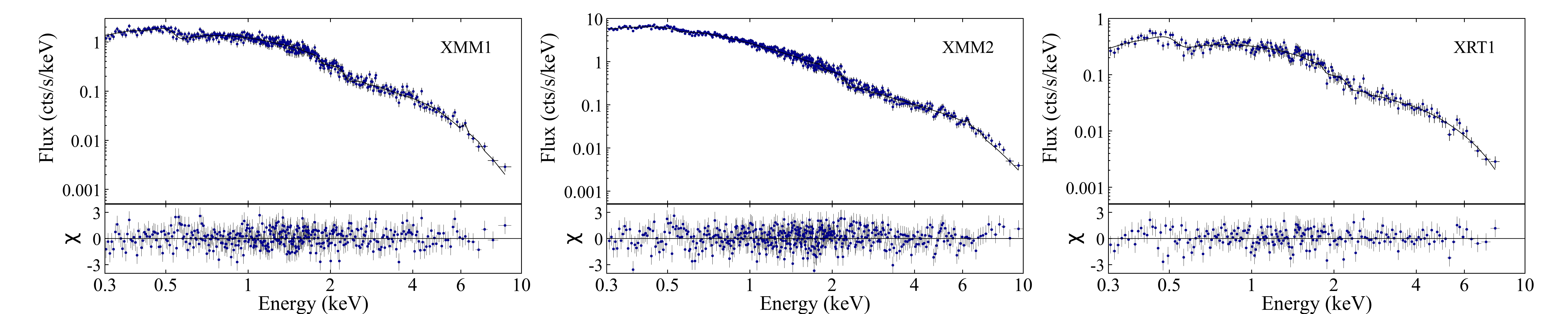}
    \vspace{0.2cm}
    \includegraphics[width=\textwidth]{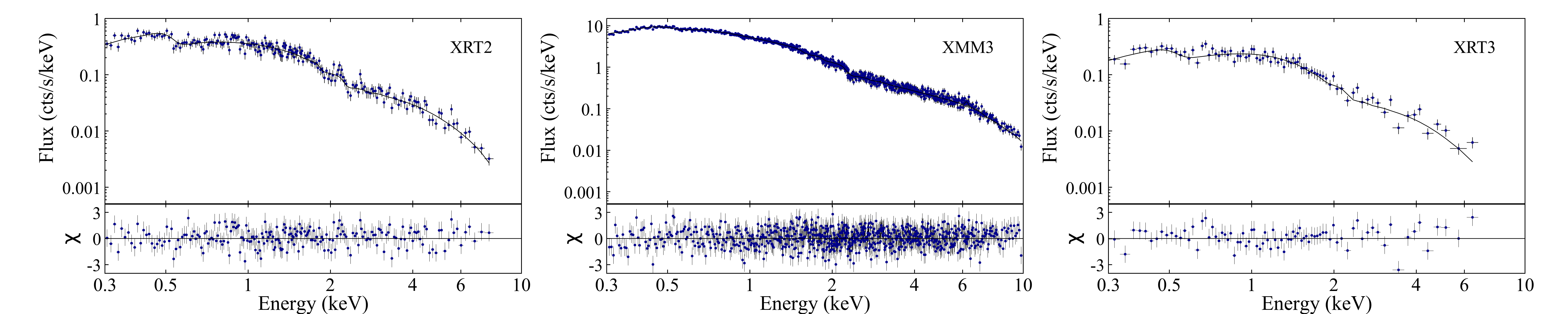}
    \vspace{0.2cm}
    \includegraphics[width=\textwidth]{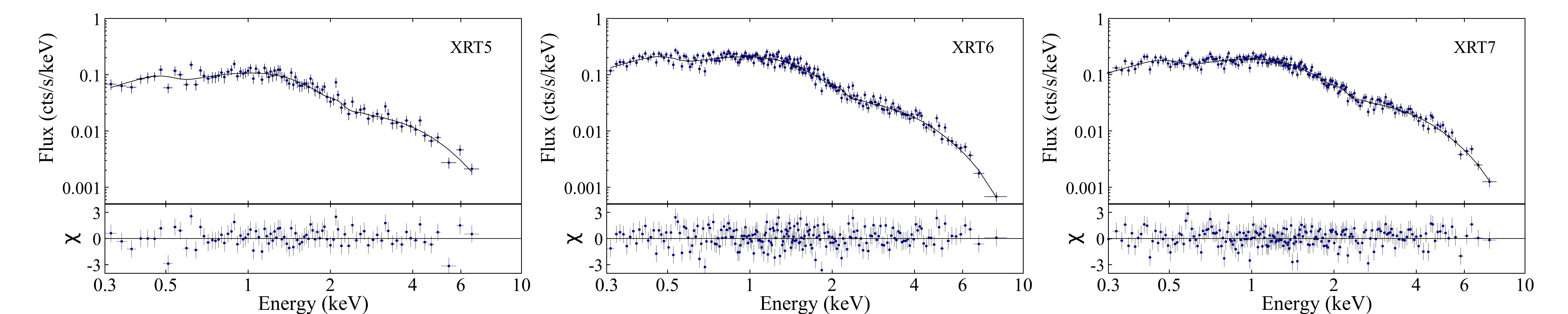}
    \caption{X-ray spectra of Mrk~530 from different epochs of observations fit with the {\tt AGNSED} model.}
    \label{fig:long_term_spectra}
\end{figure*}

Throughout the spectral analysis, we used the black hole mass of $M_{BH}=1.15\times10^8 M_{\odot}$ \citep{2012AJ....143...49W} and a comoving distance of 118.3 Mpc ($z$=0.029) \citep{2011ApJ...731...68L}. The outer radius and the inclination angle were fixed at their default values of log~$r_{out}=-1$ and $cos~i=0.5$, respectively. In the analysis of the 2012 {\it Suzaku} spectrum, \cite{2018MNRAS.478.4214E} reported a maximally spinning black hole providing the best spectral fits. In our analysis, varying the dimensionless spin parameter over the range $a=0.9-0.998$ does not lead to statistically significant changes in the fit quality for the Swift/XRT spectra, owing to the limited photon statistics. We therefore fix the spin parameter at $a=0.95$ for the \textit{Swift}/XRT observations. In contrast, for the \textit{XMM-Newton} observations, which have substantially higher photon counts, the spin parameter is left free and constrained during the fitting. For the three {\it XMM-Newton} observations, the {\tt reprocess} parameter was set to 1 (ON) to account for the presence of Fe K$\alpha$ emission lines. For  {\it Swift}/XRT observations, the {\tt reprocess} parameter was fixed at 0 (OFF), as none of the spectra show statistically significant reflection features (see Section~\ref{sec:pheno}). The model normalisation was fixed at unity for all observations. The best-fitting spectra and corresponding residuals for all epochs are shown in Figure~\ref{fig:long_term_spectra}.

\begin{figure}
    \centering
    \includegraphics[width=\linewidth]{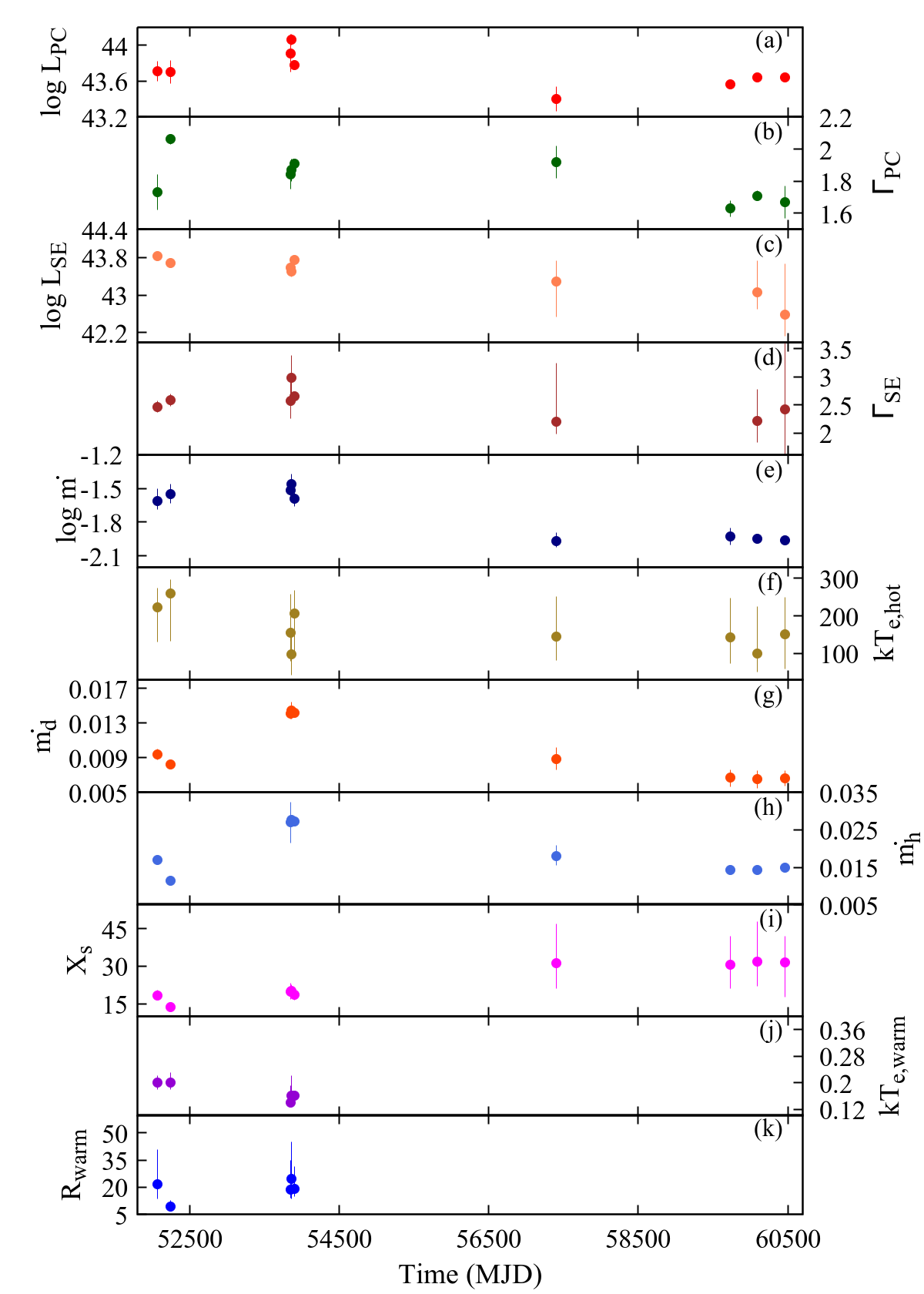}
    \caption{Long-term variation of parameters obtained from the spectral analysis of $\sim$ 24 years of X-ray monitoring of Mrk~530.}
    \label{fig:longterm}
\end{figure}

In 2001, both {\it XMM-Newton} observations (XMM1 and XMM2) exhibit a prominent soft-excess component (Figure~\ref{fig:softexcess}), characterized by comparable warm coronal electron temperatures of $kT_{e,\mathrm{warm}}\sim0.2$ keV. Between these two epochs, the size of the warm corona decreases from $R_{\mathrm{warm}}=22^{+19}_{-8}$ in XMM1 to $10^{+3}_{-1}~R_g$ in XMM2. The hot corona temperatures remain consistent within uncertainties, with $kT_{e,\mathrm{hot}}=224^{+51}_{-94}$ keV (XMM1) and $259^{+38}_{-126}$ keV (XMM2), while the corresponding coronal size decreases from $R_{hot}=15^{+10}_{-5}$ $R_g$ to $7^{+2}_{-1}$ $R_g$. Over the same period, both photon indices steepen, with $\Gamma_{\mathrm{hot}}$ increasing from $1.99^{+0.04}_{-0.06}$ to $2.23^{+0.02}_{-0.03}$ and $\Gamma_{\mathrm{warm}}$ from $2.02^{+0.03}_{-0.01}$ to $2.39^{+0.17}_{-0.15}$. The accretion rate remains consistent within uncertainties at $log~ \dot{m}=-1.61^{+0.11}_{-0.08}$ (XMM1) and $log~ \dot{m}=-1.55^{+0.09}_{-0.08}$ (XMM2). The black hole spin parameter is unconstrained toward the upper bound in both observations, with $a=0.94^{p}_{-0.76}$ for XMM1, and $a=0.93^{p}_{-0.30}$. for XMM2. A summary of all best-fit parameters is given in Table~\ref{tab:physical}, and the long-term evolution of the {\tt AGNSED} parameters is shown in Figure~\ref{fig:longterm}.

In 2006, Mrk~530 was observed with \textit{Swift}/XRT (XRT1 and XRT2) and \textit{XMM-Newton} (XMM3), showing relatively modest variability in the hot coronal properties. During XRT1, the hot corona is characterized by $kT_{e,\mathrm{hot}}=156^{+103}_{-76}$ keV, $R_{\mathrm{hot}}=9^{+3}_{-2}~R_g$, and $\Gamma_{\mathrm{hot}}=2.02^{+0.06}_{-0.04}$, consistent with the earlier epochs within uncertainties. Similar values are obtained in XRT2 ($kT_{e,\mathrm{hot}}=97^{+59}_{-55}$ keV, $R_{\mathrm{hot}}=9^{+4}_{-2}~R_g$, $\Gamma_{\mathrm{hot}}=2.00\pm0.05$). In contrast, the warm corona shows notable changes relative to XMM2. In XRT1, it appears more extended, with ${R_{\mathrm{warm}}=19^{+16}_{-5}}~R_g$, at a lower temperature of $kT_{e,\mathrm{warm}}=0.14^{+0.05}_{-0.01}$ keV, and a steeper photon index of $\Gamma_{\mathrm{warm}}=2.61^{+0.84}_{-0.40}$. Comparable parameters are obtained for the warm corona in XRT2, with $R_{\mathrm{warm}}=24^{+21}_{-11}~R_g$, $kT_{e,\mathrm{warm}}=0.16^{+0.06}_{-0.02}$ keV and $\Gamma_{\mathrm{warm}}=2.37^{+0.49}_{-0.24}$.

Approximately one month later, the XMM3 spectrum shows hot coronal parameters consistent with the earlier 2006 measurements, with $kT_{e,\mathrm{hot}}=206^{+61}_{-100}$ keV, $R_{\mathrm{hot}}=12^{+5}_{-2}~R_g$ and a steeper slope $\Gamma_{\mathrm{hot}}=2.15^{+0.01}_{-0.01}$. The warm corona stabilizes at $kT_{e,\mathrm{warm}}=0.16\pm0.01$ keV and $R_{\mathrm{warm}}=19^{+12}_{-4}~R_g$, with $\Gamma_{\mathrm{warm}}=2.28^{+0.15}_{-0.17}$. The spin parameter remains unconstrained toward high values $a=0.87^{p}_{-0.66}$. The accretion rates across the 2006 observations are mutually consistent within uncertainties, with $log~ \dot{m}=-1.52^{+0.09}_{-0.06}$ (XRT1), $log~ \dot{m}=-1.46^{+0.09}_{-0.07}$ (XRT2) and $-1.59^{+0.08}_{-0.07}$ (XMM3). The Markov Chain Monte Carlo (MCMC) posterior distributions for the best-fitting AGNSED model of the XMM3 observation are shown in Figure~\ref{fig:mcmc}.

By 2016 (XRT3), the soft excess component was no longer detected in the spectrum (Figure~\ref{fig:softexcess}). Consequently, the warm corona could not be constrained, and the corresponding parameters were fixed to their minimal values of $kT_{e,\mathrm{warm}}=0.1$ keV, $R_{\mathrm{warm}}=6~R_g$, and $\Gamma_{\mathrm{warm}}=2$. The hot corona, however, remained measurable with $kT_{e,\mathrm{hot}}=145^{+106}_{-64}$ keV, $R_{\mathrm{hot}}=22^{+75}_{-12}~R_g$, and $\Gamma_{\mathrm{hot}}=2.04^{+0.03}_{-0.04}$. This epoch coincided with a marked decline in luminosity (Table~\ref{tab_pheno}), accompanied by a corresponding drop in the accretion rate to $\log \dot{m}=-1.97^{+0.08}_{-0.05}$.

Subsequent observations (XRT5, XRT6, and XRT7) showed no evidence of a soft excess (see Section~\ref{sec:pheno}), with the spectra being dominated by the hot corona. The photon index evolved from $\Gamma_{\mathrm{hot}}=1.64\pm0.03$ in XRT5 to a steeper $\Gamma_{\mathrm{hot}}=1.84^{+0.02}_{-0.01}$ in XRT6, before flattening again to $\Gamma_{\mathrm{hot}}=1.75\pm0.02$ in XRT7. The best-fitting hot coronal temperature varies in the opposite sense, decreasing from $kT_{e,\mathrm{hot}}=143^{+103}_{-70}$ keV in XRT5 to $101^{+124}_{-49}$ keV in XRT6, and then increasing to $151^{+98}_{-91}$ keV in XRT7. However, these values are accompanied by large statistical uncertainties. The hot coronal radius shows a qualitatively similar trend, increasing from $R_{\mathrm{hot}}=126^{+260}_{-100}~R_g$ (XRT5) to $135^{+223}_{-67}~R_g$ (XRT6), before contracting to $116^{+213}_{-63}~R_g$ (XRT7), again with substantial uncertainties. While the large error bars prevent a definitive assessment of parameter correlations, the observed trends are broadly consistent with expectations from coronal Comptonization models, in which harder spectra are associated with hotter and more compact coronae, whereas cooler or more extended configurations produce a steeper continuum.

Despite these spectral changes, the normalized accretion rate remained stable across the three epochs, with $\log \dot{m}=-1.93^{+0.08}_{-0.07}$ (XRT5), $-1.95^{+0.02}_{-0.03}$ (XRT6), and $-1.96^{+0.03}_{-0.04}$ (XRT7). These values are consistent with the relatively steady luminosities reported in Section~\ref{sec:pheno}. The temporal evolution of the key spectral parameters is presented in Figure~\ref{fig:longterm}. A summary of the physical model results is provided in Table~\ref{tab:physical}, while the model-fitted spectra and residuals are shown in Figure~\ref{fig:long_term_spectra}. 

\begin{table*}
\caption{The best fit parameters for the physical models used to fit all observations of Mrk~530.}
    \resizebox{\textwidth}{!}{\begin{tabular}{c|c c c c c c c c c c}
         \hline\\
         & Parameter & XMM1 & XMM2 & XRT1 & XRT2 & XMM3 & XRT3 & XRT5 & XRT6 & XRT7  \\
         \hline \\
         \vspace{0.2cm}
         Pcfabs & $N_H~(\times10^{22}~cm^{-2})$ & $26^f$ & $26^f$ & $26^f$ & $26^f$ & $26^f$ & - & -  & - & - \\
         \vspace{0.2cm}
         & $C_f$ & $0.34^f$ & $0.34^f$ & $0.34^f$ & $0.34^f$ & $0.34^f$ & - & - & - & - \\
         \hline
         \vspace{0.2cm}
         & a & $0.94^{p}_{-0.76}$ & $0.93^{p}_{-0.30}$ & $0.95^f$ & $0.95^f$ & $0.87^{p}_{-0.66}$ & $ 0.95^f$ & $0.95^f$ & $0.95^f$ & $0.95^f$ \\
         \vspace{0.2cm}
         & $log~\dot{m}$ & $-1.61^{+0.11}_{-0.08}$ & $-1.55^{+0.09}_{-0.08}$ & $-1.52^{+0.09}_{-0.06}$ & $-1.46^{+0.09}_{-0.07}$ & $-1.59^{+0.08}_{-0.07}$ & $-1.97^{+0.08}_{-0.05}$ & $-1.93^{+0.08}_{-0.07}$ & $-1.95^{+0.02}_{-0.03}$ & $-1.96^{+0.03}_{-0.04}$ \\
         \vspace{0.2cm}
         & $KT_{e,hot}~(keV)$ & $224^{+51}_{-94}$ & $259^{+38}_{-126}$ & $156^{+103}_{-76}$ & $97^{+59}_{-55}$ & $206^{+61}_{-100}$ & $145^{+106}_{-64}$ & $143^{+103}_{-70}$ & $101^{+124}_{-49}$ & $151^{+98}_{-91}$ \\
         \vspace{0.2cm}
  {\tt AGNSED} & $KT_{e,warm}~(keV)$ & $0.20^{+0.02}_{-0.02}$ & $0.20^{+0.03}_{-0.02}$ & $0.14^{+0.05}_{-0.01}$ & $0.16^{+0.06}_{-0.02}$ & $0.16^{+0.01}_{-0.01}$ & $0.1^f$ & $0.1^f$ & $0.1^f$ & $0.1^f$ \\
        \vspace{0.2cm}
        & $\Gamma_{hot}$ & $1.99^{+0.04}_{-0.06}$ & $2.23^{+0.02}_{-0.03}$ & $2.02^{+0.06}_{-0.04}$ & $2.00^{+0.05}_{-0.05}$ & $2.15^{+0.01}_{-0.01}$ & $2.04^{+0.03}_{-0.04}$ & $1.64^{+0.03}_{-0.03}$ & $1.84^{+0.02}_{-0.01}$ & $1.75^{+0.02}_{-0.02}$ \\
        \vspace{0.2cm}
        & $\Gamma_{warm}$ & $2.02^{+0.03}_{-0.01}$ & $2.39^{+0.17}_{-0.15}$ & $2.61^{+0.84}_{-0.40}$ & $2.37^{+0.49}_{-0.24}$ & $2.28^{+0.15}_{-0.17}$ & $2^f$ & $2^f$ & $2^f$ & $2^f$ \\
        \vspace{0.2cm}
        & $R_{hot}~(R_g)$ & $15^{+10}_{-5}$ & $7^{+2}_{-1}$ & $9^{+3}_{-2}$ & $9^{+4}_{-2}$ & $12^{+5}_{-2}$ & $22^{+75}_{-12}$ & $126^{+260}_{-100}$ & $135^{+223}_{-67}$ & $116^{+213}_{-63}$ \\
        \vspace{0.2cm}
        & $R_{warm}~(R_g)$ & $22^{+19}_{-8}$ & $10^{+3}_{-1}$ & $19^{+16}_{-5}$ & $24^{+21}_{-11}$ & $19^{+12}_{-4}$ & $6^f$ & $6^f$ & $6^f$ & $6^f$ \\
        \hline
        & $\chi^2/dof$ & 428.49/415 & 582.51/496 & 197.27/208 & 229.88/211 & 824.93/802 & 91.97/81 & 90.57/90 & 248.84/212 & 197.29/209 \\
                \hline
        \\
        \vspace{0.2cm}
        & $\dot{m}_d~(\times10^{-3} ~\dot{m}_{Edd})$ & $9.37^{+0.59}_{-0.61}$ & $8.20^{+0.29}_{-0.30}$ & $14.10^{+0.65}_{-0.63}$ & $14.40^{+0.99}_{-0.96}$ & $14.20^{+0.34}_{-0.35}$ & $8.80^{+1.37}_{-1.24}$ & $6.67^{+0.93}_{-0.99}$ & $6.57^{+0.98}_{-1.09}$ & $6.61^{+0.88}_{-0.91}$ \\
        \vspace{0.2cm}
        & $\dot{m}_h~(\times10^{-2} ~\dot{m}_{Edd})$ & $1.70^{+0.11}_{-0.12}$ & $1.15^{+0.04}_{-0.04}$ & $2.70^{+0.53}_{-0.54}$ & $2.77^{+0.12}_{-0.17}$ & $2.72^{+0.06}_{-0.06}$ & $1.80^{+0.29}_{-0.25}$ & $1.44^{+0.08}_{-0.08}$ & $1.44^{+0.09}_{-0.09}$ & $1.49^{+0.09}_{-0.09}$\\
        \vspace{0.2cm}
  {\tt TCAF}  & $X_s ~(R_g)$ & $18^{+2}_{-2}$ & $14^{+2}_{-2}$ & $20^{+3}_{-3}$ & $20^{+2}_{-2}$ & $19^{+1}_{-2}$ & $31^{+16}_{-10}$ & $30^{+12}_{-10}$ & $32^{+16}_{-10}$ & $32^{+10}_{-14}$ \\
        \vspace{0.2cm}
        & R & $1.85^{+0.11}_{-0.10}$ & $1.85^{+0.09}_{-0.07}$ & $1.85^{+0.52}_{-0.55}$ & $1.85^{+0.50}_{-0.51}$ & $1.86^{+0.11}_{-0.10}$ & $1.85^{+2.68}_{-0.85}$ & $1.77^{+0.79}_{-0.72}$ & $1.71^{+1.88}_{-0.84}$ & $1.75^{+1.92}_{-0.75}$ \\
        \vspace{0.2cm}
        & Norm$^\dagger ~(\times 10^{-4})$ & $10.00^{+5.63}_{-3.18}$ & $10.10^{+1.65}_{-1.66}$ & $9.95^{+3.65}_{-3.55}$ & $9.93^{+5.74}_{-5.26}$ & $10.00^{+2.74}_{-2.77}$ & $10.00^{+3.77}_{-2.06}$ & $9.97^{+1.87}_{-1.05}$ & $9.99^{+1.97}_{-2.22}$ & $9.97^{+1.00}_{-1.21}$\\
        \vspace{0.2cm}
        &$T_c$ ($\times10^4$s) & $0.59\pm0.03$ & $0.43\pm0.02$ & $0.45\pm0.03$ & $0.45\pm0.02$ & $0.44\pm0.02$ & $1.65\pm0.28$ & $2.03\pm0.41$ & $2.22\pm0.36$ & $2.14\pm0.33$ \\
        \vspace{0.2cm}
        &$T_{fall}$ ($\times10^4$s) & $2.60\pm0.06$ & $1.92\pm0.05$ & $2.82\pm0.19$ & $2.86\pm0.17$ & $2.82\pm0.07$ & $4.39\pm0.92$ & $4.31\pm0.79$ & $4.49\pm0.79$ & $4.47\pm0.77$ \\
        \vspace{0.2cm}
        &$\tau$ ($T_c/T_{fall}$) & $0.23\pm0.01$ & $0.22\pm0.01$ & $0.16\pm0.02$ & $0.16\pm0.01$ & $0.16\pm0.01$ & $0.38\pm0.10$ & $0.47\pm0.13$ & $0.49\pm0.12$ & $0.48\pm0.11$ \\
        \hline
        & $\chi^2/dof$ & 451.42/423 & 548.67/518 & 210.17/217 & 201.65/218 & 878.63/815 & 82.76/76 & 91.72/92 & 229.15/218 & 215.47/210 \\
        \hline 
\end{tabular}}
\leftline{\textdagger in the units of photons/keV/$cm^{2}$/s. $f$ indicates a frozen parameter, $p$ indicates that the parameter was pegged at the hard limit.}
\label{tab:physical}
\end{table*}

\subsubsection{TCAF}
While the \textsc{AGNSED} model provides a comprehensive parameterization of the relative contributions from the accretion disk, warm corona, and hot corona, it does not explicitly capture the global accretion flow geometry. To obtain deeper physical insight into the accretion structure of Mrk~530, we apply the Two-Component Advective Flow (\textsc{TCAF}) model. Rooted in hydrodynamical accretion theory, \textsc{TCAF} self-consistently incorporates a geometrically thin Keplerian disk and a sub-Keplerian advective halo. Their interaction gives rise to a Comptonizing region or CENBOL, which naturally accounts for the observed X-ray spectral properties, including the soft excess component which arises from warm Comptonization of disc photons within the outer CENBOL. The \textsc{TCAF} model can constrain the accretion rates of the Keplerian and sub-Keplerian flows, as well as the location and compression ratio of the shock, providing a physically motivated picture of the accretion geometry in this source.

The \textsc{TCAF} model is defined by four primary parameters: (i) the Keplerian disk accretion rate $(\dot{m}_d)$ in units of the Eddington rate $(\dot{m}_{\rm Edd})$, (ii) the sub-Keplerian halo accretion rate $(\dot{m}_h/\dot{m}_{\rm Edd})$, (iii) the shock compression ratio $(R)$, and (iv) the shock location $(X_s)$ in units of $R_g$. The compression ratio $R$ is defined as the ratio of the post-shock to pre-shock matter density at the centrifugal pressure-supported shock and quantifies the strength of the shock. Physically, larger values of $R$ correspond to stronger shocks that produce a hotter and denser Comptonizing region (CENBOL), thereby enhancing inverse-Compton scattering of soft photons from the Keplerian disk. An additional intrinsic parameter is the black hole mass $(M_{\rm BH})$ in solar units. The allowed parameter ranges, along with their default values, are stored in {\it lmodel.dat} and summarized in Table~\ref{tcaf}. We implement the model in {\tt XSPEC} using the {\tt initpackage} and {\tt lmod} routines, which load the table model and fit the spectra. Each fit requires $\sim10^5$ iterations of the source code, with the best-fit solution selected through $\chi^2$ minimization. The corresponding parameters are reported in Table~\ref{tab:physical}.

In the \textsc{TCAF} framework, the oscillation of the Compton cloud boundary is governed by a competition between the infall timescale ($T_{\rm fall}$) of the post-shock region and the cooling timescale ($T_{\rm c}$) set by inverse Comptonization of intercepted disc photons \citep{Chakrabarti2015}. The total thermal energy of the cloud can be approximated as $E_{\rm t} \propto n_{\rm e} T_{\rm e} X_{\rm s} H_{\rm s}$, where $n_{\rm e}$ is the post-shock electron density, $T_{\rm e}$ is the mean electron temperature, and $X_{\rm s}$ and $H_{\rm s}$ are the shock location and height, respectively. The cooling power, $\dot{E}_{\rm cool}$, depends on the intercepted disc flux scaled by the fraction of photons interacting with the cloud and their average energy gain. This yields the cooling timescale $T_{\rm c}=E_{\rm t}/\dot{E}_{\rm cool}$. In contrast, the infall timescale is given by $T_{\rm fall}=X_{\rm s}/V_{+}$, where $V_{+}$ is the post-shock velocity. Their ratio, $\tau = T_{\rm c}/T_{\rm fall}$, can be expressed directly in terms of the fitted \textsc{TCAF} parameters, namely the Keplerian and sub-Keplerian accretion rates $(\dot{m}_{\rm d}, \dot{m}_{\rm h})$, shock location $(X_{\rm s})$, and compression ratio $(R)$ \citep{Chakrabarti2015}. Quasi-periodic oscillations (QPOs) are expected to occur when $\tau \sim 1$, i.e., when cooling and infall operate on comparable timescales, while significant departures from this resonance suppress coherent oscillations.

\begin{table}
\centering
\caption{The TCAF parameter space is defined in the file \texttt{lmod.dat}. The two columns for minima and maxima are provided for the range of iterations. Between them, the 1st column indicates the soft bound and the 2nd column gives the hard bound of the parameters.}
\resizebox{\columnwidth}{!}{\begin{tabular}{ccccccc}
\hline
\hline
Parameters & Default & Min. & Min. & Max. & Max. & Increment \\
 & value & & & & &\\
\hline
$M_{BH}$ ($M_\odot$) & $1.0 \times 10^7$ & $1 \times 10^6$ & $1 \times 10^6$ & $1 \times 10^9$ & $1 \times 10^9$ & 10.0 \\
$\dot{m}_d$ ($\dot{m}_{Edd}$) & 0.001 & 0.0001 & 0.0001 & 10.0 & 10.0 & 0.0001 \\
$\dot{m}_h$ ($\dot{m}_{Edd}$) & 0.01 & 0.0001 & 0.0001 & 10.0 & 10.0 & 0.0001 \\
$X_s$ ($R_g$) & 50.0 & 8.0 & 8.0 & 1000.0 & 1000.0 & 1.0 \\
$R$ & 2.5 & 1.01 & 1.01 & 6.8 & 6.8 & 0.1 \\
\hline
\end{tabular}}
\label{tcaf}
\end{table}

The XMM1 observation in 2001 indicates comparable contributions from both accretion components, with the Keplerian disc accreting at $\dot{m}_{d}=9.37^{+0.59}_{-0.61}\times10^{-3}~\dot{m}_{Edd}$ and the sub-Keplerian halo at $\dot{m}_{\rm h}=1.70^{+0.11}_{-0.12}\times10^{-2}~\dot{m}_{Edd}$. The relatively higher halo accretion rate implies the presence of a substantial Comptonizing region, consistent with the inferred shock location at $X_{\rm s}=18\pm2~R_{\rm g}$. The compression ratio is constrained to $R=1.85\pm0.11$, indicative of a moderately strong shock capable of sustaining efficient inverse Comptonization. During the XMM2 observation, later in the same year, both the disc and halo accretion rates marginally decreased to $\dot{m}_{\rm d}=8.20^{+0.29}_{-0.30}\times10^{-3}~\dot{m}_{Edd}$ and $\dot{m}_{\rm h}=1.15\pm0.04\times10^{-2}~\dot{m}_{\rm Edd}$, respectively. The corresponding shock location contracts to $X_{\rm s}=14\pm2~R_{\rm g}$, while the compression ratio remains statistically unchanged at $R=1.85^{+0.09}_{-0.07}$, despite the reduced accretion rates and coronal extent. The temporal evolution of the fitted \textsc{TCAF} parameters is shown in Figure~\ref{fig:longterm}.

The observations obtained in 2006 reveal a systematic increase in the mass accretion rates. In XRT1 and XRT2, the Keplerian disc accretes at $\dot{m}_{\rm d}=14.10^{+0.65}_{-0.63}\times10^{-3}~\dot{m}_{\rm Edd}$ and $14.40^{+0.99}_{-0.96}\times10^{-3}~\dot{m}_{\rm Edd}$, while the halo accretion rates remain consistent within uncertainties at $\dot{m}_{\rm h}=2.70^{+0.53}_{-0.54}\times10^{-2}~\dot{m}_{\rm Edd}$ (XRT1) and $2.77^{+0.12}_{-0.17}\times10^{-2}~\dot{m}_{\rm Edd}$ (XRT2). These results suggest that the supply of low-angular momentum matter remained steady during this period. The shock location expands relative to the 2001 observations, reaching $X_{\rm s}=20\pm3~R_{\rm g}$ in XRT1 and $20\pm2~R_{\rm g}$ in XRT2, while the compression ratio remains nearly constant at $R\sim1.85$. The XMM3 observation later in the same year yields comparable accretion rates and coronal parameters, with $\dot{m}_{\rm d}=14.20^{+0.34}_{-0.35}\times10^{-3}~\dot{m}_{\rm Edd}$ and $\dot{m}_{\rm h}=2.72\pm0.06\times10^{-2}~\dot{m}_{\rm Edd}$, and a stable shock configuration at $X_s\sim19~R_g$ with $R\sim1.86$. These results indicate that, although the characteristic size of the Comptonizing region increases at higher accretion rates relative to the 2001 epoch, the shock strength and the density contrast across the shock front remain approximately constant.

After a decade-long gap, the 2016 XRT3 observation revealed a significant reduction in both accretion components, with the Keplerian disk rate decreasing to $\dot{m}_{\rm d}=8.80^{+1.37}_{-1.24}\times10^{-3}~\dot{m}_{\rm Edd}$ and the halo rate declining to $\dot{m}_{\rm h}=1.80^{+0.29}_{-0.25}\times10^{-2}~\dot{m}_{\rm Edd}$. This reduction is consistent with the luminosity drop reported in Table~\ref{tab_pheno}. Simultaneously, the Comptonizing region expanded to $X_{\rm s}=31^{+16}_{-10}~R_{\rm g}$, suggesting that the hot flow became more extended as the accretion supply weakened. The compression ratio remained nearly unchanged at $R=1.85^{+2.68}_{-0.85}$, indicating that the shock strength was insensitive to the lower inflow rates.

In 2022, the XRT5 observation indicated a marginal decline in accretion activity, with the Keplerian disk rate reduced to $\dot{m}_{\rm d}=6.67^{+0.93}_{-0.99}\times10^{-3}~\dot{m}_{\rm Edd}$ and the halo rate to $\dot{m}_{\rm h}=1.44\pm0.08\times10^{-2}~\dot{m}_{\rm Edd}$. The size of the Comptonizing region remained nearly constant at $X_{\rm s}=30^{+12}_{-10}~R_{\rm g}$, suggesting a stable coronal geometry. The compression ratio was found to be within uncertainties at $R=1.77^{+0.79}_{-0.72}$, implying that both the spatial extent as well as the shock strength remained unchanged during this epoch.

The spectral parameters from the XRT6 (2023) and XRT7 (2024) observations exhibited no significant variability, indicating a stable accretion configuration. The Keplerian disk accretion rates were consistent at $\dot{m}_{\rm d}=6.57^{+0.98}_{-1.09}\times10^{-3}~\dot{m}_{\rm Edd}$ (XRT6) and $\dot{m}_{\rm d}=6.61^{+0.88}_{-0.91}\times10^{-3}~\dot{m}_{\rm Edd}$ (XRT7), while the halo accretion rates remained nearly unchanged at $\dot{m}_{\rm h}=1.44\pm0.09\times10^{-2}~\dot{m}_{\rm Edd}$ and $\dot{m}_{\rm h}=1.49\pm0.09\times10^{-2}~\dot{m}_{\rm Edd}$, respectively. The shock compression ratio was also steady, with $R=1.71^{+1.88}_{-0.84}$ (XRT6) and $R=1.75^{+1.92}_{-0.75}$ (XRT7). The Comptonizing region showed no measurable variability, with $X_{\rm s}=32^{+16}_{-10}~R{\rm g}$ and $X_{\rm s}=32^{+10}_{-14}~R{\rm g}$ for the two epochs. These results point to a prolonged phase of quasi-stable accretion flow geometry, following the gradual decline from the earlier epochs. The details of the fitted parameters are summarized in Table~\ref{tab:physical}, and their temporal evolution is illustrated in Figure~\ref{fig:longterm}.

\subsection{Study on 2018 Observations}
\label{sec:2018_spectra}
Timing analysis of the \textit{Swift}/XRT and \textit{Swift}/UVOT monitoring of Mrk~530 in 2018 reveals candidate QPO signals in both the UV and X-ray bands (see Section~\ref{sec:2018_lc}). To investigate the physical origin of this variability, we carried out a systematic spectral study of all 45 observations obtained during this campaign. The \textit{Swift}/XRT spectra, which are limited by relatively short exposure times, were grouped to a minimum of five counts per bin and analysed using Cash statistics \citep{1979ApJ...228..939C, Kaastra2017}. This approach allows us to examine whether the observed periodic modulations can plausibly be associated with changes in the accretion flow, evolution of coronal properties, or other intrinsic processes in the central engine.

For the initial spectral study, we modeled all 2018 \textit{Swift}/XRT spectra with a Galactic absorbed power law in the 2–10 keV band. No significant residuals corresponding to a fluorescent Fe~K$\alpha$ line were detected in any observation, likely due to the limited spectral resolution and short exposures. Extending the fit to lower energies revealed that most spectra lacked excess emission below 2 keV, indicating the absence of a detectable soft excess during this period. Physically, this suggests that the warm Comptonizing region was either weak or absent, or that the source was in a state dominated by the hot corona. To test for additional line-of-sight absorption, we included a partial covering absorber (\texttt{Pcfabs}), but the fits showed no significant improvement. Consequently, the baseline model was reduced to \texttt{Tbabs $\times$ powerlaw}, with the corresponding results summarized in Table~\ref{tab:2018_pheno}. The absence of both a soft excess and intrinsic absorption rules out variability driven by the warm corona or absorption changes. Instead, the observed correlation between the photon index and luminosity (PCC $\approx0.62$, $p < 0.01$) indicates that the variability originates in the hot corona and accretion flow, making them the most plausible drivers of the quasi-periodic behaviour in Mrk~530.

\begin{figure}
    \centering
    \includegraphics[width=\linewidth]{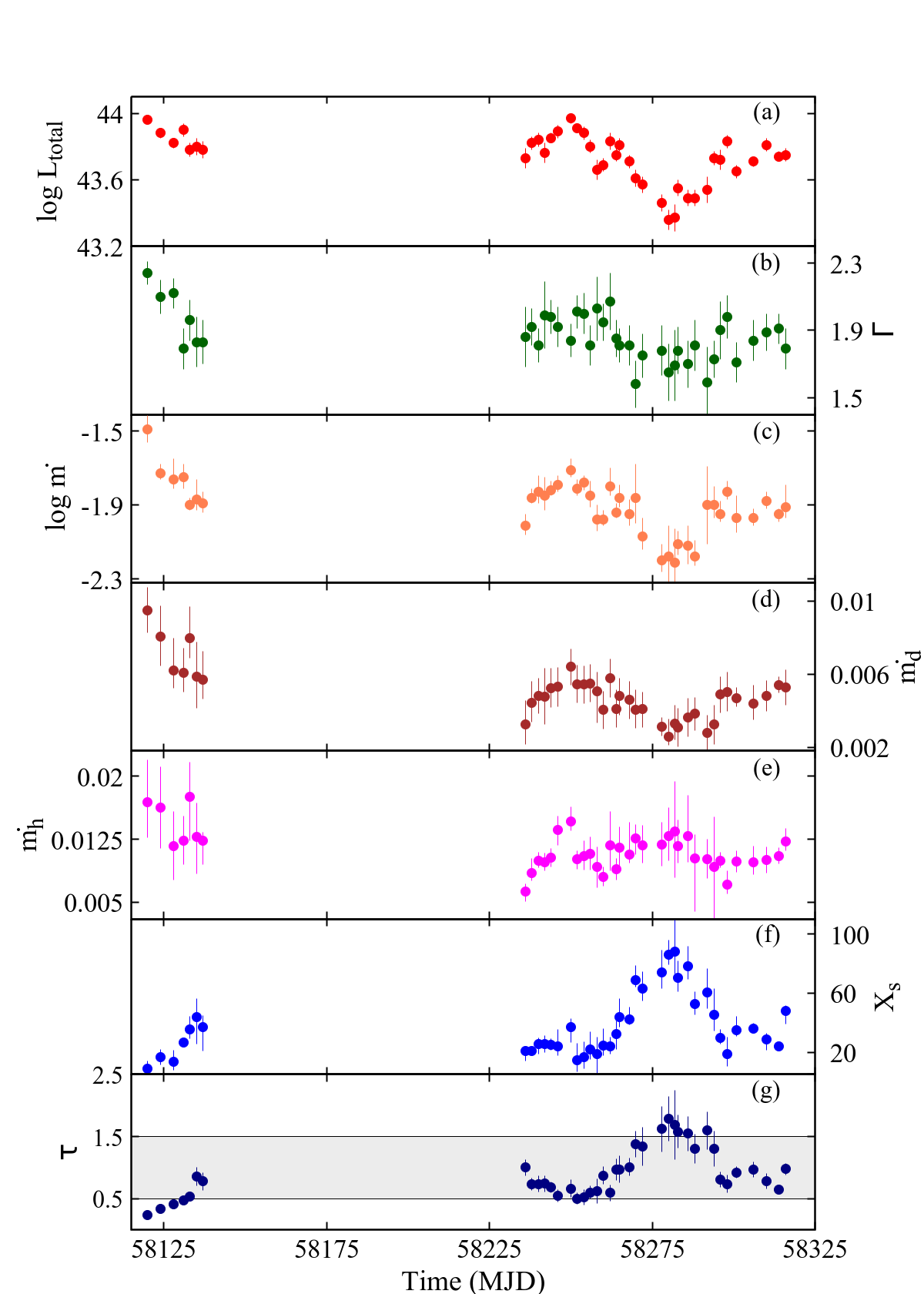}
    \caption{Variation of spectral parameters derived from the {\it Swift}/XRT observations in the year 2018.}
    \label{fig:2018_param}
\end{figure}

From the spectral analysis, we found that the 0.3–10 keV luminosity exhibited strong variability (Figure~\ref{fig:2018_param}a), ranging from $\log L_{\rm Total}=43.36\pm0.06$ to $44.07\pm0.12$. The soft (0.3--2 keV) and hard (2--10 keV) bands followed the same trend, with luminosities spanning $\log L_{\rm Soft}=42.95\pm0.06$ to $43.77\pm0.02$ and $\log L_{\rm Hard}=43.14\pm0.13$ to $43.89\pm0.18$, respectively (Table~\ref{tab:2018_pheno}). The photon index varied between $1.52\pm0.30$ and $2.24\pm0.07$, indicating transitions between comparatively harder and softer spectral states during 2018 (Figure~\ref{fig:2018_param}b). The temporal evolution of luminosity closely followed the total count rate presented in Section~\ref{sec:2018_lc}, confirming that the observed variability is intrinsic to the source rather than an instrumental artifact.

After the observational gap beginning at MJD 58236 (2018 April 28), the luminosity rose from $\log L_{\rm Total}=43.73\pm0.06$ to $43.97\pm0.03$ on MJD 58250 (2018 May 12), decreased to $43.36\pm0.06$ on MJD 58280 (2018 June 11), and subsequently peaked at $44.07\pm0.12$ on MJD 58312 (2018 July 13) (Table~\ref{tab:2018_pheno}). The soft and hard band luminosities mirrored this behavior. Notably, epochs of low luminosity were associated with flatter photon indices (e.g., $\Gamma=1.65\pm0.17$ on MJD 58280), consistent with the source residing in a harder spectral state (Figure~\ref{fig:2018_param}b). This behavior suggests that the hot corona dominates during low-luminosity intervals, whereas higher accretion rates produce softer spectra and enhanced X-ray output. A full list of best-fit parameters is provided in Table~\ref{tab:2018_pheno}, and their evolution is shown in Figure~\ref{fig:2018_param}.

To investigate whether the observed variability arises from changes in the accretion flow or coronal structure, we applied the physical models \texttt{AGNSED} and \texttt{TCAF} (Section~\ref{sec:physical}) to the X-ray spectra. Only spectra with $\geq 25$ degrees of freedom were considered for reliable fits, excluding observations from MJD 58141 (2018 Jan 23), 58308 (2018 July 09), 58312 (2018 July 13), 58320 (2018 July 21), and 58324 (2018 July 25). As no prominent soft excess was detected, the warm corona parameters in \texttt{AGNSED} were fixed at their minimal values. The fits indicate that the hot coronal photon index, $\Gamma_{\rm hot}$, closely tracks the phenomenological $\Gamma$. The normalized accretion rate varied substantially, from $\log \dot{m}=-2.21^{+0.18}_{-0.11}$ at the faintest epoch to $-1.49^{+0.08}_{-0.07}$ at the brightest (Table~\ref{tab:2018_physical}, Figure~\ref{fig:2018_param}c). This correlation between accretion rate and luminosity is physically expected: higher accretion rates supply more seed photons for inverse Comptonization, producing brighter and softer X-ray spectra. We note that the tighter constraints on the normalized mass accretion rate, which are comparable to the constraints derived in the long-term spectral fits with higher net exposures, are a consequence of the reduced parameter degeneracy in the individual {\it Swift} spectral fits for the year 2018. Although $kT_{e,{\rm hot}}$ and $R_{\rm hot}$ are poorly constrained, their variations suggest fluctuations in the heating and spatial extent of the hot electron population. The derived parameters from the physical model {\tt AGNSED} are listed in Table~\ref{tab:2018_physical}, with their temporal evolution shown in Figure~\ref{fig:2018_param}.

The \texttt{TCAF} results are broadly consistent with the trends inferred from the \texttt{AGNSED} modelling. The Keplerian disk accretion rate varies from $\dot{m}_d=2.57^{+0.97}_{-0.47}\times10^{-3}$ to $9.48^{+1.28}_{-1.21}\times10^{-3}~\dot{m}_{\rm Edd}$, reflecting the $\dot{m}$ variations inferred from \texttt{AGNSED} (Figure~\ref{fig:2018_param}d). In contrast, the sub-Keplerian halo accretion rate remains relatively stable, spanning $\dot{m}_h=0.63^{+0.09}_{-0.12}\times10^{-2}$ to $1.75^{+0.41}_{-0.52}\times10^{-2}~\dot{m}_{\rm Edd}$ (Figure~\ref{fig:2018_param}e). The compression ratio remained near $R\sim1.85$, indicating no substantial change in the post-shock density contrast. On the other side, the shock location, $X_s$, exhibits significant variability, ranging from $19^{+05}_{-03}~R_g$ to $87^{+19}_{-13}~R_g$, showing an anti-correlation with luminosity (Figure~\ref{fig:2018_param}f). This behaviour is consistent with a scenario in which lower-luminosity states are associated with an expanded Comptonizing region and harder spectra ($\Gamma\sim1.6$), while higher accretion rates lead to a more compact corona and softer continua accompanied by enhanced X-ray emission. The cooling and infall timescales inferred from the \textsc{TCAF} parameters follow a similar evolutionary pattern, with the cooling timescale varying from $T_c=0.61\pm0.05\times10^4~s$ to $20.69\pm2.10\times10^4~s$ and the infall timescale from $T_{fall}=2.64\pm0.18\times10^4~s$ to $12.25\pm3.87\times10^4~s$. The ratio of the cooling to infall timescales, $\tau$, remained in the range $0.5\le\tau\le1.5$ for nearly all observations (Figure \ref{fig:2018_param}g) after MJD 58236 (2018 April 28), coincident with the epoch during which quasi-periodic variability is suggested by the timing analysis (Section~\ref{sec:2018_lc}). All the best-fitting \textsc{TCAF} parameters are listed in Table~\ref{tab:2018_physical}, with their temporal evolution shown in Figure~\ref{fig:2018_param}.

To evaluate the impact of the UV data on the physical interpretation of the 2018 observations, we performed broadband UV–X-ray spectral fits including the UV photometric point for all available \textit{Swift} epochs. The Galactic extinction was accounted for using the {\tt zredden} component with $E(B-V)=0.0391$, obtained from the Infrared Science Archive.\footnote{\url{http://irsa.ipac.caltech.edu/applications/DUST/}}. The resulting parameters are listed in Table~\ref{tab:2018_physical_uv}. Including a single UV point systematically increases the fit statistic relative to X-ray–only modeling. In these cases, the spectrum converges toward a solution dominated by the hot Comptonizing component. The hot-corona photon index softens to $\Gamma_{hot}\sim2.4$ with minimal variability, and the inferred hot-coronal radius increases to $\sim200-300~r_g$, although the large uncertainties prevent any firm conclusions regarding its evolution.

The broadband fits show systematically higher Eddington ratios than the X-ray–only analysis, spanning $\log \dot{m} \sim -1.7$ to $-1.2$, although they are found to be within the same order. Crucially, the temporal behavior remains unchanged, with variations in X-ray luminosity states corresponding to the change in accretion rates (Figure \ref{fig:2018_param_uv}). The UV-inclusive Eddington ratios retain a strong positive correlation with the 0.3–10 keV luminosity (PCC $= 0.86, ~p<0.01$), indicating that the variability interpretation is driven by the X-ray spectral evolution rather than the UV normalization. This limited leverage is inherent to the energy-conserving {\tt AGNSED} framework \citep{2018MNRAS.480.1247K}, in which the disc and coronal components are coupled through the global accretion power. A single UV measurement constrains only the luminosity scale and does not probe the disc spectral shape predicted by thin-disc theory \citep{1973A&A....24..337S}, thereby increasing the parameter value of the Eddington ratio without altering its temporal trend.

For consistency, we also performed the broadband UV-X-ray modelling with the {\tt TCAF} model. Although the UV inclusion modestly increases the fit statistic, the derived parameters, $\dot{m}_d, \dot{m}_h, X_s$ and $R$ remain consistent with the X-ray–only results within uncertainties. The evolution of accretion rate continues to track the X-ray luminosity with the same strong correlation (PCC $= 0.86, ~p<0.01$), indicating that the inferred accretion rate dynamics are robust. The temporal variation of relevant parameters is presented in Figure \ref{fig:2018_param_uv}.

The 2018 dataset lacks sufficient EUV–UV coverage to independently constrain the disc spectral shape. The dominant physical constraints, therefore, arise from the X-ray Comptonization spectrum. While the timing analysis does not establish a coherent QPO, the spectral evolution suggests that the coronal cooling and infall timescales approach comparable values during the modulation phase, consistent with expectations from advective-flow models. We therefore interpret the spectral evolution as providing a physically motivated context for the observed quasi-periodic modulation, without claiming a definitive QPO detection.

\begin{table*}
\caption{The best fit parameters for the physical models TCAF and AGNSED for the \textit{Swift}/XRT observations of Mrk 530 in 2018.}
    \resizebox{\textwidth}{!}{\begin{tabular}{c|c|c|c|c|c|c|c|c|c|c|c|c|c|c}
        \hline 
        & & & & & TCAF & & & & & & & AGNSED & & \\
        \hline
        Date & $\dot{m}_d$ & $\dot{m}_h$ & $X_s$ & $R$ & $Norm^\dagger$ & $T_c$ & $T_{fall}$ & $\tau$ & C-stat/dof & log $\dot{m}$ & $\Gamma_{hot}$ & $KT_{e,hot}$ & $R_{hot}$ & C-stat/dof \\
         (MJD) & ($ 10^{-3} ~\dot{m}_{Edd}$) & $(10^{-2} ~\dot{m}_{Edd})$ & ($R_g$) &  & $(10^{-4})$ & $( 10^4~s)$  & ($ 10^{4}~s$) & $(T_c/T_{fall})$ & & & & (keV) & $(R_g)$ & \\
         \hline
         \vspace{0.2cm}
         58120 & $9.48^{+1.28}_{-1.21}$ & $1.69^{+0.51}_{-0.42}$ & $19^{+5}_{-3}$ & $1.79^{+0.48}_{-0.49}$ & $10.50^{+1.45}_{-1.51}$ & $0.61\pm0.05$ & $2.64\pm0.18$ & $0.23\pm0.03$ & 213.14/188 & $-1.49^{+0.08}_{-0.07}$ & $2.23^{+0.04}_{-0.04}$ & $160^{+97}_{-97}$ & $249^{+169}_{-163}$ & 212.76/189 \\ \vspace{0.2cm}
         
         58124 & $8.06^{+1.67}_{-1.60}$ & $1.63^{+0.48}_{-0.49}$ & $25^{+5}_{-4}$ & $1.86^{+0.69}_{-0.68}$ & $9.95^{+1.98}_{-1.84}$ & $1.21\pm0.10$ & $3.59\pm0.29$ & $0.34\pm0.04$ & 85.76/90 & $-1.73^{+0.05}_{-0.03}$ & $2.10^{+0.07}_{-0.05}$ & $289^{+6}_{-134}$ & $26^{+60}_{-1}$ & 81.11/92 \\ \vspace{0.2cm}
         
         58128 & $6.22^{+1.77}_{-1.01}$ & $1.17^{+0.41}_{-0.41}$ & $23^{+7}_{-5}$ & $1.95^{+0.34}_{-0.11}$ & $10.10^{+1.00}_{-1.01}$ & $1.34\pm0.15$ & $3.23\pm0.18$ & $0.41\pm0.05$ & 145.77/145 & $-1.76^{+0.11}_{-0.05}$ & $2.12^{+0.06}_{-0.06}$ & $252^{+5}_{-152}$ & $63^{+281}_{-50}$ & 169.00/147 \\ \vspace{0.2cm}
         
         58131 & $6.06^{+1.36}_{-1.05}$ & $1.23^{+2.99}_{-2.86}$ & $34^{+2}_{-1}$ & $1.79^{+0.15}_{-0.55}$ & $10.10^{+1.47}_{-1.34}$ & $2.23\pm0.08$ & $4.78\pm0.19$ & $0.47\pm0.02$ & 54.84/58 & $-1.75^{+0.07}_{-0.06}$ & $1.79^{+0.07}_{-0.07}$ & $140^{+107}_{-79}$ & $122^{+246}_{-91}$ & 60.59/68 \\ \vspace{0.2cm}
         
         58133 & $7.95^{+1.75}_{-1.08}$ & $1.75^{+0.41}_{-0.52}$ & $42^{+8}_{-5}$ & $1.90^{+0.67}_{-0.89}$ & $10.00^{+1.99}_{-2.08}$ & $3.16\pm0.23$ & $5.91\pm0.51$ & $0.53\pm0.06$ & 77.33/66 & $-1.90^{+0.04}_{-0.03}$ & $1.95^{+0.07}_{-0.07}$ & $163^{+82}_{-88}$ & $168^{+227}_{-126}$ & 91.52/77 \\ \vspace{0.2cm}
         
         58135 & $5.87^{+1.88}_{-1.75}$ & $1.28^{+0.40}_{-0.44}$ & $49^{+11}_{-16}$ & $1.90^{+0.42}_{-1.80}$ & $10.10^{+2.94}_{-3.88}$ & $5.88\pm0.70$ & $6.90\pm0.89$ & $0.85\pm0.15$ & 66.66/43 & $-1.87^{+0.11}_{-0.06}$ & $1.83^{+0.10}_{-0.11}$ & $90^{+123}_{-39}$ & $80^{+232}_{-42}$ & 76.84/50 \\ \vspace{0.2cm}
         
         58137 & $5.70^{+1.55}_{-1.07}$ & $1.23^{+0.10}_{-0.28}$ & $43^{+7}_{-14}$ & $1.82^{+0.99}_{-1.84}$ & $10.00^{+1.72}_{-1.57}$ & $4.71\pm0.47$ & $6.07\pm0.98$ & $0.78\pm0.15$ & 56.86/54 & $-1.89^{+0.06}_{-0.05}$ & $1.83^{+0.09}_{-0.08}$ & $147^{+101}_{-85}$ & $149^{+232}_{-108}$ & 57.75/55 \\ \vspace{0.2cm}
         
         58236 & $3.25^{+1.29}_{-1.11}$ & $0.63^{+0.09}_{-0.12}$ & $29^{+1}_{-6}$ & $1.86^{+1.22}_{-1.01}$ & $10.20^{+5.73}_{-5.41}$ & $4.13\pm0.24$ & $4.14\pm0.51$ & $0.99\pm0.14$ & 27.9/29 & $-2.01^{+0.05}_{-0.06}$ & $1.93^{+0.15}_{-0.15}$ & $135^{+109}_{-87}$ & $90^{+251}_{-62}$ & 27.79/30 \\ \vspace{0.2cm}
         
         58238 & $4.41^{+1.20}_{-1.04}$ & $0.85^{+0.18}_{-0.09}$ & $29^{+3}_{-3}$ & $1.81^{+1.53}_{-0.74}$ & $10.00^{+1.82}_{-1.08}$ & $3.02\pm0.14$ & $4.11\pm0.52$ & $0.73\pm0.10$ & 88.07/79 & $-1.86^{+0.05}_{-0.03}$ & $1.91^{+0.07}_{-0.08}$ & $119^{+114}_{-46}$ & $144^{+233}_{-115}$ & 100.07/80 \\ \vspace{0.2cm}
         
         58240 & $4.82^{+0.95}_{-0.98}$ & $0.99^{+0.11}_{-0.07}$ & $33^{+3}_{-6}$ & $1.89^{+1.89}_{-1.20}$ & $9.98^{+2.22}_{-3.79}$ & $3.44\pm0.20$ & $4.69\pm0.78$ & $0.73\pm0.13$ & 94.6/92 & $-1.83^{+0.09}_{-0.06}$ & $1.82^{+0.06}_{-0.06}$ & $115^{+89}_{-65}$ & $62^{+243}_{-44}$ & 97.98/93 \\ \vspace{0.2cm}
        
         58242 & $4.77^{+1.55}_{-1.52}$ & $0.98^{+0.11}_{-0.10}$ & $33^{+5}_{-5}$ & $1.86^{+1.90}_{-1.05}$ & $10.10^{+2.92}_{-2.98}$ & $3.47\pm0.23$ & $4.68\pm0.76$ & $0.74\pm0.13$ & 20.44/23 & $-1.85^{+0.10}_{-0.08}$ & $1.98^{+0.11}_{-0.11}$ & $131^{+118}_{-56}$ & $36^{+256}_{-26}$ & 28.10/25 \\ \vspace{0.2cm}
        
         58244 & $5.24^{+1.08}_{-1.07}$ & $1.03^{+0.11}_{-0.11}$ & $32^{+4}_{-3}$ & $1.85^{+1.09}_{-1.08}$ & $11.10^{+1.47}_{-1.41}$ & $3.13\pm0.14$ & $4.60\pm0.55$ & $0.68\pm0.09$ & 83.78/93  & $-1.82^{+0.05}_{-0.03}$ & $1.99^{+0.07}_{-0.07}$ & $171^{+62}_{-101}$ & $120^{+232}_{-87}$ & 81.63/94 \\ \vspace{0.2cm}
        
         58246 & $5.32^{+1.04}_{-1.09}$ & $1.36^{+0.16}_{-0.19}$ & $32^{+10}_{-5}$ & $1.84^{+1.03}_{-1.72}$ & $10.10^{+1.79}_{-1.38}$ & $2.48\pm0.24$ & $4.51\pm0.71$ & $0.55\pm0.10$ & 82.25/75 & $-1.79^{+0.05}_{-0.04}$ & $1.91^{+0.08}_{-0.07}$ & $144^{+69}_{-93}$ & $98^{+260}_{-65}$ & 91.99/76 \\ \vspace{0.2cm}
        
         58250 & $6.41^{+0.99}_{-1.01}$ & $1.46^{+1.79}_{-1.05}$ & $43^{+5}_{-9}$ & $1.89^{+1.84}_{-1.88}$ & $10.00^{+0.15}_{-0.15}$ & $4.01\pm0.27$ & $6.05\pm1.21$ & $0.66\pm0.14$ & 96.08/98 & $-1.71^{+0.06}_{-0.04}$ & $1.84^{+0.06}_{-0.05}$ & $141^{+118}_{-78}$ & $146^{+229}_{-106}$ & 119.20/100 \\ \vspace{0.2cm} 
        
         58252 & $5.42^{+1.08}_{-1.01}$ & $1.01^{+0.11}_{-0.11}$ & $24^{+10}_{-7}$ & $1.86^{+0.96}_{-1.01}$ & $10.10^{+1.71}_{-1.83}$ & $1.68\pm0.24$ & $3.36\pm0.43$ & $0.50\pm0.09$ & 98.01/97 & $-1.81^{+0.05}_{-0.04}$ & $1.95^{+0.07}_{-0.07}$ & $161^{+85}_{-90}$ & $41^{+208}_{-20}$ & 105.94/98 \\ \vspace{0.2cm}
        
         58254 & $5.44^{+0.99}_{-1.06}$ & $1.05^{+0.18}_{-0.17}$ & $26^{+9}_{-7}$ & $1.89^{+1.94}_{-1.81}$ & $10.00^{+1.86}_{-1.67}$ & $1.88\pm0.23$ & $3.61\pm0.75$ & $0.52\pm0.13$ & 76.55/68 & $-1.78^{+0.04}_{-0.04}$ & $2.00^{+0.01}_{-0.01}$ & $117^{+65}_{-69}$ & $55^{+179}_{-38}$ & 75.94/67 \\ \vspace{0.2cm}
        
         58256 & $5.49^{+1.05}_{-1.00}$ & $1.08^{+0.20}_{-0.19}$ & $30^{+10}_{-6}$ & $1.85^{+1.01}_{-1.22}$ & $9.99^{+1.93}_{-1.01}$ & $2.55\pm0.28$ & $4.25\pm0.56$ & $0.60\pm0.10$ & 75.64/69 & $-1.85^{+0.08}_{-0.06}$ & $1.81^{+0.08}_{-0.08}$ & $146^{+102}_{-56}$ & $64^{+249}_{-39}$ & 77.01/71 \\ \vspace{0.2cm}
        
         58258 & $5.05^{+1.08}_{-1.70}$ & $0.92^{+0.24}_{-0.25}$ & $27^{+10}_{-11}$ & $1.81^{+2.82}_{-1.81}$ & $9.97^{+4.81}_{-3.18}$ & $2.38\pm0.38$ & $3.83\pm1.02$ & $0.62\pm0.19$ & 28.99/28 & $-1.98^{+0.08}_{-0.06}$ & $2.03^{+0.10}_{-0.10}$ & $148^{+83}_{-92}$ & $72^{+253}_{-46}$ & 19.86/30 \\ \vspace{0.2cm}
        
         58260 & $4.04^{+1.01}_{-1.04}$ & $0.80^{+0.12}_{-0.11}$ & $32^{+10}_{-6}$ & $1.85^{+1.11}_{-1.04}$ & $10.10^{+1.01}_{-1.00}$ & $3.98\pm0.40$ & $4.57\pm0.58$ & $0.87\pm0.14$ & 91.76/80 & $-1.98^{+0.05}_{-0.03}$ & $1.95^{+0.06}_{-0.06}$ & $160^{+98}_{-79}$ & $51^{+225}_{-26}$ & 110.01/81 \\ \vspace{0.2cm}
        
         58262 & $5.76^{+1.08}_{-1.05}$ & $1.18^{+0.40}_{-0.31}$ & $32^{+5}_{-5}$ & $1.89^{+1.99}_{-1.80}$ & $10.20^{+3.18}_{-4.81}$ & $2.67\pm0.20$ & $4.51\pm0.92$ & $0.59\pm0.13$ & 31.76/32 & $-1.80^{+0.10}_{-0.05}$ & $2.07^{+0.11}_{-0.13}$ & $184^{+60}_{-114}$ & $85^{+250}_{-38}$ & 25.23/34 \\ \vspace{0.2cm}
         
         58264 & $4.09^{+1.04}_{-1.02}$ & $0.89^{+0.13}_{-0.13}$ & $39^{+8}_{-9}$ & $1.81^{+1.65}_{-1.04}$ & $10.00^{+1.19}_{-1.18}$ & $5.33\pm0.47$ & $5.48\pm0.85$ & $0.97\pm0.17$ & 84.97/85 & $-1.94^{+0.05}_{-0.03}$ & $1.85^{+0.06}_{-0.06}$ & $127^{+113}_{-59}$ & $182^{+221}_{-145}$ & 86.73/87 \\ \vspace{0.2cm}
         
         58265 & $4.80^{+0.99}_{-1.05}$ & $1.15^{+0.25}_{-0.28}$ & $49^{+11}_{-6}$ & $1.85^{+2.09}_{-1.83}$ & $10.10^{+2.98}_{+2.52}$ & $6.70\pm0.51$ & $6.89\pm1.48$ & $0.97\pm0.22$ & 87.75/89 & $-1.86^{+0.07}_{-0.05}$ & $1.81^{+0.07}_{-0.06}$ & $129^{+109}_{-58}$ & $129^{+263}_{-82}$ & 87.40/91 \\ \vspace{0.2cm}         
         
         58268 & $4.59^{+1.00}_{-1.06}$ & $1.07^{+0.38}_{-0.11}$ & $47^{+7}_{-3}$ & $1.84^{+1.42}_{-1.00}$ & $9.99^{+1.00}_{-1.01}$ & $6.77\pm0.37$ & $6.70\pm0.89$ & $1.01\pm0.15$ & 72.4/71 & $-1.95^{+0.08}_{-0.06}$ & $1.81^{+0.08}_{-0.06}$ & $144^{+101}_{-67}$ & $68^{+283}_{-34}$ & 80.17/73 \\ \vspace{0.2cm}
         
         58270 & $4.05^{+1.10}_{-1.03}$ & $1.26^{+0.16}_{-0.19}$ & $70^{+8}_{-4}$ & $1.84^{+1.01}_{-1.69}$ & $9.95^{+2.07}_{-2.18}$ & $13.67\pm0.59$ & $9.94\pm1.47$ & $1.38\pm0.21$ & 43.01/48 & $-1.86^{+0.18}_{-0.14}$ & $1.60^{+0.08}_{-0.08}$ & $92^{+118}_{-46}$ & $91^{+256}_{-62}$ & 43.79/50 \\ \vspace{0.2cm}
         
         58272 & $4.09^{+0.92}_{-1.00}$ & $1.18^{+0.23}_{-0.21}$ & $65^{+10}_{-7}$ & $1.87^{+3.05}_{-1.07}$ & $10.00^{+2.89}_{-2.00}$ & $12.34\pm0.75$ & $9.23\pm2.05$ & $1.34\pm0.31$ & 56.4/57 & $-2.07^{+0.10}_{-0.07}$ & $1.75^{+0.08}_{-0.08}$ & $174^{+83}_{-113}$ & $159^{+215}_{-89}$ & 64.86/59 \\ \vspace{0.2cm} 
         
         58278 & $3.12^{+0.51}_{-0.49}$ & $1.19^{+0.26}_{-0.26}$ & $75^{+13}_{-9}$ & $1.84^{+2.58}_{-1.33}$ & $10.10^{+1.97}_{-1.58}$ & $17.16\pm1.18$ & $10.58\pm2.27$ & $1.62\pm0.37$ & 37.55/42 & $-2.20^{+0.09}_{-0.06}$ & $1.78^{+0.07}_{-0.09}$ & $181^{+59}_{-115}$ & $126^{+221}_{-81}$ & 34.01/44 \\ \vspace{0.2cm}
         
         58280 & $2.57^{+0.97}_{-0.47}$ & $1.29^{+0.33}_{-0.30}$ & $85^{+9}_{-5}$ & $1.85^{+2.09}_{-1.48}$ & $10.00^{+1.63}_{-1.42}$ & $21.37\pm1.15$ & $11.98\pm2.32$ & $1.78\pm0.36$ & 40.67/37 & $-2.18^{+0.17}_{-0.14}$ & $1.66^{+0.11}_{-0.10}$ & $129^{+68}_{-61}$ & $114^{+254}_{-62}$ & 56.54/38 \\ \vspace{0.2cm}
         
         58282 & $3.31^{+0.99}_{-0.89}$ & $1.34^{+0.60}_{-0.55}$ & $87^{+19}_{-13}$ & $1.84^{+3.97}_{-1.80}$ & $10.20^{+6.03}_{-5.98}$ & $20.69\pm2.10$ & $12.25\pm3.87$ & $1.69\pm0.56$ & 19.29/21 & $-2.21^{+0.18}_{-0.11}$ & $1.70^{+0.12}_{-0.12}$ & $179^{+79}_{-117}$ & $99^{+246}_{-35}$ & 19.87/23 \\ \vspace{0.2cm}
         
         58283 & $3.06^{+1.08}_{-1.02}$ & $1.17^{+3.07}_{-2.00}$ & $72^{+10}_{-8}$ & $1.86^{+1.85}_{-1.08}$ & $10.10^{+1.87}_{-2.04}$ & $16.01\pm1.00$ & $10.13\pm1.62$ & $1.58\pm0.27$ & 43.46/48 & $-2.11^{+0.07}_{-0.06}$ & $1.78^{+0.09}_{-0.08}$ & $155^{+92}_{-96}$ & $187^{+198}_{-124}$ & 58.06/50 \\ \vspace{0.2cm}
         
         58286 & $3.62^{+1.07}_{-1.04}$ & $1.29^{+0.48}_{-0.32}$ & $78^{+11}_{-8}$ & $1.84^{+2.04}_{-1.00}$ & $10.00^{+3.49}_{-2.74}$ & $17.15\pm1.20$ & $11.09\pm1.85$ & $1.55\pm0.28$ & 43.82/44 & $-2.12^{+0.11}_{-0.10}$ & $1.72^{+0.08}_{-0.08}$ & $172^{+72}_{-107}$ & $139^{+234}_{-70}$ & 39.34/46 \\ \vspace{0.2cm}
        
         58288 & $3.58^{+0.86}_{-0.94}$ & $1.02^{+0.28}_{-0.63}$ & $56^{+7}_{-6}$ & $1.84^{+1.24}_{-1.72}$ & $10.00^{+1.91}_{-1.48}$ & $10.36\pm0.85$ & $7.95\pm1.29$ & $1.30\pm0.29$ & 39.44/43 & $-2.18^{+0.09}_{-0.05}$ & $1.82^{+0.09}_{-0.10}$ & $141^{+99}_{-84}$ & $130^{+217}_{-78}$ & 45.98/45 \\ \vspace{0.2cm}
         
         58292 & $2.77^{+1.00}_{-1.08}$ & $1.01^{+0.24}_{-0.23}$ & $63^{+14}_{-9}$ & $1.85^{+1.44}_{-1.67}$ & $9.99^{+2.10}_{-1.97}$ & $14.22\pm1.18$ & $8.92\pm1.53$ & $1.60\pm0.30$ & 20.28/20 & $-1.90^{+0.21}_{-0.21}$ & $1.59^{+0.13}_{-0.11}$ & $125^{+124}_{-53}$ & $101^{+202}_{-62}$ & 31.17/22 \\ \vspace{0.2cm}
         
         58294 & $3.25^{+1.06}_{-1.09}$ & $0.93^{+0.59}_{-0.78}$ & $50^{+15}_{-9}$ & $1.83^{+1.06}_{-1.80}$ & $10.10^{+4.08}_{-7.80}$ & $9.22\pm1.36$ & $7.08\pm1.16$ & $1.30\pm0.29$ & 71.88/80 & $-1.90^{+0.10}_{-0.07}$ & $1.74^{+0.07}_{-0.06}$ & $116^{+129}_{-75}$ & $50^{+301}_{-26}$ & 84.30/82 \\ \vspace{0.2cm}
         
         58296 & $4.88^{+0.99}_{-0.98}$ & $1.00^{+0.10}_{-0.09}$ & $37^{+5}_{-3}$ & $1.82^{+1.89}_{-0.80}$ & $9.99^{+1.08}_{-1.07}$ & $4.16\pm0.21$ & $5.18\pm0.77$ & $0.80\pm0.13$ & 32.89/36 & $-1.95^{+0.07}_{-0.04}$ & $1.89^{+0.10}_{-0.11}$ & $123^{+119}_{-57}$ & $79^{+256}_{-47}$ & 44.13/38 \\ \vspace{0.2cm}
         
         58298 & $5.03^{+1.07}_{-1.05}$ & $0.71^{+0.16}_{-0.11}$ & $27^{+10}_{-7}$ & $1.81^{+1.57}_{-0.80}$ & $10.30^{+3.89}_{-3.87}$ & $2.84\pm0.36$ & $3.87\pm0.56$ & $0.73\pm0.14$ & 60.01/58 & $-1.83^{+0.06}_{-0.04}$ & $1.98^{+0.07}_{-0.08}$ & $177^{+79}_{-99}$ & $83^{+256}_{-61}$ & 87.95/60 \\ \vspace{0.2cm}
         
         58301 & $4.68^{+0.58}_{-0.46}$ & $0.98^{+0.13}_{-0.13}$ & $41^{+8}_{-3}$ & $1.84^{+1.06}_{-0.34}$ & $10.00^{+2.99}_{-2.07}$ & $5.38\pm0.32$ & $5.82\pm0.47$ & $0.92\pm0.09$ & 55.69/66 & $-1.97^{+0.12}_{-0.08}$ & $1.73^{+0.07}_{-0.09}$ & $113^{+124}_{-46}$ & $124^{+250}_{-85}$ & 69.59/69 \\ \vspace{0.2cm}
         
         58306 & $4.39^{+0.99}_{-0.87}$ & $0.97^{+0.19}_{-0.14}$ & $42^{+3}_{-3}$ & $1.84^{+1.69}_{-0.51}$ & $10.10^{+2.94}_{-2.78}$ & $5.80\pm0.22$ & $5.97\pm0.72$ & $0.97\pm0.12$ & 74.14/74 & $-1.97^{+0.05}_{-0.04}$ & $1.84^{+0.08}_{-0.07}$ & $117^{+108}_{-60}$ & $147^{+240}_{-108}$ & 87.77/76 \\ \vspace{0.2cm}
         
         58310 & $4.80^{+0.84}_{-0.83}$ & $1.00^{+0.16}_{-0.15}$ & $36^{+4}_{-6}$ & $1.84^{+1.99}_{-0.57}$ & $9.99^{+3.40}_{-3.88}$ & $3.94\pm0.24$ & $5.03\pm0.71$ & $0.78\pm0.12$ & 67.35/74 & $-1.88^{+0.05}_{-0.04}$ & $1.89^{+0.07}_{-0.07}$ & $119^{+108}_{-67}$ & $144^{+243}_{-106}$ & 71.06/76 \\ \vspace{0.2cm}
         
         58314 & $5.39^{+0.47}_{-0.41}$ & $1.05^{+0.10}_{-0.09}$ & $32^{+2}_{-2}$ & $1.84^{+0.92}_{-0.40}$ & $10.00^{+1.09}_{-1.89}$ & $2.93\pm0.08$ & $4.49\pm0.33$ & $0.65\pm0.05$ & 120.61/116 & $-1.95^{+0.03}_{-0.04}$ & $1.91^{+0.04}_{-0.05}$ & $135^{+101}_{-65}$ & $98^{+234}_{-67}$ & 112.31/118 \\ \vspace{0.2cm}
         
         58316 & $5.29^{+0.97}_{-0.98}$ & $1.22^{+0.16}_{-0.11}$ & $52^{+3}_{-8}$ & $1.84^{+0.70}_{-0.81}$ & $9.96^{+2.03}_{-2.98}$ & $7.24\pm0.34$ & $7.42\pm0.63$ & $0.98\pm0.09$ & 67.25/66 & $-1.91^{+0.12}_{-0.06}$ & $1.80^{+0.08}_{-0.08}$ & $187^{+64}_{-91}$ & $131^{+221}_{-104}$ & 69.03/67 \\
         \hline
    \end{tabular}}
    \label{tab:2018_physical}
\end{table*}

\section{Discussion}
\label{discussion}
Over the past two decades (2001–2024), Mrk~530 has been systematically monitored by multiple X-ray missions, allowing us to track the long-term evolution of its accretion flow. Our spectral analysis has revealed variability in both the primary continuum and the soft excess, with the latter gradually weakening and eventually disappearing in the most recent epochs. The coronal properties, inferred from both phenomenological and physical models, demonstrate that fluctuations in the inflow are accompanied by systematic changes in the size and structure of the Comptonizing region. Timing analysis further strengthens this picture by detecting quasi-periodic oscillations during 2018, which appear physically connected to the observed spectral variability. Together, these results indicate that variations in mass accretion rate and coronal geometry regulate the luminosity and spectral state of the source. This combined timing–spectral approach provides new insights into the long-term accretion dynamics of Mrk~530. In the following, we discuss the principal findings summarized in Sections~\ref{timing} and \ref{spectral}.

\subsection{Long-term variation of spectral properties}

The 2001 \textit{XMM-Newton} observations (XMM1 and XMM2) provide important constraints on the interplay between the accretion flow and the coronal structure in Mrk~530. During XMM1, the source exhibits its strongest soft excess, with a luminosity of $\log L_{SE}\sim43.84$, alongside a moderate primary continuum ($\log L_{PC}\sim43.71$) and distinct spectral slopes ($\Gamma_{PC}\sim1.73$, $\Gamma_{SE}\sim2.47$). A weak Fe~K$\alpha$ line was detected at $6.38\pm0.11$ keV (EW $=168^{+106}_{-107}$ eV). Five months later, in XMM2, the primary continuum luminosity remains approximately constant, while the spectrum steepens ($\log L_{PC}\sim43.70$, $\Gamma_{PC}\sim2.06$), and the soft excess weakens ($\log L_{SE}\sim43.68$, $\Gamma_{SE}\sim2.59$). Consequently, the Fe~K$\alpha$ feature becomes weaker, with an equivalent width of (EW $=65^{+52}_{-51}$ eV at $6.43\pm0.07$ keV).

Physical modeling of the X-ray spectrum supports phenomenological evolution. In XMM1, the soft excess is reproduced by emission from a warm corona with an electron temperature of $kT_{e,\rm warm}\sim0.2$ keV and a characteristic size of $R_{\rm warm}\sim22~R_g$, while the hard continuum arises in a compact hot corona ($kT_{e,\rm hot}\sim224$ keV, $R_{\rm hot}\sim15~R_g$). The \textsc{TCAF} model fittings indicate a Keplerian disc accreting at $\dot{m}_d\sim9.37\times10^{-3}~\dot{m}_{\rm Edd}$ embedded within a sub-Keplerian halo at $\dot{m}_h\sim1.7\times10^{-2}~\dot{m}_{\rm Edd}$, forming a Comptonizing region bounded by a shock at $X_s\sim19~R_g$. By XMM2, the warm corona contracts to $\sim10~R_g$, naturally reducing the soft-excess luminosity. At the same time, both the hot coronal size and shock location decrease, leading to a steeper primary continuum, while the primary luminosity remains nearly unchanged owing to the approximately constant accretion rate log $\dot{m}\sim-1.6$.

Thus, the short-term evolution between XMM1 and XMM2 is best explained by dynamical changes in the warm and hot coronal structures occurring at nearly constant mass accretion rate. Such behaviour is consistent with the broader Seyfert population, in which both warm and hot corona are known to exhibit complex variability on relatively short timescales \citep{1997ApJ...476..620H,2020A&A...634A..92U,Hu2022}.

The 2006 monitoring campaign with \textit{Swift}/XRT (XRT1, XRT2) and \textit{XMM-Newton} (XMM3) shows comparatively stable spectral and coronal properties. The photon indices of both the primary continuum ($\Gamma_{PC}\sim1.8$--$1.9$) and the soft excess ($\Gamma_{SE}\sim2.5$-$3.0$) remain consistent within uncertainties, while luminosity variations are modest. A weak Fe~K$\alpha$ line detected at $6.39\pm0.07$ keV (EW $=44^{+27}_{-23}$ eV) suggests reflection from the disc or circumnuclear material. Physical modelling indicates a persistently compact hot corona ($R_{\rm hot}\sim10~R_g$), with temperatures spanning $kT_{e,hot}\sim100-200$ keV (Figure~\ref{fig:longterm}f) while the warm corona remains stable with $kT_{e,warm}\sim0.16$ keV and $R_{warm}\sim20~R_g$ (Figures~\ref{fig:longterm}j,k). Both \textsc{AGNSED} and \textsc{TCAF} fits indicate stable accretion rates during this period, accompanied by a nearly constant shock location $(X_s\sim20~R_g)$ (Figure~\ref{fig:longterm}i) and compression ratio ($R\sim1.85$). These results point to a scenario in which a stable hot corona dominates the X-ray emission, while an approximately invariant warm corona produces the soft excess.

In 2016 (XRT3), Mrk~530 entered a low-flux state in which both the primary continuum and soft excess were significantly weakened. The soft excess becomes undetectable and the spectrum is dominated by emission from the hot corona, with $kT_{e,\rm hot}\sim145$ keV, $R_{\rm hot}\sim22~R_g$, and $\Gamma_{\rm hot}\sim2.04$, consistent with a reduced accretion rate ($\log\dot{m}\sim-1.97$). Correspondingly, the X-ray luminosities decline, and weak spectral features such as the Fe~K$\alpha$ line and intrinsic absorption are no longer detected, likely due to the reduced flux. The \textsc{TCAF} analysis confirms declines in both disc and halo accretion rates and indicates an expansion of the Comptonizing region to $X_s\sim31~R_g$ (Figure~\ref{fig:longterm}i), while the compression ratio remains approximately constant. This behaviour is consistent with a transition toward a corona-dominated accretion state driven by changes in the inflow properties \citep{Hagen2024}.

In 2018 \textit{Swift} monitoring campaign, Mrk~530 exhibits pronounced quasi-periodic variability. Spectral modelling demonstrates that the observed luminosity variations are primarily driven by changes in the mass accretion rate. In low-flux states, a reduced seed-photon supply allows the corona to expand ($X_s\le87R_g$), producing harder spectra, whereas higher accretion rates lead to coronal contraction, spectral softening, and enhanced X-ray emission. Similar accretion–corona coupling has been reported in other AGNs, including Mrk~50 \citep{2025ApJ...981...74L}, Mrk~1018 \citep{noda2018explaining}, and NGC~1566 \citep{2022cosp...44.2313T}. While these results firmly establish accretion-driven coronal variability as the origin of the spectral changes, the physical mechanism underlying the quasi-periodicity remains uncertain.

Between 2022 and 2024, Mrk~530 shows signatures of reduced accretion accompanied by evolving coronal geometry. In 2022 (XRT5), the spectrum is hard and lacks a detectable soft excess, consistent with a diminished disc contribution. The \textsc{AGNSED} fits indicate a hot, extended corona, while the \textsc{TCAF} modelling reveals low disc and halo accretion rates and an expanded Comptonizing region. By 2023–2024 (XRT6 and XRT7), the spectrum softens modestly, while the soft excess continues to weaken. Both modelling approaches indicate relatively stable accretion rates, with the observed variability primarily reflecting coronal reconfiguration rather than major changes in the inflow.

Overall, the long-term X-ray spectral evolution of Mrk~530 reveals substantial variability in both the accretion flow and coronal properties. The prominent soft excess observed during earlier epochs (2001–2006) is naturally explained by the presence of a warm corona with $kT_{\mathrm{e,warm}} \sim 0.2 $ keV. Its disappearance in later epochs likely reflects changes in coronal geometry associated with reduced mass accretion rates. While a warm-corona scenario provides a consistent explanation for the origin and evolution of the soft excess, blurred ionized reflection models cannot be definitively ruled out. Previous work has shown that both scenarios can account for the soft excess in the 2012 Suzaku observation of Mrk~530 \citep{2018MNRAS.478.4214E}. However, the present analysis is limited by the absence of simultaneous hard X-ray coverage above 10 keV and by the weakness of the Fe~K$\alpha$ line, which prevents robust constraints on the reflection continuum and Compton hump. Consequently, given these observational limitations, we do not attempt to further explore blurred reflection models in this work.

\subsection{Possible explanation of QPO in 2018}
\label{sec:qpo}

The 2018 {\it Swift}/XRT provides a notable case in which timing and spectral diagnostics point to a closely related origin of the observed variability. Both the UV and X-ray light curves exhibit quasi-periodic modulations, occurring on different characteristic timescales (approximately 90 days in the UV and 60 days in the X-rays), with the UV variations leading those in the X-ray band. This behaviour suggests variability arising in physically coupled but spatially distinct regions of the accretion flow. Previous work by \citet{2018MNRAS.478.4214E} reported a sinusoidal feature in the 0.3--10 keV band with a characteristic period of $T\sim1$ day, which was found to be statistically insignificant and was most prominent in the 1--3 keV range, without a clear physical interpretation. In contrast, the quasi-periodic modulations reported here are detected independently in both the UV and X-ray bands and are statistically significant, although they exhibit low coherence. In addition, the variability is broadly distributed across the full 0.3--10 keV energy range. The large disparity between the characteristic periods indicates that the two phenomena are likely associated with different physical mechanisms. Within this context, the observed UV–X-ray variability can be plausibly interpreted as arising from accretion-rate perturbations that first affect the outer regions of the Comptonizing flow, producing modulation in the UV band, and subsequently propagate inward to the more compact X-ray–emitting region on characteristic dynamical timescales \citep{AU2006, Pahari2020, Hagen2024}. Such a scenario is consistent with models invoking an extended corona with radial gradients in optical depth and temperature, in which UV photons originate at larger radii while X-ray emission is produced in deeper, hotter regions closer to the black hole \citep{Wilkins2016, Chainakun2019, Zhang2023}

Spectral analysis supports this picture. A strong correlation was found between luminosity and spectral slope. The low-luminosity states corresponded to harder spectra ($\Gamma \sim 1.6$) and high-luminosity states are accompanied by softer spectra ($\Gamma \sim 2.2$; Figures~\ref{fig:2018_param}a–b). \textsc{AGNSED} modeling attributed this directly to changes in the accretion rate ($\log \dot{m} \sim -2.21$ to $-1.49$), where higher accretion increased the seed photon supply, producing softer and brighter emission (Figure~\ref{fig:2018_param}c). \textsc{TCAF} analysis further showed that this variability was driven by changes in the Keplerian disc rate, while the sub-Keplerian halo remained relatively stable (Figure~\ref{fig:2018_param}d–e). The shock location shifted between $18$–$87~R_{\rm g}$, anti-correlated with luminosity (Figure~\ref{fig:2018_param}f). These results suggest a scenario where lower accretion rates allowed the corona to expand outward, hardening the spectrum. Whereas higher disc accretion compressed the corona, yielding softer and more luminous states.

The temporal and spectral analyses converge on a coherent scenario in which the 2018 quasi-periodic signals of Mrk~530 arise from accretion-rate oscillations that periodically reconfigure the hot corona. The longer UV period traces the modulation of the outer corona, while the shorter X-ray period reflects fluctuations of the inner Compton cloud. The observed phase lag is naturally explained by inward propagation of perturbations through the disk–corona system. Estimates of cooling and infall timescales from the TCAF model satisfy the resonance condition ($0.5 \lesssim \tau \lesssim 1.5$) expected for low-frequency QPOs (LF-QPOs), directly linking the variability to oscillations of the Comptonizing region (Figure~\ref{fig:2018_param}g). In contrast, we note that the resonance condition was not satisfied during any other observation in the long-term X-ray monitoring of Mrk~530 (Table \ref{tab:physical}). Thus, the 2018 behavior can be interpreted as a resonance-driven QPO triggered by fluctuations in the disk accretion rate that modulated coronal size and structure. This combined temporal–spectral framework explains both the oscillatory light curves and the correlated evolution of spectral parameters, positioning Mrk~530 as a rare AGN analogue of LF-QPOs seen in stellar-mass black hole binaries \citep{Ingram2009,Ingram_2019}.

We can independently estimate the characteristic location from which the observed variability originates using measurements of the time periods. Since the variability is associated with the Comptonizing flow, rather than the cold accretion disk, the relevant physical timescale is the thermal timescale ($T_{th}$) of the hot inner flow. The oscillations of the Compton cloud are well known to produce QPO-like variability with characteristic frequencies linked to the thermal and dynamical timescales of the hot flow \citep{2000ApJ...531L..41C, 2010MNRAS.404..738C, 2011MNRAS.415.2323I, Ingram_2019}.
In a geometrically thick, radiatively inefficient corona, this thermal timescale increases with distance from the black hole and depends on the efficiency of angular momentum transport, commonly parameterized by the viscosity parameter $\alpha$ \citep{1973A&A....24..337S, 1994ApJ...428L..13N, 1995ApJ...455..623C}. 
\begin{equation}
    T_{th}(r)=\frac{T_{dyn}(r)}{\alpha}=\frac{1}{\alpha}\frac{2\pi GM}{c^3}r^{3/2};~\text{where} ~r=\frac{R}{R_g}
\end{equation}
By assuming that the observed quasi-periodic variability timescale corresponds to the local thermal timescale of the Comptonizing region ($T_{QPO}\sim T_{th}(r)$), we can infer the radial location at which the variability is produced.
\begin{equation}
    r\sim\left(\frac{\alpha T_{QPO}c^3}{2\pi GM}\right)^{2/3}
\end{equation}
Adopting a black hole mass of $M_{BH}=1.15\times10^8~M_\odot$ (see Section~\ref{sec:physical}) and using the characteristic periods measured in the UV and X-ray bands ($T_{X-ray}=59.34\pm2.44$ days; $T_{UV}=87.76\pm2.44$ days; Section~\ref{sec:2018_lc}), we estimate the corresponding variability locations. Considering the viscosity parameter within a physical range ($0.1<\alpha<0.3$), we found that for the UV band, the inferred variability originates at radii of approximately $35-75~R_g$, while the X-ray variability arises from a somewhat smaller radius of $30-60~R_g$. These estimates are consistent with the independently inferred spatial extent of the hot corona from our spectral modelling (Section~\ref{sec:2018_spectra}), reinforcing the interpretation that the observed UV and X-ray quasi-periodic signals originate from different regions of the same radially extended Comptonizing flow.

QPOs in AGNs span a wide range of characteristic timescales, extending from hours to years \citep{zhou2014universal, 2016ApJ...819L..19P, 2024A&A...691A...7Y}. While high-frequency QPOs, such as that reported in RE~J1034+396 \citep{2008Natur.455..369G}, are generally associated with processes occurring in the innermost accretion flow, the variability observed in Mrk~530 falls within the emerging class of low-frequency QPO (LF-QPO) candidates on week-to-year timescales \citep{2013MNRAS.436L.114K, Smith2023, 2023A&A...672A..86R, 2025PASJ...77..381W}. Within the TCAF framework, such variability can be interpreted in terms of expansion–contraction cycles of the Comptonizing region. These variations are consistent with the accretion-rate fluctuations inferred from the \textsc{AGNSED} modelling and are analogous to LF-QPO phenomenology observed in black hole X-ray binaries \citep{2016AN....337..398M, Ingram_2019}. In this picture, quasi-periodic behaviour may arise when the cooling and infall timescales of the Compton cloud become comparable, leading to oscillatory behaviour of the post-shock region \citep{Chakrabarti2015}. While this resonance mechanism provides a physically motivated explanation for the observed QPO candidate in Mrk~530, alternative scenarios, such as global Lense–Thirring precession of a misaligned hot inner flow \citep{1999ApJ...513..827M}, cannot be excluded and may also contribute to the observed variability. In this context, Mrk~530 can be placed within the broader LF-QPO phenomenology, in which relatively slow accretion-rate fluctuations modulate the coronal geometry. Such modulations can imprint quasi-periodic signatures on longer timescales in the UV emitting regions and on shorter timescales in the X-ray emitting zones, while preserving the coupled disc–corona variability observed across both AGNs and Galactic X-ray binaries \citep{1995ApJ...452..379A, 2009ASPC..408..296A}.

\subsection{Relation between Parameters}
The $\sim24$ year spectral study of Mrk~530 reveals pronounced long-term variability, including quasi-periodic oscillations, and establishes a tight link between accretion dynamics and coronal structure. The parameters obtained from different models are not independent; rather, they are interconnected. Our comparative analysis of phenomenological and physical fits highlights consistent physical trends. By relating the photon index, accretion rates, coronal size, and luminosity, we uncover correlations that clarify how the disk and corona co-evolve to regulate the X-ray output.

\begin{figure*}
    \includegraphics[width=\textwidth]{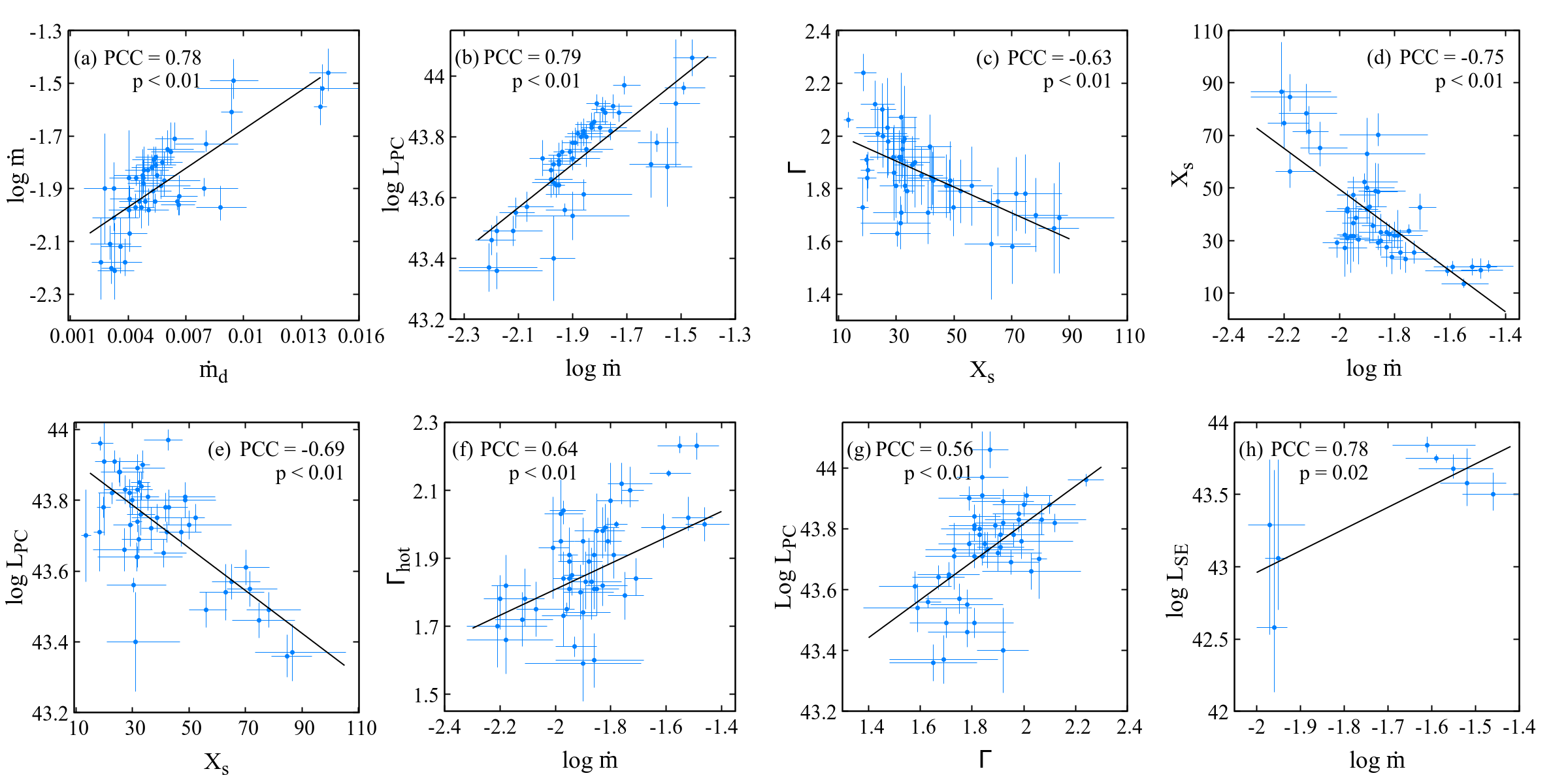}
    \caption{Correlations observed between derived parameters from the spectral analysis of $\sim$24 years X-ray observations of Mrk~530.}
    \label{fig:correlations}
\end{figure*}

The hot-corona photon index from \textsc{AGNSED} ($\Gamma_{\rm hot}$) strongly agrees with the phenomenological slope ($\Gamma$; $PCC \approx 0.90$, $p < 0.01$), confirming that both probe the same Comptonized continuum and that the spectral slope is a robust tracer of the hot electron distribution. The accretion rates from \textsc{TCAF} and \textsc{AGNSED} are also tightly correlated ($PCC \approx 0.78$, $p < 0.01$) and positively track the X-ray luminosity ($PCC \approx 0.79$, $p < 0.01$), consistent with higher disk inflow supplying more seed photons that enhance Comptonized emission (Figure~\ref{fig:correlations}a-b). Thus, variability in Mrk~530 is fundamentally accretion-driven.

Coronal size plays a key role in regulating the spectral response. The shock location in \textsc{TCAF} ($X_s$) anti-correlates with both $\Gamma$ ($PCC \approx -0.63$, $p < 0.01$) and $\dot{m}$ ($PCC \approx -0.75$, $p < 0.01$), indicating that an expanded corona is associated with harder and less luminous spectral states (Figure~\ref{fig:correlations}c). Conversely, higher accretion rates correspond to a more compact hot flow, resulting in softer and more luminous spectra (Figure~\ref{fig:correlations}d). This behaviour is further supported by the strong anti-correlation between $X_s$ and luminosity ($PCC \approx -0.69$, $p < 0.01$). Overall, spectral softening accompanies increasing accretion rate and luminosity, with both the hot-corona photon index $\Gamma_{\rm hot}$ ($PCC \approx 0.64$, $p < 0.01$) and the broadband photon index $\Gamma$ ($PCC \approx 0.56$, $p < 0.01$) showing positive correlations with $\dot{m}$ (Figure~\ref{fig:correlations}f) and with the primary continuum luminosity $L_{PC}$ (Figure~\ref{fig:correlations}g).

During epochs in which a soft excess is detected (2001–2006), the warm corona showed different behavior. Its electron temperature remains approximately constant at $\sim0.2$ keV and shows no significant correlation with other spectral or accretion parameters, suggesting a quasi-stable internal regulation largely decoupled from global accretion-rate variations. In contrast, the soft-excess luminosity displays a strong positive correlation with the mass accretion rate ($PCC \approx 0.78$, $p=0.02$, Figure~\ref{fig:correlations}h), consistent with expectations from warm Comptonisation models commonly invoked to explain the soft excess in Seyfert galaxies \citep{2009MNRAS.394..250M, Done2012, Petrucci2020}.

Overall, these correlations establish a coherent framework in which accretion rate controls both radiative output and coronal structure, while the geometry of the hot flow dictates the spectral hardness. The persistent $\Gamma$–$\dot{m}$ and $X_s$–$L_{PC}$ relations show that the long-term spectral evolution of Mrk~530 can be understood as the interplay between a variable disk supply and a thermally adjusting corona.

\subsection{Properties of Absorbers}
From 2001 to 2006, the soft X-ray spectra below $\sim2$ keV were significantly affected by an intrinsic partial-covering absorber with a column density $N_H \sim 2.6 \times 10^{23}$ cm$^{-2}$ and the covering fraction of the order of $\sim0.34$. The inclusion of this component resulted in a substantial improvement in the fit quality (see Figure~\ref{fig:chi_plot} and Section~\ref{sec:pheno}), demonstrating that it is statistically required by these observations. In contrast, the spectra obtained from 2016 onward are well described by a simple absorbed power law. The inclusion of the same intrinsic absorber in these later epochs provides no statistically significant improvement to the fit. To ensure consistency, we initially fitted all the post-2016 spectra by fixing the absorber parameters at $N_H$ = $2.6 \times 10^{23}$ cm$^{-2}$ and covering factor = $0.34$. This resulted in an increase in the fit statistic and the systematic deviations between the model and the observed soft X-ray data. Allowing the absorber parameters to vary freely led them to converge toward the lowest permitted values, indicating that the data do not support the presence of such an absorption component in these epochs. We therefore conclude that intrinsic absorption is not required in the post-2016 observations and exclude this component from the corresponding spectral fits.

The appearance and/or disappearance of intrinsic X-ray absorption is not unexpected in Seyfert AGNs. Variability on timescales of years to decades is commonly observed and is often attributed to a clumpy circumnuclear medium, such as clouds associated with the broad-line region or the dusty torus moving into and out of the line of sight \citep{Risaliti2002, Risaliti2009, Nenkova2008, Markowitz2014}. For a supermassive black hole, cloud crossing times at broad-line region scales naturally to span several years, consistent with the temporal separation between the XMM3 (2006) and XRT3 (2016) observations. Similar transient absorption behaviour has been reported in several nearby Seyferts, where absorbers are present during some epochs and absent during others without requiring a global change in the accretion structure \citep{Risaliti2009,2025ApJ...994..216L}.

Importantly, the long-term evolution of the soft excess component is not driven by changes in the absorption model. We verified that forcing the intrinsic absorber into the post-2016 spectral fits does not recover a statistically significant soft excess, nor does it alter the overall spectral trends. The observed weakening or disappearance of the soft excess therefore reflects an intrinsic change in the emission components rather than a modelling artefact. This interpretation is further supported by the persistence of the photon index–luminosity correlation across all epochs, which remains unchanged regardless of whether the intrinsic absorber is included. We thus interpret the disappearance of the absorber after 2016 as a genuine physical change in the line-of-sight environment, while the long-term spectral variability is dominated by intrinsic changes in the accretion flow and coronal emission.

\section{Conclusions}
\label{conclusions}
We carried out a comprehensive multi-epoch spectral and temporal study of Mrk~530 using observations spanning nearly $\sim24$ years (2001--2024) from multiple space-based X-ray observatories. Our key findings are summarized as follows.

\begin{enumerate}
    \item Mrk~530 exhibits strong spectral variability across the long-term observational baseline. Both the photon index and the X-ray luminosity show systematic variations. In addition, combined spectral and temporal analyses in 2018 revealed the presence of quasi-periodic signals in both the UV and X-ray wavelength regimes.

    \item A significant soft-excess component is detected in the early observations between 2001 and 2006, whereas this component is absent in later epochs. This behaviour points to an evolving warm corona, plausibly linked to changes in the accretion state. The gradual weakening and eventual disappearance of the soft excess are robust against different absorber treatments, indicating that the evolution is intrinsic to the source rather than driven by line-of-sight effects.

    \item Spectral fitting with physically motivated frameworks such as \textsc{AGNSED} and \textsc{TCAF} demonstrates that variations in the mass accretion rate play a primary role in shaping the coronal geometry. Higher accretion rate, particularly from the disc, is associated with a more compact corona and softer spectra, whereas lower accretion rates correspond to an expanded hot flow and harder spectral states.

    \item Correlation analysis reveals a strong coupling between accretion dynamics and coronal structure. The disc accretion rate correlates positively with luminosity, while the characteristic coronal size anti-correlates with both luminosity and photon index. These trends support a picture in which radiative cooling of the corona by disc photons regulates the observed spectral states.

    \item We identify a tentative QPO candidate in the 2018 data, detected simultaneously in the UV and X-ray bands. This modulation is characterized by low coherence and fewer than two observed cycles, and therefore cannot be classified as a confirmed QPO. Longer-duration, high-cadence monitoring will be required to establish the persistence of the modulation. The correlated UV–X-ray variability observed during this epoch is consistent with a common underlying physical driver. A plausible interpretation involves accretion-rate fluctuations propagating inward on dynamical timescales, first modulating the outer Comptonizing regions responsible for the UV emission and subsequently influencing the more compact X-ray emitting corona. Such behaviour naturally arises in an extended, radially stratified Comptonizing medium and does not require strictly coherent oscillations.
       
\end{enumerate}

Overall, Mrk~530 presents a coherent picture of accretion-driven variability, in which the coupled evolution of the accretion disc, warm corona, and hot corona governs the long-term spectral and temporal behaviour of the source. Our results emphasize that both spectral and timing properties in AGNs are shaped by the interplay between accretion dynamics and coronal geometry, with changes in the size and structure of the Comptonizing regions playing a central role. Future high-sensitivity missions such as Athena and XRISM will provide the spectral resolution and throughput required to place stronger constraints on these processes in Mrk~530 and in the broader Seyfert population.

\section*{Acknowledgements}
We sincerely thank the reviewer for providing insightful comments and constructive suggestions that helped us improve the manuscript. The research work at the Physical Research Laboratory, Ahmedabad, is funded by the Department of Space, Government of India. PN and SKC acknowledge support from the Indian Centre for Space Physics, Kolkata, India. The data and software used in this work are accessed from the High Energy Astrophysics Science Archive Research Center (HEASARC), serviced by Astrophysics Science Division at NASA/GSFC and the High Energy Astrophysics Division of the Smithsonian Astrophysical Observatory. This work has used {\it Swift} data supplied by the UK Swift Science Data Center, University of Leicester. This work has used observational data from {\it XMM-Newton}, a European Space Agency (ESA) mission with instruments and contributions directly funded by ESA Member States and NASA.
{\it Facilities:} HEASARC, XMM-Newton, Swift.

\section*{Data Availability}
We used archival data of {\it XMM-Newton} and {\it Swift} observatories for this work. These data are publicly available on their corresponding websites. Appropriate links are given in the text.



\bibliographystyle{mnras}
\bibliography{reference.bib} 

@ARTICLE{1973A&A....24..337S,
       author = {{Shakura}, N.~I. and {Sunyaev}, R.~A.},
        title = "{Black holes in binary systems. Observational appearance.}",
      journal = {\aap},
         year = 1973,
        month = jan,
       volume = {24},
        pages = {337-355},
       adsurl = {https://ui.adsabs.harvard.edu/abs/1973A&A....24..337S}
}

@INPROCEEDINGS{1973blho.conf..343N,
       author = {{Novikov}, I.~D. and {Thorne}, K.~S.},
        title = "{Astrophysics of black holes.}",
    booktitle = {Black Holes (Les Astres Occlus)},
         year = 1973,
       editor = {{Dewitt}, C. and {Dewitt}, B.~S.},
        month = jan,
        pages = {343-450},
       adsurl = {https://ui.adsabs.harvard.edu/abs/1973blho.conf..343N}
}

@article{lynden1969galactic,
  title={Galactic nuclei as collapsed old quasars.},
  author={Lynden-Bell, Donald},
  journal={Nature},
  volume={223},
  number={5207},
  year={1969}
}

@ARTICLE{2004ApJ...611.1005G,
       author = {{Gehrels}, N. and {Chincarini}, G. and {Giommi}, P. and {Mason}, K.~O. and {Nousek}, J.~A. and {Wells}, A.~A. and {White}, N.~E. and {Barthelmy}, S.~D. and {Burrows}, D.~N. and {Cominsky}, L.~R. and {Hurley}, K.~C. and {Marshall}, F.~E. and {M{\'e}sz{\'a}ros}, P. and {Roming}, P.~W.~A. and {Angelini}, L. and {Barbier}, L.~M. and {Belloni}, T. and {Campana}, S. and {Caraveo}, P.~A. and {Chester}, M.~M. and {Citterio}, O. and {Cline}, T.~L. and {Cropper}, M.~S. and {Cummings}, J.~R. and {Dean}, A.~J. and {Feigelson}, E.~D. and {Fenimore}, E.~E. and {Frail}, D.~A. and {Fruchter}, A.~S. and {Garmire}, G.~P. and {Gendreau}, K. and {Ghisellini}, G. and {Greiner}, J. and {Hill}, J.~E. and {Hunsberger}, S.~D. and {Krimm}, H.~A. and {Kulkarni}, S.~R. and {Kumar}, P. and {Lebrun}, F. and {Lloyd-Ronning}, N.~M. and {Markwardt}, C.~B. and {Mattson}, B.~J. and {Mushotzky}, R.~F. and {Norris}, J.~P. and {Osborne}, J. and {Paczynski}, B. and {Palmer}, D.~M. and {Park}, H. -S. and {Parsons}, A.~M. and {Paul}, J. and {Rees}, M.~J. and {Reynolds}, C.~S. and {Rhoads}, J.~E. and {Sasseen}, T.~P. and {Schaefer}, B.~E. and {Short}, A.~T. and {Smale}, A.~P. and {Smith}, I.~A. and {Stella}, L. and {Tagliaferri}, G. and {Takahashi}, T. and {Tashiro}, M. and {Townsley}, L.~K. and {Tueller}, J. and {Turner}, M.~J.~L. and {Vietri}, M. and {Voges}, W. and {Ward}, M.~J. and {Willingale}, R. and {Zerbi}, F.~M. and {Zhang}, W.~W.},
        title = "{The Swift Gamma-Ray Burst Mission}",
      journal = {\apj},
         year = 2004,
        month = aug,
       volume = {611},
       number = {2},
        pages = {1005-1020},
          doi = {10.1086/422091},
archivePrefix = {arXiv},
       eprint = {astro-ph/0405233},
 primaryClass = {astro-ph},
       adsurl = {https://ui.adsabs.harvard.edu/abs/2004ApJ...611.1005G}
}

@ARTICLE{2005SSRv..120...95R,
       author = {{Roming}, Peter W.~A. and {Kennedy}, Thomas E. and {Mason}, Keith O. and {Nousek}, John A. and {Ahr}, Lindy and {Bingham}, Richard E. and {Broos}, Patrick S. and {Carter}, Mary J. and {Hancock}, Barry K. and {Huckle}, Howard E. and {Hunsberger}, S.~D. and {Kawakami}, Hajime and {Killough}, Ronnie and {Koch}, T. Scott and {McLelland}, Michael K. and {Smith}, Kelly and {Smith}, Philip J. and {Soto}, Juan Carlos and {Boyd}, Patricia T. and {Breeveld}, Alice A. and {Holland}, Stephen T. and {Ivanushkina}, Mariya and {Pryzby}, Michael S. and {Still}, Martin D. and {Stock}, Joseph},
        title = "{The Swift Ultra-Violet/Optical Telescope}",
      journal = {\ssr},
         year = 2005,
        month = oct,
       volume = {120},
       number = {3-4},
        pages = {95-142},
          doi = {10.1007/s11214-005-5095-4},
archivePrefix = {arXiv},
       eprint = {astro-ph/0507413},
 primaryClass = {astro-ph},
       adsurl = {https://ui.adsabs.harvard.edu/abs/2005SSRv..120...95R}
}

@ARTICLE{2005SSRv..120..165B,
       author = {{Burrows}, David N. and {Hill}, J.~E. and {Nousek}, J.~A. and {Kennea}, J.~A. and {Wells}, A. and {Osborne}, J.~P. and {Abbey}, A.~F. and {Beardmore}, A. and {Mukerjee}, K. and {Short}, A.~D.~T. and {Chincarini}, G. and {Campana}, S. and {Citterio}, O. and {Moretti}, A. and {Pagani}, C. and {Tagliaferri}, G. and {Giommi}, P. and {Capalbi}, M. and {Tamburelli}, F. and {Angelini}, L. and {Cusumano}, G. and {Br{\"a}uninger}, H.~W. and {Burkert}, W. and {Hartner}, G.~D.},
        title = "{The Swift X-Ray Telescope}",
      journal = {\ssr},
         year = 2005,
        month = oct,
       volume = {120},
       number = {3-4},
        pages = {165-195},
          doi = {10.1007/s11214-005-5097-2},
archivePrefix = {arXiv},
       eprint = {astro-ph/0508071},
 primaryClass = {astro-ph},
       adsurl = {https://ui.adsabs.harvard.edu/abs/2005SSRv..120..165B}
}

@ARTICLE{2009MNRAS.397.1177E,
       author = {{Evans}, P.~A. and {Beardmore}, A.~P. and {Page}, K.~L. and {Osborne}, J.~P. and {O'Brien}, P.~T. and {Willingale}, R. and {Starling}, R.~L.~C. and {Burrows}, D.~N. and {Godet}, O. and {Vetere}, L. and {Racusin}, J. and {Goad}, M.~R. and {Wiersema}, K. and {Angelini}, L. and {Capalbi}, M. and {Chincarini}, G. and {Gehrels}, N. and {Kennea}, J.~A. and {Margutti}, R. and {Morris}, D.~C. and {Mountford}, C.~J. and {Pagani}, C. and {Perri}, M. and {Romano}, P. and {Tanvir}, N.},
        title = "{Methods and results of an automatic analysis of a complete sample of Swift-XRT observations of GRBs}",
      journal = {\mnras},
         year = 2009,
        month = aug,
       volume = {397},
       number = {3},
        pages = {1177-1201},
          doi = {10.1111/j.1365-2966.2009.14913.x},
archivePrefix = {arXiv},
       eprint = {0812.3662},
 primaryClass = {astro-ph},
       adsurl = {https://ui.adsabs.harvard.edu/abs/2009MNRAS.397.1177E}
}

@ARTICLE{2001A&A...365L...1J,
       author = {{Jansen}, F. and {Lumb}, D. and {Altieri}, B. and {Clavel}, J. and {Ehle}, M. and {Erd}, C. and {Gabriel}, C. and {Guainazzi}, M. and {Gondoin}, P. and {Much}, R. and {Munoz}, R. and {Santos}, M. and {Schartel}, N. and {Texier}, D. and {Vacanti}, G.},
        title = "{XMM-Newton observatory. I. The spacecraft and operations}",
      journal = {\aap},
         year = 2001,
        month = jan,
       volume = {365},
        pages = {L1-L6},
          doi = {10.1051/0004-6361:20000036},
       adsurl = {https://ui.adsabs.harvard.edu/abs/2001A&A...365L...1J}
}

@article{ricci2023changing,
  title={Changing-look active galactic nuclei},
  author={Ricci, Claudio and Trakhtenbrot, Benny},
  journal={Nature Astronomy},
  volume={7},
  number={11},
  pages={1282--1294},
  year={2023},
  publisher={Nature Publishing Group UK London}
}

@article{noda2018explaining,
  title={Explaining changing-look AGN with state transition triggered by rapid mass accretion rate drop},
  author={Noda, Hirofumi and Done, Chris},
  journal={Monthly Notices of the Royal Astronomical Society},
  volume={480},
  number={3},
  pages={3898--3906},
  year={2018},
  publisher={Oxford University Press}
}

@ARTICLE{1995ApJ...440..141G,
       author = {{Goodrich}, Robert W.},
        title = "{Dust in the Broad-Line Regions of Seyfert Galaxies}",
      journal = {\apj},
         year = 1995,
        month = feb,
       volume = {440},
        pages = {141},
          doi = {10.1086/175256},
       adsurl = {https://ui.adsabs.harvard.edu/abs/1995ApJ...440..141G}
}

@ARTICLE{2018MNRAS.478.4214E,
       author = {{Ehler}, H.~J.~S. and {Gonzalez}, A.~G. and {Gallo}, L.~C.},
        title = "{Exploring the spectral variability of the Seyfert 1.5 galaxy Markarian 530 with Suzaku}",
      journal = {\mnras},
         year = 2018,
        month = aug,
       volume = {478},
       number = {3},
        pages = {4214-4224},
          doi = {10.1093/mnras/sty1306},
archivePrefix = {arXiv},
       eprint = {1805.06742},
 primaryClass = {astro-ph.HE},
       adsurl = {https://ui.adsabs.harvard.edu/abs/2018MNRAS.478.4214E}
}

@INPROCEEDINGS{1996ASPC..101...17A,
       author = {{Arnaud}, K.~A.},
        title = "{XSPEC: The First Ten Years}",
    booktitle = {Astronomical Data Analysis Software and Systems V},
         year = 1996,
       editor = {{Jacoby}, George H. and {Barnes}, Jeannette},
       series = {Astronomical Society of the Pacific Conference Series},
       volume = {101},
        month = jan,
        pages = {17},
       adsurl = {https://ui.adsabs.harvard.edu/abs/1996ASPC..101...17A}
}

@ARTICLE{2002MNRAS.331L..35F,
       author = {{Fabian}, A.~C. and {Ballantyne}, D.~R. and {Merloni}, A. and {Vaughan}, S. and {Iwasawa}, K. and {Boller}, Th.},
        title = "{How the X-ray spectrum of a narrow-line Seyfert 1 galaxy may be reflection-dominated}",
      journal = {\mnras},
         year = 2002,
        month = apr,
       volume = {331},
       number = {3},
        pages = {L35-L39},
          doi = {10.1046/j.1365-8711.2002.05419.x},
archivePrefix = {arXiv},
       eprint = {astro-ph/0202297},
 primaryClass = {astro-ph},
       adsurl = {https://ui.adsabs.harvard.edu/abs/2002MNRAS.331L..35F}
}

@ARTICLE{1997ApJ...476...70N,
       author = {{Nandra}, K. and {George}, I.~M. and {Mushotzky}, R.~F. and {Turner}, T.~J. and {Yaqoob}, T.},
        title = "{ASCA Observations of Seyfert 1 Galaxies. I. Data Analysis, Imaging, and Timing}",
      journal = {\apj},
         year = 1997,
        month = feb,
       volume = {476},
       number = {1},
        pages = {70-82},
          doi = {10.1086/303600},
       adsurl = {https://ui.adsabs.harvard.edu/abs/1997ApJ...476...70N}
}

@ARTICLE{2003MNRAS.345.1271V,
       author = {{Vaughan}, S. and {Edelson}, R. and {Warwick}, R.~S. and {Uttley}, P.},
        title = "{On characterizing the variability properties of X-ray light curves from active galaxies}",
      journal = {\mnras},
         year = 2003,
        month = nov,
       volume = {345},
       number = {4},
        pages = {1271-1284},
          doi = {10.1046/j.1365-2966.2003.07042.x},
archivePrefix = {arXiv},
       eprint = {astro-ph/0307420},
 primaryClass = {astro-ph},
       adsurl = {https://ui.adsabs.harvard.edu/abs/2003MNRAS.345.1271V}
}

@ARTICLE{1991ApJ...380L..51H,
       author = {{Haardt}, F. and {Maraschi}, L.},
        title = "{A Two-Phase Model for the X-Ray Emission from Seyfert Galaxies}",
      journal = {\apjl},
         year = 1991,
        month = oct,
       volume = {380},
        pages = {L51},
          doi = {10.1086/186171},
       adsurl = {https://ui.adsabs.harvard.edu/abs/1991ApJ...380L..51H}
}

@ARTICLE{1980A&A....86..121S,
       author = {{Sunyaev}, R.~A. and {Titarchuk}, L.~G.},
        title = "{Comptonization of X-Rays in Plasma Clouds - Typical Radiation Spectra}",
      journal = {\aap},
         year = 1980,
        month = jun,
       volume = {86},
        pages = {121},
       adsurl = {https://ui.adsabs.harvard.edu/abs/1980A&A....86..121S}
}

@ARTICLE{1991MNRAS.249..352G,
       author = {{George}, I.~M. and {Fabian}, A.~C.},
        title = "{X-ray reflection from cold matter in Active Galactic Nuclei and X-ray binaries.}",
      journal = {\mnras},
         year = 1991,
        month = mar,
       volume = {249},
        pages = {352},
          doi = {10.1093/mnras/249.2.352},
       adsurl = {https://ui.adsabs.harvard.edu/abs/1991MNRAS.249..352G}
}

@ARTICLE{1984ApJ...281...90H,
       author = {{Halpern}, J.~P.},
        title = "{Variable X-ray absorption in the QSO MR 2251-178.}",
      journal = {\apj},
         year = 1984,
        month = jun,
       volume = {281},
        pages = {90-94},
          doi = {10.1086/162077},
       adsurl = {https://ui.adsabs.harvard.edu/abs/1984ApJ...281...90H}
}

@ARTICLE{1989ApJ...346...68S,
       author = {{Sun}, Wei-Hsin and {Malkan}, Matthew A.},
        title = "{Fitting Improved Accretion Disk Models to the Multiwavelength Continua of Quasars and Active Galactic Nuclei}",
      journal = {\apj},
         year = 1989,
        month = nov,
       volume = {346},
        pages = {68},
          doi = {10.1086/167986},
       adsurl = {https://ui.adsabs.harvard.edu/abs/1989ApJ...346...68S}
}

@ARTICLE{2012AJ....143...49W,
       author = {{Woo}, Jong-Hak and {Kim}, Ji Hoon and {Imanishi}, Masatoshi and {Park}, Dawoo},
        title = "{The Connection between 3.3 {\ensuremath{\mu}}m Polycyclic Aromatic Hydrocarbon Emission and Active Galactic Nucleus Activity}",
      journal = {\aj},
         year = 2012,
        month = feb,
       volume = {143},
       number = {2},
          eid = {49},
        pages = {49},
          doi = {10.1088/0004-6256/143/2/49},
archivePrefix = {arXiv},
       eprint = {1112.1461},
 primaryClass = {astro-ph.CO},
       adsurl = {https://ui.adsabs.harvard.edu/abs/2012AJ....143...49W}
}

@ARTICLE{2000A&A...361..901K,
       author = {{Kollatschny}, W. and {Bischoff}, K. and {Dietrich}, M.},
        title = "{Strong spectral variability in NGC 7603 over 20 years}",
      journal = {\aap},
         year = 2000,
        month = sep,
       volume = {361},
        pages = {901-912},
       adsurl = {https://ui.adsabs.harvard.edu/abs/2000A&A...361..901K}
}

@ARTICLE{1994ApJ...428L..13N,
       author = {{Narayan}, Ramesh and {Yi}, Insu},
        title = "{Advection-dominated Accretion: A Self-similar Solution}",
      journal = {\apjl},
         year = 1994,
        month = jun,
       volume = {428},
        pages = {L13},
          doi = {10.1086/187381},
archivePrefix = {arXiv},
       eprint = {astro-ph/9403052},
 primaryClass = {astro-ph},
       adsurl = {https://ui.adsabs.harvard.edu/abs/1994ApJ...428L..13N}
}

@ARTICLE{1995ApJ...455..623C,
       author = {{Chakrabarti}, Sandip and {Titarchuk}, Lev G.},
        title = "{Spectral Properties of Accretion Disks around Galactic and Extragalactic Black Holes}",
      journal = {\apj},
         year = 1995,
        month = dec,
       volume = {455},
        pages = {623},
          doi = {10.1086/176610},
archivePrefix = {arXiv},
       eprint = {astro-ph/9510005},
 primaryClass = {astro-ph},
       adsurl = {https://ui.adsabs.harvard.edu/abs/1995ApJ...455..623C}
}

@ARTICLE{2004A&A...425...99L,
       author = {{Lal}, D.~V. and {Shastri}, P. and {Gabuzda}, D.~C.},
        title = "{Milliarcsec-scale radio structure of a matched sample of Seyfert 1 and Seyfert 2 galaxies}",
      journal = {\aap},
         year = 2004,
        month = oct,
       volume = {425},
        pages = {99-108},
          doi = {10.1051/0004-6361:20040486},
archivePrefix = {arXiv},
       eprint = {astro-ph/0406597},
 primaryClass = {astro-ph},
       adsurl = {https://ui.adsabs.harvard.edu/abs/2004A&A...425...99L}
}

@ARTICLE{2016ApJ...822...45T,
       author = {{Theios}, Rachel L. and {Malkan}, Matthew A. and {Ross}, Nathaniel R.},
        title = "{H{\ensuremath{\alpha}} Imaging of Nearby Seyfert Host Galaxies}",
      journal = {\apj},
         year = 2016,
        month = may,
       volume = {822},
       number = {1},
          eid = {45},
        pages = {45},
          doi = {10.3847/0004-637X/822/1/45},
archivePrefix = {arXiv},
       eprint = {1604.00089},
 primaryClass = {astro-ph.GA},
       adsurl = {https://ui.adsabs.harvard.edu/abs/2016ApJ...822...45T}
}

@INPROCEEDINGS{1997ASSL..218..163A,
       author = {{Alexander}, Tal},
        title = "{Is AGN Variability Correlated with Other AGN Properties? ZDCF Analysis of Small Samples of Sparse Light Curves}",
    booktitle = {Astronomical Time Series},
         year = 1997,
       editor = {{Maoz}, D. and {Sternberg}, A. and {Leibowitz}, E.~M.},
       series = {Astrophysics and Space Science Library},
       volume = {218},
        month = jan,
        pages = {163},
          doi = {10.1007/978-94-015-8941-3_14},
       adsurl = {https://ui.adsabs.harvard.edu/abs/1997ASSL..218..163A}
}

@article{alexander2013improved,
  title={Improved AGN light curve analysis with the z-transformed discrete correlation function},
  author={Alexander, Tal},
  journal={arXiv preprint arXiv:1302.1508},
  year={2013}
}

@ARTICLE{Uttley2003,
       author = {{Uttley}, Philip and {Fruscione}, Antonella and {McHardy}, Ian and {Lamer}, Georg},
        title = "{Catching NGC 4051 in the Low State with Chandra}",
      journal = {\apj},
         year = 2003,
        month = oct,
       volume = {595},
       number = {2},
        pages = {656-664},
          doi = {10.1086/377468},
archivePrefix = {arXiv},
       eprint = {astro-ph/0306234},
 primaryClass = {astro-ph},
       adsurl = {https://ui.adsabs.harvard.edu/abs/2003ApJ...595..656U}
}

@ARTICLE{Paolillo2004,
       author = {{Paolillo}, M. and {Schreier}, E.~J. and {Giacconi}, R. and {Koekemoer}, A.~M. and {Grogin}, N.~A.},
        title = "{Prevalence of X-Ray Variability in the Chandra Deep Field-South}",
      journal = {\apj},
         year = 2004,
        month = aug,
       volume = {611},
       number = {1},
        pages = {93-106},
          doi = {10.1086/421967},
archivePrefix = {arXiv},
       eprint = {astro-ph/0404418},
 primaryClass = {astro-ph},
       adsurl = {https://ui.adsabs.harvard.edu/abs/2004ApJ...611...93P}
}

@ARTICLE{Done2012,
       author = {{Done}, Chris and {Davis}, S.~W. and {Jin}, C. and {Blaes}, O. and {Ward}, M.},
        title = "{Intrinsic disc emission and the soft X-ray excess in active galactic nuclei}",
      journal = {\mnras},
         year = 2012,
        month = mar,
       volume = {420},
       number = {3},
        pages = {1848-1860},
          doi = {10.1111/j.1365-2966.2011.19779.x},
archivePrefix = {arXiv},
       eprint = {1107.5429},
 primaryClass = {astro-ph.HE},
       adsurl = {https://ui.adsabs.harvard.edu/abs/2012MNRAS.420.1848D}
}

@ARTICLE{Petrucci2020,
       author = {{Petrucci}, P. -O. and {Gronkiewicz}, D. and {Rozanska}, A. and {Belmont}, R. and {Bianchi}, S. and {Czerny}, B. and {Matt}, G. and {Malzac}, J. and {Middei}, R. and {De Rosa}, A. and {Ursini}, F. and {Cappi}, M.},
        title = "{Radiation spectra of warm and optically thick coronae in AGNs}",
      journal = {\aap},
         year = 2020,
        month = feb,
       volume = {634},
          eid = {A85},
        pages = {A85},
          doi = {10.1051/0004-6361/201937011},
archivePrefix = {arXiv},
       eprint = {2001.02026},
 primaryClass = {astro-ph.HE},
       adsurl = {https://ui.adsabs.harvard.edu/abs/2020A&A...634A..85P}
}

@article{lomb1976least,
  title={Least-squares frequency analysis of unequally spaced data},
  author={Lomb, Nicholas R},
  journal={Astrophysics and space science},
  volume={39},
  number={2},
  pages={447--462},
  year={1976},
  publisher={Springer}
}

@article{scargle1982studies,
  title={Studies in astronomical time series analysis. II-Statistical aspects of spectral analysis of unevenly spaced data},
  author={Scargle, Jeffrey D},
  journal={Astrophysical Journal, Part 1, vol. 263, Dec. 15, 1982, p. 835-853.},
  volume={263},
  pages={835--853},
  year={1982}
}

@article{edelson1988discrete,
  title={The discrete correlation function-A new method for analyzing unevenly sampled variability data},
  author={Edelson, RA and Krolik, JH},
  journal={Astrophysical Journal, Part 1 (ISSN 0004-637X), vol. 333, Oct. 15, 1988, p. 646-659.},
  volume={333},
  pages={646--659},
  year={1988}
}

@ARTICLE{2018MNRAS.480.1247K,
       author = {{Kubota}, Aya and {Done}, Chris},
        title = "{A physical model of the broad-band continuum of AGN and its implications for the UV/X relation and optical variability}",
      journal = {\mnras},
         year = 2018,
        month = oct,
       volume = {480},
       number = {1},
        pages = {1247-1262},
          doi = {10.1093/mnras/sty1890},
archivePrefix = {arXiv},
       eprint = {1804.00171},
 primaryClass = {astro-ph.HE},
       adsurl = {https://ui.adsabs.harvard.edu/abs/2018MNRAS.480.1247K}
}

@ARTICLE{2011ApJ...731...68L,
       author = {{Lal}, Dharam V. and {Shastri}, Prajval and {Gabuzda}, Denise C.},
        title = "{Seyfert Galaxies: Nuclear Radio Structure and Unification}",
      journal = {\apj},
         year = 2011,
        month = apr,
       volume = {731},
       number = {1},
          eid = {68},
        pages = {68},
          doi = {10.1088/0004-637X/731/1/68},
archivePrefix = {arXiv},
       eprint = {1102.3955},
 primaryClass = {astro-ph.CO},
       adsurl = {https://ui.adsabs.harvard.edu/abs/2011ApJ...731...68L}
}

@ARTICLE{1996AJ....112.1709F,
       author = {{Foster}, Grant},
        title = "{Wavelets for period analysis of unevenly sampled time series}",
      journal = {\aj},
         year = 1996,
        month = oct,
       volume = {112},
        pages = {1709-1729},
          doi = {10.1086/118137},
       adsurl = {https://ui.adsabs.harvard.edu/abs/1996AJ....112.1709F}
}

@ARTICLE{Hu2022,
       author = {{Hu}, Jingwei and {Jin}, Chichuan and {Cheng}, Huaqing and {Yuan}, Weimin},
        title = "{A Systematic Study of the Short-term X-Ray Variability of Seyfert Galaxies. I. Diversity of the X-Ray rms Spectra}",
      journal = {\apj},
         year = 2022,
        month = sep,
       volume = {936},
       number = {2},
          eid = {105},
        pages = {105},
          doi = {10.3847/1538-4357/ac83ba},
archivePrefix = {arXiv},
       eprint = {2208.05921},
 primaryClass = {astro-ph.HE},
       adsurl = {https://ui.adsabs.harvard.edu/abs/2022ApJ...936..105H}
}

@ARTICLE{Beckmann2007,
       author = {{Beckmann}, V. and {Barthelmy}, S.~D. and {Courvoisier}, T.~J. -L. and {Gehrels}, N. and {Soldi}, S. and {Tueller}, J. and {Wendt}, G.},
        title = "{Hard X-ray variability of active galactic nuclei}",
      journal = {\aap},
         year = 2007,
        month = dec,
       volume = {475},
       number = {3},
        pages = {827-835},
          doi = {10.1051/0004-6361:20078355},
archivePrefix = {arXiv},
       eprint = {0709.2230},
 primaryClass = {astro-ph},
       adsurl = {https://ui.adsabs.harvard.edu/abs/2007A&A...475..827B}
}

@ARTICLE{Panagiotou2022,
       author = {{Panagiotou}, Christos and {Kara}, Erin and {Dov{\v{c}}iak}, Michal},
        title = "{Explaining the Moderate UV/X-Ray Correlation in AGN}",
      journal = {\apj},
         year = 2022,
        month = dec,
       volume = {941},
       number = {1},
          eid = {57},
        pages = {57},
          doi = {10.3847/1538-4357/aca2a4},
archivePrefix = {arXiv},
       eprint = {2211.06963},
 primaryClass = {astro-ph.GA},
       adsurl = {https://ui.adsabs.harvard.edu/abs/2022ApJ...941...57P}
}

@ARTICLE{Torrence1998,
       author = {{Torrence}, Christopher and {Compo}, Gilbert P.},
        title = "{A Practical Guide to Wavelet Analysis.}",
      journal = {Bulletin of the American Meteorological Society},
         year = 1998,
        month = jan,
       volume = {79},
       number = {1},
        pages = {61-78},
          doi = {10.1175/1520-0477(1998)079<0061:APGTWA>2.0.CO;2},
       adsurl = {https://ui.adsabs.harvard.edu/abs/1998BAMS...79...61T}
}

@article{zhou2014universal,
  title={Universal scaling of the 3: 2 twin-peak quasi-periodic oscillation frequencies with black hole mass and spin revisited},
  author={Zhou, Xin-Lin and Yuan, Weimin and Pan, Hai-Wu and Liu, Zhu},
  journal={The Astrophysical Journal Letters},
  volume={798},
  number={1},
  pages={L5},
  year={2014},
  publisher={IOP Publishing}
}

@ARTICLE{2008Natur.455..369G,
       author = {{Gierli{\'n}ski}, Marek and {Middleton}, Matthew and {Ward}, Martin and {Done}, Chris},
        title = "{A periodicity of \raisebox{-0.5ex}\textasciitilde1hour in X-ray emission from the active galaxy RE J1034+396}",
      journal = {\nat},
         year = 2008,
        month = sep,
       volume = {455},
       number = {7211},
        pages = {369-371},
          doi = {10.1038/nature07277},
       adsurl = {https://ui.adsabs.harvard.edu/abs/2008Natur.455..369G}
}

@ARTICLE{2016ApJ...819L..19P,
       author = {{Pan}, Hai-Wu and {Yuan}, Weimin and {Yao}, Su and {Zhou}, Xin-Lin and {Liu}, Bifang and {Zhou}, Hongyan and {Zhang}, Shuang-Nan},
        title = "{Detection of a Possible X-Ray Quasi-periodic Oscillation in the Active Galactic Nucleus 1H 0707-495}",
      journal = {\apjl},
         year = 2016,
        month = mar,
       volume = {819},
       number = {2},
          eid = {L19},
        pages = {L19},
          doi = {10.3847/2041-8205/819/2/L19},
archivePrefix = {arXiv},
       eprint = {1601.07639},
 primaryClass = {astro-ph.HE},
       adsurl = {https://ui.adsabs.harvard.edu/abs/2016ApJ...819L..19P}
}

@ARTICLE{2018ApJ...853..193Z,
       author = {{Zhang}, Peng-fei and {Zhang}, Peng and {Liao}, Neng-hui and {Yan}, Jing-zhi and {Fan}, Yi-zhong and {Liu}, Qing-zhong},
        title = "{Two Transient X-Ray Quasi-periodic Oscillations Separated by an Intermediate State in 1H 0707-495}",
      journal = {\apj},
         year = 2018,
        month = feb,
       volume = {853},
       number = {2},
          eid = {193},
        pages = {193},
          doi = {10.3847/1538-4357/aaa29a},
archivePrefix = {arXiv},
       eprint = {1703.07186},
 primaryClass = {astro-ph.HE},
       adsurl = {https://ui.adsabs.harvard.edu/abs/2018ApJ...853..193Z}
}

@ARTICLE{2017ApJ...849....9Z,
       author = {{Zhang}, Peng and {Zhang}, Peng-fei and {Yan}, Jing-zhi and {Fan}, Yi-zhong and {Liu}, Qing-zhong},
        title = "{An X-Ray Periodicity of {\ensuremath{\sim}}1.8 hr in Narrow-line Seyfert 1 Galaxy Mrk 766}",
      journal = {\apj},
         year = 2017,
        month = nov,
       volume = {849},
       number = {1},
          eid = {9},
        pages = {9},
          doi = {10.3847/1538-4357/aa8d6e},
archivePrefix = {arXiv},
       eprint = {1707.03586},
 primaryClass = {astro-ph.HE},
       adsurl = {https://ui.adsabs.harvard.edu/abs/2017ApJ...849....9Z}
}

@ARTICLE{2024A&A...691A...7Y,
       author = {{Yan}, Y.~K. and {Zhang}, P. and {Liu}, Q.~Z. and {Chang}, Z. and {Liu}, G.~C. and {Yan}, J.~Z. and {Zeng}, X.~Y.},
        title = "{An X-ray high-frequency quasi-periodic oscillation in NGC 1365}",
      journal = {\aap},
         year = 2024,
        month = nov,
       volume = {691},
          eid = {A7},
        pages = {A7},
          doi = {10.1051/0004-6361/202450875},
       adsurl = {https://ui.adsabs.harvard.edu/abs/2024A&A...691A...7Y}
}

@ARTICLE{2025PASJ...77..381W,
       author = {{Wang}, Liang and {Yi}, Tingfeng and {Zhang}, Shun and {Dhiman}, Vinit and {Dong}, Liang},
        title = "{Search for quasi-periodic oscillations of AGNs in the Swift BAT 157-month hard X-ray survey}",
      journal = {\pasj},
         year = 2025,
        month = apr,
       volume = {77},
       number = {2},
        pages = {381-388},
          doi = {10.1093/pasj/psaf005},
       adsurl = {https://ui.adsabs.harvard.edu/abs/2025PASJ...77..381W}
}

@ARTICLE{2013MNRAS.436L.114K,
       author = {{King}, O.~G. and {Hovatta}, T. and {Max-Moerbeck}, W. and {Meier}, D.~L. and {Pearson}, T.~J. and {Readhead}, A.~C.~S. and {Reeves}, R. and {Richards}, J.~L. and {Shepherd}, M.~C.},
        title = "{A quasi-periodic oscillation in the blazar J1359+4011.}",
      journal = {\mnras},
         year = 2013,
        month = nov,
       volume = {436},
        pages = {L114-L117},
          doi = {10.1093/mnrasl/slt125},
archivePrefix = {arXiv},
       eprint = {1309.1158},
 primaryClass = {astro-ph.HE},
       adsurl = {https://ui.adsabs.harvard.edu/abs/2013MNRAS.436L.114K}
}

@ARTICLE{1999ApJ...513..827M,
       author = {{Morsink}, Sharon M. and {Stella}, Luigi},
        title = "{Relativistic Precession around Rotating Neutron Stars: Effects Due to Frame Dragging and Stellar Oblateness}",
      journal = {\apj},
         year = 1999,
        month = mar,
       volume = {513},
       number = {2},
        pages = {827-844},
          doi = {10.1086/306876},
archivePrefix = {arXiv},
       eprint = {astro-ph/9808227},
 primaryClass = {astro-ph},
       adsurl = {https://ui.adsabs.harvard.edu/abs/1999ApJ...513..827M}
}

@ARTICLE{2011MNRAS.415.2323I,
       author = {{Ingram}, Adam and {Done}, Chris},
        title = "{A physical model for the continuum variability and quasi-periodic oscillation in accreting black holes}",
      journal = {\mnras},
         year = 2011,
        month = aug,
       volume = {415},
       number = {3},
        pages = {2323-2335},
          doi = {10.1111/j.1365-2966.2011.18860.x},
archivePrefix = {arXiv},
       eprint = {1101.2336},
 primaryClass = {astro-ph.SR},
       adsurl = {https://ui.adsabs.harvard.edu/abs/2011MNRAS.415.2323I}
}

@ARTICLE{2025ApJ...981...74L,
       author = {{Layek}, Narendranath and {Nandi}, Prantik and {Naik}, Sachindra and {Jana}, Arghajit},
        title = "{A Long-term Study of Mrk 50: Appearance and Disappearance of Soft Excess}",
      journal = {\apj},
         year = 2025,
        month = mar,
       volume = {981},
       number = {1},
          eid = {74},
        pages = {74},
          doi = {10.3847/1538-4357/adabc0},
archivePrefix = {arXiv},
       eprint = {2501.09300},
 primaryClass = {astro-ph.HE},
       adsurl = {https://ui.adsabs.harvard.edu/abs/2025ApJ...981...74L}
}

@INPROCEEDINGS{2022cosp...44.2313T,
       author = {{Tripathi}, Prakash and {Dewangan}, Gulab},
        title = "{State transition and Thermal Comptonization in the changing-look active galaxy NGC 1566}",
    booktitle = {44th COSPAR Scientific Assembly. Held 16-24 July},
         year = 2022,
       volume = {44},
        month = jul,
        pages = {2313},
       adsurl = {https://ui.adsabs.harvard.edu/abs/2022cosp...44.2313T}
}

@ARTICLE{2006ARA&A..44...49R,
       author = {{Remillard}, Ronald A. and {McClintock}, Jeffrey E.},
        title = "{X-Ray Properties of Black-Hole Binaries}",
      journal = {\araa},
         year = 2006,
        month = sep,
       volume = {44},
       number = {1},
        pages = {49-92},
          doi = {10.1146/annurev.astro.44.051905.092532},
archivePrefix = {arXiv},
       eprint = {astro-ph/0606352},
 primaryClass = {astro-ph},
       adsurl = {https://ui.adsabs.harvard.edu/abs/2006ARA&A..44...49R}
}

@ARTICLE{2001A&A...374L..19A,
       author = {{Abramowicz}, M.~A. and {Klu{\'z}niak}, W.},
        title = "{A precise determination of black hole spin in GRO J1655-40}",
      journal = {\aap},
         year = 2001,
        month = aug,
       volume = {374},
        pages = {L19-L20},
          doi = {10.1051/0004-6361:20010791},
archivePrefix = {arXiv},
       eprint = {astro-ph/0105077},
 primaryClass = {astro-ph},
       adsurl = {https://ui.adsabs.harvard.edu/abs/2001A&A...374L..19A}
}

@ARTICLE{Chakrabarti2015,
       author = {{Chakrabarti}, Sandip K. and {Mondal}, Santanu and {Debnath}, Dipak},
        title = "{Resonance condition and low-frequency quasi-periodic oscillations of the outbursting source H1743-322}",
      journal = {\mnras},
         year = 2015,
        month = oct,
       volume = {452},
       number = {4},
        pages = {3451-3456},
          doi = {10.1093/mnras/stv1566},
archivePrefix = {arXiv},
       eprint = {1507.02831},
 primaryClass = {astro-ph.HE},
       adsurl = {https://ui.adsabs.harvard.edu/abs/2015MNRAS.452.3451C}
}

@ARTICLE{Smith2023,
       author = {{Smith}, Evan and {Oramas}, Lani and {Perlman}, Eric},
        title = "{A QPO in Mkn 421 from Archival RXTE Data}",
      journal = {\apj},
         year = 2023,
        month = jun,
       volume = {950},
       number = {2},
          eid = {174},
        pages = {174},
          doi = {10.3847/1538-4357/acd171},
archivePrefix = {arXiv},
       eprint = {2305.07510},
 primaryClass = {astro-ph.HE},
       adsurl = {https://ui.adsabs.harvard.edu/abs/2023ApJ...950..174S}
}

@ARTICLE{Kaastra2017,
       author = {{Kaastra}, J.~S.},
        title = "{On the use of C-stat in testing models for X-ray spectra}",
      journal = {\aap},
         year = 2017,
        month = sep,
       volume = {605},
          eid = {A51},
        pages = {A51},
          doi = {10.1051/0004-6361/201629319},
archivePrefix = {arXiv},
       eprint = {1707.09202},
 primaryClass = {astro-ph.HE},
       adsurl = {https://ui.adsabs.harvard.edu/abs/2017A&A...605A..51K}
}

@ARTICLE{1979ApJ...228..939C,
       author = {{Cash}, W.},
        title = "{Parameter estimation in astronomy through application of the likelihood ratio.}",
      journal = {\apj},
         year = 1979,
        month = mar,
       volume = {228},
        pages = {939-947},
          doi = {10.1086/156922},
       adsurl = {https://ui.adsabs.harvard.edu/abs/1979ApJ...228..939C}
}

@ARTICLE{Hagen2024,
       author = {{Hagen}, Scott and {Done}, Chris and {Edelson}, Rick},
        title = "{What drives the variability in AGN? Explaining the UV-Xray disconnect through propagating fluctuations}",
      journal = {\mnras},
         year = 2024,
        month = jun,
       volume = {530},
       number = {4},
        pages = {4850-4867},
          doi = {10.1093/mnras/stae1177},
archivePrefix = {arXiv},
       eprint = {2401.03452},
 primaryClass = {astro-ph.HE},
       adsurl = {https://ui.adsabs.harvard.edu/abs/2024MNRAS.530.4850H}
}

@ARTICLE{AU2006,
       author = {{Ar{\'e}valo}, P. and {Uttley}, P.},
        title = "{Investigating a fluctuating-accretion model for the spectral-timing properties of accreting black hole systems}",
      journal = {\mnras},
         year = 2006,
        month = apr,
       volume = {367},
       number = {2},
        pages = {801-814},
          doi = {10.1111/j.1365-2966.2006.09989.x},
archivePrefix = {arXiv},
       eprint = {astro-ph/0512394},
 primaryClass = {astro-ph},
       adsurl = {https://ui.adsabs.harvard.edu/abs/2006MNRAS.367..801A}
}

@ARTICLE{Pahari2020,
       author = {{Pahari}, Mayukh and {McHardy}, I.~M. and {Vincentelli}, Federico and {Cackett}, Edward and {Peterson}, Bradley M. and {Goad}, Mike and {G{\"u}ltekin}, Kayhan and {Horne}, Keith},
        title = "{Evidence for variability time-scale-dependent UV/X-ray delay in Seyfert 1 AGN NGC 7469}",
      journal = {\mnras},
         year = 2020,
        month = may,
       volume = {494},
       number = {3},
        pages = {4057-4068},
          doi = {10.1093/mnras/staa1055},
archivePrefix = {arXiv},
       eprint = {2004.07901},
 primaryClass = {astro-ph.HE},
       adsurl = {https://ui.adsabs.harvard.edu/abs/2020MNRAS.494.4057P}
}

@ARTICLE{Wilkins2016,
       author = {{Wilkins}, D.~R. and {Cackett}, E.~M. and {Fabian}, A.~C. and {Reynolds}, C.~S.},
        title = "{Towards modelling X-ray reverberation in AGN: piecing together the extended corona}",
      journal = {\mnras},
         year = 2016,
        month = may,
       volume = {458},
       number = {1},
        pages = {200-225},
          doi = {10.1093/mnras/stw276},
archivePrefix = {arXiv},
       eprint = {1602.00022},
 primaryClass = {astro-ph.HE},
       adsurl = {https://ui.adsabs.harvard.edu/abs/2016MNRAS.458..200W}
}

@ARTICLE{Chainakun2019,
       author = {{Chainakun}, P. and {Watcharangkool}, A. and {Young}, A.~J. and {Hancock}, S.},
        title = "{X-ray time lags in AGN: inverse-Compton scattering and spherical corona model}",
      journal = {\mnras},
         year = 2019,
        month = jul,
       volume = {487},
       number = {1},
        pages = {667-680},
          doi = {10.1093/mnras/stz1319},
archivePrefix = {arXiv},
       eprint = {1905.03683},
 primaryClass = {astro-ph.HE},
       adsurl = {https://ui.adsabs.harvard.edu/abs/2019MNRAS.487..667C}
}

@ARTICLE{Zhang2023,
       author = {{Zhang}, W. and {Papadakis}, I.~E. and {Dov{\v{c}}iak}, M. and {Bursa}, M. and {Karas}, V.},
        title = "{A theoretical study of the time lags due to Comptonization and the constraints on the X-ray corona in AGNs}",
      journal = {\mnras},
         year = 2023,
        month = mar,
       volume = {519},
       number = {4},
        pages = {4951-4965},
          doi = {10.1093/mnras/stac3625},
archivePrefix = {arXiv},
       eprint = {2212.02734},
 primaryClass = {astro-ph.HE},
       adsurl = {https://ui.adsabs.harvard.edu/abs/2023MNRAS.519.4951Z}
}

@ARTICLE{1998MNRAS.301..179M,
       author = {{Magdziarz}, Pawel and {Blaes}, Omer M. and {Zdziarski}, Andrzej A. and {Johnson}, W. Neil and {Smith}, David A.},
        title = "{A spectral decomposition of the variable optical, ultraviolet and X-ray continuum of NGC 5548}",
      journal = {\mnras},
         year = 1998,
        month = nov,
       volume = {301},
       number = {1},
        pages = {179-192},
          doi = {10.1046/j.1365-8711.1998.02015.x},
       adsurl = {https://ui.adsabs.harvard.edu/abs/1998MNRAS.301..179M}
}

@ARTICLE{1985A&A...143..374S,
       author = {{Sunyaev}, R.~A. and {Titarchuk}, L.~G.},
        title = "{Comptonization of low-frequency radiation in accretion disks Angular distribution and polarization of hard radiation}",
      journal = {\aap},
         year = 1985,
        month = feb,
       volume = {143},
       number = {2},
        pages = {374-388},
       adsurl = {https://ui.adsabs.harvard.edu/abs/1985A&A...143..374S}
}

@ARTICLE{2005MNRAS.358..211R,
       author = {{Ross}, R.~R. and {Fabian}, A.~C.},
        title = "{A comprehensive range of X-ray ionized-reflection models}",
      journal = {\mnras},
         year = 2005,
        month = mar,
       volume = {358},
       number = {1},
        pages = {211-216},
          doi = {10.1111/j.1365-2966.2005.08797.x},
archivePrefix = {arXiv},
       eprint = {astro-ph/0501116},
 primaryClass = {astro-ph},
       adsurl = {https://ui.adsabs.harvard.edu/abs/2005MNRAS.358..211R}
}

@ARTICLE{2006MNRAS.365.1067C,
       author = {{Crummy}, J. and {Fabian}, A.~C. and {Gallo}, L. and {Ross}, R.~R.},
        title = "{An explanation for the soft X-ray excess in active galactic nuclei}",
      journal = {\mnras},
         year = 2006,
        month = feb,
       volume = {365},
       number = {4},
        pages = {1067-1081},
          doi = {10.1111/j.1365-2966.2005.09844.x},
archivePrefix = {arXiv},
       eprint = {astro-ph/0511457},
 primaryClass = {astro-ph},
       adsurl = {https://ui.adsabs.harvard.edu/abs/2006MNRAS.365.1067C}
}

@ARTICLE{2010ApJ...718..695G,
       author = {{Garc{\'\i}a}, J. and {Kallman}, T.~R.},
        title = "{X-ray Reflected Spectra from Accretion Disk Models. I. Constant Density Atmospheres}",
      journal = {\apj},
         year = 2010,
        month = aug,
       volume = {718},
       number = {2},
        pages = {695-706},
          doi = {10.1088/0004-637X/718/2/695},
archivePrefix = {arXiv},
       eprint = {1006.0485},
 primaryClass = {astro-ph.HE},
       adsurl = {https://ui.adsabs.harvard.edu/abs/2010ApJ...718..695G}
}

@ARTICLE{2009MNRAS.394..250M,
       author = {{Middleton}, Matthew and {Done}, Chris and {Ward}, Martin and {Gierli{\'n}ski}, Marek and {Schurch}, Nick},
        title = "{RE J1034+396: the origin of the soft X-ray excess and quasi-periodic oscillation}",
      journal = {\mnras},
         year = 2009,
        month = mar,
       volume = {394},
       number = {1},
        pages = {250-260},
          doi = {10.1111/j.1365-2966.2008.14255.x},
archivePrefix = {arXiv},
       eprint = {0807.4847},
 primaryClass = {astro-ph},
       adsurl = {https://ui.adsabs.harvard.edu/abs/2009MNRAS.394..250M}
}

@ARTICLE{1996ApJ...470..364E,
       author = {{Edelson}, R.~A. and {Alexander}, T. and {Crenshaw}, D.~M. and {Kaspi}, S. and {Malkan}, M.~A. and {Peterson}, B.~M. and {Warwick}, R.~S. and {Clavel}, J. and {Filippenko}, A.~V. and {Horne}, K. and {Korista}, K.~T. and {Kriss}, G.~A. and {Krolik}, J.~H. and {Maoz}, D. and {Nandra}, K. and {O'Brien}, P.~T. and {Penton}, S.~V. and {Yaqoob}, T. and {Albrecht}, P. and {Alloin}, D. and {Ayres}, T.~R. and {Balonek}, T.~J. and {Barr}, P. and {Barth}, A.~J. and {Bertram}, R. and {Bromage}, G.~E. and {Carini}, M. and {Carone}, T.~E. and {Cheng}, F. -Z. and {Chuvaev}, K.~K. and {Dietrich}, M. and {Dultzin-Hacyan}, D. and {Gaskell}, C.~M. and {Glass}, I.~S. and {Goad}, M.~R. and {Hemar}, S. and {Ho}, L.~C. and {Huchra}, J.~P. and {Hutchings}, J. and {Johnson}, W.~N. and {Kazanas}, D. and {Kollatschny}, W. and {Koratkar}, A.~P. and {Kovo}, O. and {Laor}, A. and {MacAlpine}, G.~M. and {Magdziarz}, P. and {Martin}, P.~G. and {Matheson}, T. and {McCollum}, B. and {Miller}, H.~R. and {Morris}, S.~L. and {Oknyanskij}, V.~L. and {Penfold}, J. and {Perez}, E. and {Perola}, G.~C. and {Pike}, G. and {Pogge}, R.~W. and {Ptak}, R.~L. and {Qian}, B. -C. and {Recondo-Gonzalez}, M.~C. and {Reichert}, G.~A. and {Rodriguez-Espinoza}, J.~M. and {Rodriguez-Pascual}, P.~M. and {Rokaki}, E.~L. and {Roland}, J. and {Sadun}, A.~C. and {Salamanca}, I. and {Santos-Lleo}, M. and {Shields}, J.~C. and {Shull}, J.~M. and {Smith}, D.~A. and {Smith}, S.~M. and {Snijders}, M.~A.~J. and {Stirpe}, G.~M. and {Stoner}, R.~E. and {Sun}, W. -H. and {Ulrich}, M. -H. and {van Groningen}, E. and {Wagner}, R.~M. and {Wagner}, S. and {Wanders}, I. and {Welsh}, W.~F. and {Weymann}, R.~J. and {Wilkes}, B.~J. and {Wu}, H. and {Wurster}, J. and {Xue}, S. -J. and {Zdziarski}, A.~A. and {Zheng}, W. and {Zou}, Z. -L.},
        title = "{Multiwavelength Observations of Short-Timescale Variability in NGC 4151. IV. Analysis of Multiwavelength Continuum Variability}",
      journal = {\apj},
         year = 1996,
        month = oct,
       volume = {470},
        pages = {364},
          doi = {10.1086/177872},
archivePrefix = {arXiv},
       eprint = {astro-ph/9605082},
 primaryClass = {astro-ph},
       adsurl = {https://ui.adsabs.harvard.edu/abs/1996ApJ...470..364E}
}

@article{Ingram_2019,
   title={A review of quasi-periodic oscillations from black hole X-ray binaries: Observation and theory},
   volume={85},
   ISSN={1387-6473},
   url={http://dx.doi.org/10.1016/j.newar.2020.101524},
   DOI={10.1016/j.newar.2020.101524},
   journal={New Astronomy Reviews},
   publisher={Elsevier BV},
   author={Ingram, Adam R. and Motta, Sara E.},
   year={2019},
   month=sep, pages={101524} }

@article{Duras_2020,
   title={Universal bolometric corrections for active galactic nuclei over seven luminosity decades},
   volume={636},
   ISSN={1432-0746},
   url={http://dx.doi.org/10.1051/0004-6361/201936817},
   DOI={10.1051/0004-6361/201936817},
   journal={Astronomy &amp; Astrophysics},
   publisher={EDP Sciences},
   author={Duras, F. and Bongiorno, A. and Ricci, F. and Piconcelli, E. and Shankar, F. and Lusso, E. and Bianchi, S. and Fiore, F. and Maiolino, R. and Marconi, A. and Onori, F. and Sani, E. and Schneider, R. and Vignali, C. and La Franca, F.},
   year={2020},
   month=apr, pages={A73} }

@ARTICLE{2015MNRAS.447.3368B,
       author = {{Banerji}, Manda and {Alaghband-Zadeh}, S. and {Hewett}, Paul C. and {McMahon}, Richard G.},
        title = "{Heavily reddened type 1 quasars at z > 2 - I. Evidence for significant obscured black hole growth at the highest quasar luminosities}",
      journal = {\mnras},
         year = 2015,
        month = mar,
       volume = {447},
       number = {4},
        pages = {3368-3389},
          doi = {10.1093/mnras/stu2649},
archivePrefix = {arXiv},
       eprint = {1501.00815},
 primaryClass = {astro-ph.GA},
       adsurl = {https://ui.adsabs.harvard.edu/abs/2015MNRAS.447.3368B}
}

@ARTICLE{2015MNRAS.451.4375F,
       author = {{Fabian}, A.~C. and {Lohfink}, A. and {Kara}, E. and {Parker}, M.~L. and {Vasudevan}, R. and {Reynolds}, C.~S.},
        title = "{Properties of AGN coronae in the NuSTAR era}",
      journal = {\mnras},
         year = 2015,
        month = aug,
       volume = {451},
       number = {4},
        pages = {4375-4383},
          doi = {10.1093/mnras/stv1218},
archivePrefix = {arXiv},
       eprint = {1505.07603},
 primaryClass = {astro-ph.HE},
       adsurl = {https://ui.adsabs.harvard.edu/abs/2015MNRAS.451.4375F}
}

@BOOK{1999agnc.book.....K,
       author = {{Krolik}, Julian H.},
        title = "{Active galactic nuclei : from the central black hole to the galactic environment}",
         year = 1999,
       adsurl = {https://ui.adsabs.harvard.edu/abs/1999agnc.book.....K}
}

@ARTICLE{1993MNRAS.262..179M,
       author = {{Matt}, G. and {Fabian}, A.~C. and {Ross}, R.~R.},
        title = "{Iron K-alpha lines from X-ray photoionized accretion discs.}",
      journal = {\mnras},
         year = 1993,
        month = may,
       volume = {262},
        pages = {179-186},
          doi = {10.1093/mnras/262.1.179},
       adsurl = {https://ui.adsabs.harvard.edu/abs/1993MNRAS.262..179M}
}

@ARTICLE{2013A&A...554A..85S,
       author = {{Singh}, V. and {Shastri}, P. and {Ishwara-Chandra}, C.~H. and {Athreya}, R.},
        title = "{Low-frequency radio observations of Seyfert galaxies: A test of the unification scheme}",
      journal = {\aap},
         year = 2013,
        month = jun,
       volume = {554},
          eid = {A85},
        pages = {A85},
          doi = {10.1051/0004-6361/201221003},
archivePrefix = {arXiv},
       eprint = {1304.0720},
 primaryClass = {astro-ph.CO},
       adsurl = {https://ui.adsabs.harvard.edu/abs/2013A&A...554A..85S}
}

@ARTICLE{2017ApJ...846..102M,
       author = {{Malkan}, Matthew A. and {Jensen}, Lisbeth D. and {Rodriguez}, David R. and {Spinoglio}, Luigi and {Rush}, Brian},
        title = "{Emission Line Properties of Seyfert Galaxies in the 12 {\ensuremath{\mu}}m Sample}",
      journal = {\apj},
         year = 2017,
        month = sep,
       volume = {846},
       number = {2},
          eid = {102},
        pages = {102},
          doi = {10.3847/1538-4357/aa8302},
archivePrefix = {arXiv},
       eprint = {1708.08563},
 primaryClass = {astro-ph.GA},
       adsurl = {https://ui.adsabs.harvard.edu/abs/2017ApJ...846..102M}
}

@ARTICLE{Poutanen1997,
       author = {{Poutanen}, Juri and {Krolik}, Julian H. and {Ryde}, Felix},
        title = "{The nature of spectral transitions in accreting black holes: the case of CYG X-1}",
      journal = {\mnras},
         year = 1997,
        month = nov,
       volume = {292},
       number = {1},
        pages = {L21-L25},
          doi = {10.1093/mnras/292.1.L21},
archivePrefix = {arXiv},
       eprint = {astro-ph/9709007},
 primaryClass = {astro-ph},
       adsurl = {https://ui.adsabs.harvard.edu/abs/1997MNRAS.292L..21P}
}

@ARTICLE{Ingram2009,
       author = {{Ingram}, Adam and {Done}, Chris and {Fragile}, P. Chris},
        title = "{Low-frequency quasi-periodic oscillations spectra and Lense-Thirring precession}",
      journal = {\mnras},
         year = 2009,
        month = jul,
       volume = {397},
       number = {1},
        pages = {L101-L105},
          doi = {10.1111/j.1745-3933.2009.00693.x},
archivePrefix = {arXiv},
       eprint = {0901.1238},
 primaryClass = {astro-ph.SR},
       adsurl = {https://ui.adsabs.harvard.edu/abs/2009MNRAS.397L.101I}
}

@ARTICLE{Middei2017,
       author = {{Middei}, R. and {Vagnetti}, F. and {Bianchi}, S. and {La Franca}, F. and {Paolillo}, M. and {Ursini}, F.},
        title = "{A long-term study of AGN X-ray variability . Structure function analysis on a ROSAT-XMM quasar sample}",
      journal = {\aap},
         year = 2017,
        month = mar,
       volume = {599},
          eid = {A82},
        pages = {A82},
          doi = {10.1051/0004-6361/201629940},
archivePrefix = {arXiv},
       eprint = {1612.08547},
 primaryClass = {astro-ph.HE},
       adsurl = {https://ui.adsabs.harvard.edu/abs/2017A&A...599A..82M}
}

@ARTICLE{Ponti2012,
       author = {{Ponti}, G. and {Papadakis}, I. and {Bianchi}, S. and {Guainazzi}, M. and {Matt}, G. and {Uttley}, P. and {Bonilla}, N.~F.},
        title = "{CAIXA: a catalogue of AGN in the XMM-Newton archive. III. Excess variance analysis}",
      journal = {\aap},
         year = 2012,
        month = jun,
       volume = {542},
          eid = {A83},
        pages = {A83},
          doi = {10.1051/0004-6361/201118326},
archivePrefix = {arXiv},
       eprint = {1112.2744},
 primaryClass = {astro-ph.HE},
       adsurl = {https://ui.adsabs.harvard.edu/abs/2012A&A...542A..83P}
}

@ARTICLE{Kammoun2015,
       author = {{Kammoun}, E.~S. and {Papadakis}, I.~E. and {Sabra}, B.~M.},
        title = "{Variability of the soft X-ray excess in IRAS 13224-3809}",
      journal = {\aap},
         year = 2015,
        month = oct,
       volume = {582},
          eid = {A40},
        pages = {A40},
          doi = {10.1051/0004-6361/201526811},
archivePrefix = {arXiv},
       eprint = {1508.02589},
 primaryClass = {astro-ph.HE},
       adsurl = {https://ui.adsabs.harvard.edu/abs/2015A&A...582A..40K}
}

@ARTICLE{Ding2022,
       author = {{Ding}, Nan and {Gu}, Qiusheng and {Tang}, Yunyong and {Xiong}, Dingrong and {Guo}, Xiaotong and {Xu}, Xinpeng and {Geng}, Xiongfei and {Ge}, Xue and {Chen}, Yongyun},
        title = "{The variability and soft X-ray excess properties of narrow-line Seyfert 1 galaxies: The Swift view}",
      journal = {\aap},
         year = 2022,
        month = mar,
       volume = {659},
          eid = {A172},
        pages = {A172},
          doi = {10.1051/0004-6361/202142650},
       adsurl = {https://ui.adsabs.harvard.edu/abs/2022A&A...659A.172D}
}

@ARTICLE{Gliozzi2020,
       author = {{Gliozzi}, Mario and {Williams}, James K.},
        title = "{The soft X-ray excess: NLS1s versus BLS1s}",
      journal = {\mnras},
         year = 2020,
        month = jan,
       volume = {491},
       number = {1},
        pages = {532-543},
          doi = {10.1093/mnras/stz3005},
archivePrefix = {arXiv},
       eprint = {1910.12115},
 primaryClass = {astro-ph.HE},
       adsurl = {https://ui.adsabs.harvard.edu/abs/2020MNRAS.491..532G}
}

@ARTICLE{2016AN....337..398M,
       author = {{Motta}, S.~E.},
        title = "{Quasi periodic oscillations in black hole binaries}",
      journal = {Astronomische Nachrichten},
         year = 2016,
        month = may,
       volume = {337},
       number = {4-5},
        pages = {398},
          doi = {10.1002/asna.201612320},
archivePrefix = {arXiv},
       eprint = {1603.07885},
 primaryClass = {astro-ph.HE},
       adsurl = {https://ui.adsabs.harvard.edu/abs/2016AN....337..398M}
}

@ARTICLE{1997ApJ...476..620H,
       author = {{Haardt}, Francesco and {Maraschi}, Laura and {Ghisellini}, Gabriele},
        title = "{X-Ray Variability and Correlations in the Two-Phase Disk-Corona Model for Seyfert Galaxies}",
      journal = {\apj},
         year = 1997,
        month = feb,
       volume = {476},
       number = {2},
        pages = {620-631},
          doi = {10.1086/303656},
archivePrefix = {arXiv},
       eprint = {astro-ph/9609050},
 primaryClass = {astro-ph},
       adsurl = {https://ui.adsabs.harvard.edu/abs/1997ApJ...476..620H}
}

@ARTICLE{2020A&A...634A..92U,
       author = {{Ursini}, F. and {Petrucci}, P. -O. and {Bianchi}, S. and {Matt}, G. and {Middei}, R. and {Marcel}, G. and {Ferreira}, J. and {Cappi}, M. and {De Marco}, B. and {De Rosa}, A. and {Malzac}, J. and {Marinucci}, A. and {Ponti}, G. and {Tortosa}, A.},
        title = "{NuSTAR/XMM-Newton monitoring of the Seyfert 1 galaxy HE 1143-1810. Testing the two-corona scenario}",
      journal = {\aap},
         year = 2020,
        month = feb,
       volume = {634},
          eid = {A92},
        pages = {A92},
          doi = {10.1051/0004-6361/201936486},
archivePrefix = {arXiv},
       eprint = {1912.08720},
 primaryClass = {astro-ph.HE},
       adsurl = {https://ui.adsabs.harvard.edu/abs/2020A&A...634A..92U}
}

@ARTICLE{1995ApJ...452..379A,
       author = {{Abramowicz}, Marek A. and {Chen}, Xingming and {Taam}, Ronald E.},
        title = "{The Evolution of Accretion Disks with Coronae: A Model for the Low-Frequency Quasi-periodic Oscillations in X-Ray Binaries}",
      journal = {\apj},
         year = 1995,
        month = oct,
       volume = {452},
        pages = {379},
          doi = {10.1086/176309},
archivePrefix = {arXiv},
       eprint = {astro-ph/9504074},
 primaryClass = {astro-ph},
       adsurl = {https://ui.adsabs.harvard.edu/abs/1995ApJ...452..379A}
}

@INPROCEEDINGS{2009ASPC..408..296A,
       author = {{Ar{\'e}valo}, P.},
        title = "{Probing the Accretion Disc--Corona Connection in AGN through X-ray and Optical Variability}",
    booktitle = {The Starburst-AGN Connection},
         year = 2009,
       editor = {{Wang}, W. and {Yang}, Z. and {Luo}, Z. and {Chen}, Z.},
       series = {Astronomical Society of the Pacific Conference Series},
       volume = {408},
        month = oct,
        pages = {296},
       adsurl = {https://ui.adsabs.harvard.edu/abs/2009ASPC..408..296A}
}

@ARTICLE{1995A&A...300..707T,
       author = {{Timmer}, J. and {K{\"o}nig}, M.},
        title = "{On generating power law noise.}",
      journal = {\aap},
         year = 1995,
        month = aug,
       volume = {300},
        pages = {707},
       adsurl = {https://ui.adsabs.harvard.edu/abs/1995A&A...300..707T}
}

@ARTICLE{2024MNRAS.528.5269L,
       author = {{Layek}, Narendranath and {Nandi}, Prantik and {Naik}, Sachindra and {Kumari}, Neeraj and {Jana}, Arghajit and {Chhotaray}, Birendra},
        title = "{Long-term X-ray temporal and spectral study of a Seyfert galaxy Mrk 6}",
      journal = {\mnras},
         year = 2024,
        month = mar,
       volume = {528},
       number = {3},
        pages = {5269-5285},
          doi = {10.1093/mnras/stae299},
archivePrefix = {arXiv},
       eprint = {2401.16780},
 primaryClass = {astro-ph.HE},
       adsurl = {https://ui.adsabs.harvard.edu/abs/2024MNRAS.528.5269L}
}

@ARTICLE{2021MNRAS.506.3111N,
       author = {{Nandi}, Prantik and {Chatterjee}, Arka and {Chakrabarti}, Sandip K. and {Dutta}, Broja G.},
        title = "{Long-term X-ray observations of seyfert 1 galaxy ark 120: on the origin of soft-excess}",
      journal = {\mnras},
         year = 2021,
        month = sep,
       volume = {506},
       number = {3},
        pages = {3111-3127},
          doi = {10.1093/mnras/stab1699},
archivePrefix = {arXiv},
       eprint = {2101.08043},
 primaryClass = {astro-ph.HE},
       adsurl = {https://ui.adsabs.harvard.edu/abs/2021MNRAS.506.3111N}
}

@article{nandi2024accretion,
  title={Accretion properties of a low-mass active galactic nucleus: UGC 6728},
  author={Nandi, Prantik and Naik, Sachindra and Chatterjee, Arka and Chakrabarti, Sandip K and Safi-Harb, Samar and Kumari, Neeraj and Layek, Narendranath},
  journal={Monthly Notices of the Royal Astronomical Society},
  volume={532},
  number={1},
  pages={1185--1198},
  year={2024},
  publisher={Oxford University Press}
}

@article{nandi2023survey,
  title={Survey of Bare Active Galactic Nuclei in the Local Universe (z< 0.2). I. On the Origin of Soft Excess},
  author={Nandi, Prantik and Chatterjee, Arka and Jana, Arghajit and Chakrabarti, Sandip K and Naik, Sachindra and Safi-Harb, Samar and Chang, Hsiang-Kuang and Heyl, Jeremy},
  journal={The Astrophysical Journal Supplement Series},
  volume={269},
  number={1},
  pages={15},
  year={2023},
  publisher={IOP Publishing}
}

@article{nandi2019spectral,
  title={Spectral properties of NGC 4151 and the Estimation of black hole mass using TCAF solution},
  author={Nandi, Prantik and Chakrabarti, Sandip K and Mondal, Santanu},
  journal={The Astrophysical Journal},
  volume={877},
  number={2},
  pages={65},
  year={2019},
  publisher={IOP Publishing}
}

@ARTICLE{1997ApJ...482L.167Z,
       author = {{Zhang}, W. and {Strohmayer}, T.~E. and {Swank}, J.~H.},
        title = "{Neutron Star Masses and Radii as Inferred from Kilohertz Quasi-periodic Oscillations}",
      journal = {\apjl},
         year = 1997,
        month = jun,
       volume = {482},
       number = {2},
        pages = {L167-L170},
          doi = {10.1086/310719},
archivePrefix = {arXiv},
       eprint = {astro-ph/9703151},
 primaryClass = {astro-ph},
       adsurl = {https://ui.adsabs.harvard.edu/abs/1997ApJ...482L.167Z}
}

@ARTICLE{2003ApJ...584L..83M,
       author = {{Mukhopadhyay}, Banibrata and {Ray}, Subharthi and {Dey}, Jishnu and {Dey}, Mira},
        title = "{Origin and Interpretation of Kilohertz Quasi-periodic Oscillations from Strange Stars in an X-Ray Binary System: Theoretical Hydrodynamical Description}",
      journal = {\apjl},
         year = 2003,
        month = feb,
       volume = {584},
       number = {2},
        pages = {L83-L86},
          doi = {10.1086/373890},
archivePrefix = {arXiv},
       eprint = {astro-ph/0211611},
 primaryClass = {astro-ph},
       adsurl = {https://ui.adsabs.harvard.edu/abs/2003ApJ...584L..83M}
}

@ARTICLE{1999A&A...349.1003T,
       author = {{Tagger}, M. and {Pellat}, R.},
        title = "{An accretion-ejection instability in magnetized disks}",
      journal = {\aap},
         year = 1999,
        month = sep,
       volume = {349},
        pages = {1003-1016},
          doi = {10.48550/arXiv.astro-ph/9907267},
archivePrefix = {arXiv},
       eprint = {astro-ph/9907267},
 primaryClass = {astro-ph},
       adsurl = {https://ui.adsabs.harvard.edu/abs/1999A&A...349.1003T}
}

@ARTICLE{2003ApJ...593..980L,
       author = {{Li}, Li-Xin and {Goodman}, Jeremy and {Narayan}, Ramesh},
        title = "{Nonaxisymmetric g-Mode and p-Mode Instability in a Hydrodynamic Thin Accretion Disk}",
      journal = {\apj},
         year = 2003,
        month = aug,
       volume = {593},
       number = {2},
        pages = {980-991},
          doi = {10.1086/376695},
archivePrefix = {arXiv},
       eprint = {astro-ph/0210455},
 primaryClass = {astro-ph},
       adsurl = {https://ui.adsabs.harvard.edu/abs/2003ApJ...593..980L}
}

@article{chakrabarti2004effect,
  title={The effect of cooling on time dependent behaviour of accretion flows around black holes},
  author={Chakrabarti, Sandip K and Acharyya, K and Molteni, D},
  journal={Astronomy \& Astrophysics},
  volume={421},
  number={1},
  pages={1--8},
  year={2004},
  publisher={EDP Sciences}
}

@ARTICLE{1985ApJ...297..633S,
       author = {{Singh}, K.~P. and {Garmire}, G.~P. and {Nousek}, J.},
        title = "{Observations of Soft X-Ray Spectra from a Seyfert 1 and a Narrow Emission-Line Galaxy}",
      journal = {\apj},
         year = 1985,
        month = oct,
       volume = {297},
        pages = {633},
          doi = {10.1086/163560},
       adsurl = {https://ui.adsabs.harvard.edu/abs/1985ApJ...297..633S}
}

@ARTICLE{1993ApJ...413..507H,
       author = {{Haardt}, Francesco and {Maraschi}, Laura},
        title = "{X-Ray Spectra from Two-Phase Accretion Disks}",
      journal = {\apj},
         year = 1993,
        month = aug,
       volume = {413},
        pages = {507},
          doi = {10.1086/173020},
       adsurl = {https://ui.adsabs.harvard.edu/abs/1993ApJ...413..507H}
}

@ARTICLE{2001A&A...365L..27T,
       author = {{Turner}, M.~J.~L. and {Abbey}, A. and {Arnaud}, M. and {Balasini}, M. and {Barbera}, M. and {Belsole}, E. and {Bennie}, P.~J. and {Bernard}, J.~P. and {Bignami}, G.~F. and {Boer}, M. and {Briel}, U. and {Butler}, I. and {Cara}, C. and {Chabaud}, C. and {Cole}, R. and {Collura}, A. and {Conte}, M. and {Cros}, A. and {Denby}, M. and {Dhez}, P. and {Di Coco}, G. and {Dowson}, J. and {Ferrando}, P. and {Ghizzardi}, S. and {Gianotti}, F. and {Goodall}, C.~V. and {Gretton}, L. and {Griffiths}, R.~G. and {Hainaut}, O. and {Hochedez}, J.~F. and {Holland}, A.~D. and {Jourdain}, E. and {Kendziorra}, E. and {Lagostina}, A. and {Laine}, R. and {La Palombara}, N. and {Lortholary}, M. and {Lumb}, D. and {Marty}, P. and {Molendi}, S. and {Pigot}, C. and {Poindron}, E. and {Pounds}, K.~A. and {Reeves}, J.~N. and {Reppin}, C. and {Rothenflug}, R. and {Salvetat}, P. and {Sauvageot}, J.~L. and {Schmitt}, D. and {Sembay}, S. and {Short}, A.~D.~T. and {Spragg}, J. and {Stephen}, J. and {Str{\"u}der}, L. and {Tiengo}, A. and {Trifoglio}, M. and {Tr{\"u}mper}, J. and {Vercellone}, S. and {Vigroux}, L. and {Villa}, G. and {Ward}, M.~J. and {Whitehead}, S. and {Zonca}, E.},
        title = "{The European Photon Imaging Camera on XMM-Newton: The MOS cameras}",
      journal = {\aap},
         year = 2001,
        month = jan,
       volume = {365},
        pages = {L27-L35},
          doi = {10.1051/0004-6361:20000087},
archivePrefix = {arXiv},
       eprint = {astro-ph/0011498},
 primaryClass = {astro-ph},
       adsurl = {https://ui.adsabs.harvard.edu/abs/2001A&A...365L..27T}
}

@ARTICLE{2001A&A...365L..18S,
       author = {{Str{\"u}der}, L. and {Briel}, U. and {Dennerl}, K. and {Hartmann}, R. and {Kendziorra}, E. and {Meidinger}, N. and {Pfeffermann}, E. and {Reppin}, C. and {Aschenbach}, B. and {Bornemann}, W. and {Br{\"a}uninger}, H. and {Burkert}, W. and {Elender}, M. and {Freyberg}, M. and {Haberl}, F. and {Hartner}, G. and {Heuschmann}, F. and {Hippmann}, H. and {Kastelic}, E. and {Kemmer}, S. and {Kettenring}, G. and {Kink}, W. and {Krause}, N. and {M{\"u}ller}, S. and {Oppitz}, A. and {Pietsch}, W. and {Popp}, M. and {Predehl}, P. and {Read}, A. and {Stephan}, K.~H. and {St{\"o}tter}, D. and {Tr{\"u}mper}, J. and {Holl}, P. and {Kemmer}, J. and {Soltau}, H. and {St{\"o}tter}, R. and {Weber}, U. and {Weichert}, U. and {von Zanthier}, C. and {Carathanassis}, D. and {Lutz}, G. and {Richter}, R.~H. and {Solc}, P. and {B{\"o}ttcher}, H. and {Kuster}, M. and {Staubert}, R. and {Abbey}, A. and {Holland}, A. and {Turner}, M. and {Balasini}, M. and {Bignami}, G.~F. and {La Palombara}, N. and {Villa}, G. and {Buttler}, W. and {Gianini}, F. and {Lain{\'e}}, R. and {Lumb}, D. and {Dhez}, P.},
        title = "{The European Photon Imaging Camera on XMM-Newton: The pn-CCD camera}",
      journal = {\aap},
         year = 2001,
        month = jan,
       volume = {365},
        pages = {L18-L26},
          doi = {10.1051/0004-6361:20000066},
       adsurl = {https://ui.adsabs.harvard.edu/abs/2001A&A...365L..18S}
}

@ARTICLE{2023A&A...672A..86R,
       author = {{Ren}, Helena X. and {Cerruti}, Matteo and {Sahakyan}, Narek},
        title = "{Quasi-periodic oscillations in the {\ensuremath{\gamma}}-ray light curves of bright active galactic nuclei}",
      journal = {\aap},
         year = 2023,
        month = apr,
       volume = {672},
          eid = {A86},
        pages = {A86},
          doi = {10.1051/0004-6361/202244754},
archivePrefix = {arXiv},
       eprint = {2204.13051},
 primaryClass = {astro-ph.HE},
       adsurl = {https://ui.adsabs.harvard.edu/abs/2023A&A...672A..86R}
}

@ARTICLE{1976ApJ...210L.117T,
       author = {{Tohline}, J.~E. and {Osterbrock}, D.~E.},
        title = "{Variation of the spectrum of the Seyfert galaxy NGC 7603.}",
      journal = {\apjl},
         year = 1976,
        month = dec,
       volume = {210},
        pages = {L117-L120},
          doi = {10.1086/182317},
       adsurl = {https://ui.adsabs.harvard.edu/abs/1976ApJ...210L.117T}
}

@ARTICLE{2023MNRAS.521.5440K,
       author = {{Kumari}, Neeraj and {Jana}, Arghajit and {Naik}, Sachindra and {Nandi}, Prantik},
        title = "{Investigation of a small X-ray flaring event in NLS1 galaxy NGC 4051}",
      journal = {\mnras},
         year = 2023,
        month = jun,
       volume = {521},
       number = {4},
        pages = {5440-5452},
          doi = {10.1093/mnras/stad867},
archivePrefix = {arXiv},
       eprint = {2303.11799},
 primaryClass = {astro-ph.HE},
       adsurl = {https://ui.adsabs.harvard.edu/abs/2023MNRAS.521.5440K}
}

@ARTICLE{1990ApJ...363L...1Z,
       author = {{Zdziarski}, Andrzej A. and {Ghisellini}, Gabriele and {George}, Ian M. and {Svensson}, Roland and {Fabian}, A.~C. and {Done}, Chris},
        title = "{Electron-Positron Pairs, Compton Reflection, and the X-Ray Spectra of Active Galactic Nuclei}",
      journal = {\apjl},
         year = 1990,
        month = nov,
       volume = {363},
        pages = {L1},
          doi = {10.1086/185851},
       adsurl = {https://ui.adsabs.harvard.edu/abs/1990ApJ...363L...1Z}
}

@ARTICLE{1996ApJ...457..805M,
       author = {{Molteni}, Diego and {Sponholz}, Hanno and {Chakrabarti}, Sandip K.},
        title = "{Resonance Oscillation of Radiative Shock Waves in Accretion Disks around Compact Objects}",
      journal = {\apj},
         year = 1996,
        month = feb,
       volume = {457},
        pages = {805},
          doi = {10.1086/176775},
archivePrefix = {arXiv},
       eprint = {astro-ph/9508022},
 primaryClass = {astro-ph},
       adsurl = {https://ui.adsabs.harvard.edu/abs/1996ApJ...457..805M}
}

@ARTICLE{1996ApJ...464..760H,
       author = {{Halpern}, Jules P. and {Marshall}, Herman L.},
        title = "{A Long EUVE Observation of the Seyfert Galaxy RX J0437.4-4711}",
      journal = {\apj},
         year = 1996,
        month = jun,
       volume = {464},
        pages = {760},
          doi = {10.1086/177361},
       adsurl = {https://ui.adsabs.harvard.edu/abs/1996ApJ...464..760H}
}

@ARTICLE{2000ApJ...531L..41C,
       author = {{Chakrabarti}, Sandip K. and {Manickam}, Sivakumar G.},
        title = "{Correlation among Quasi-Periodic Oscillation Frequencies and Quiescent-State Duration in Black Hole Candidate GRS 1915+105}",
      journal = {\apjl},
         year = 2000,
        month = mar,
       volume = {531},
       number = {1},
        pages = {L41-L44},
          doi = {10.1086/312512},
archivePrefix = {arXiv},
       eprint = {astro-ph/9910012},
 primaryClass = {astro-ph},
       adsurl = {https://ui.adsabs.harvard.edu/abs/2000ApJ...531L..41C}
}

@INPROCEEDINGS{1999ASPC..161..178R,
       author = {{Reynolds}, C.~S.},
        title = "{Compton Reflection and Iron Fluorescence in Active Galactic Nuclei and Galactic Black Hole Candidates}",
    booktitle = {High Energy Processes in Accreting Black Holes},
         year = 1999,
       editor = {{Poutanen}, Juri and {Svensson}, Roland},
       series = {Astronomical Society of the Pacific Conference Series},
       volume = {161},
        month = jan,
        pages = {178},
          doi = {10.48550/arXiv.astro-ph/9810018},
archivePrefix = {arXiv},
       eprint = {astro-ph/9810018},
 primaryClass = {astro-ph},
       adsurl = {https://ui.adsabs.harvard.edu/abs/1999ASPC..161..178R}
}

@ARTICLE{2021MNRAS.507..687J,
       author = {{Jana}, Arghajit and {Kumari}, Neeraj and {Nandi}, Prantik and {Naik}, Sachindra and {Chatterjee}, Arka and {Jaisawal}, Gaurava K. and {Hayasaki}, Kimitake and {Ricci}, Claudio},
        title = "{Broad-band X-ray observations of the 2018 outburst of the changing-look active galactic nucleus NGC 1566}",
      journal = {\mnras},
         year = 2021,
        month = oct,
       volume = {507},
       number = {1},
        pages = {687-703},
          doi = {10.1093/mnras/stab2155},
archivePrefix = {arXiv},
       eprint = {2107.11127},
 primaryClass = {astro-ph.HE},
       adsurl = {https://ui.adsabs.harvard.edu/abs/2021MNRAS.507..687J}
}

@ARTICLE{2013MNRAS.430.1694D,
       author = {{Dauser}, T. and {Garcia}, J. and {Wilms}, J. and {B{\"o}ck}, M. and {Brenneman}, L.~W. and {Falanga}, M. and {Fukumura}, K. and {Reynolds}, C.~S.},
        title = "{Irradiation of an accretion disc by a jet: general properties and implications for spin measurements of black holes}",
      journal = {\mnras},
         year = 2013,
        month = apr,
       volume = {430},
       number = {3},
        pages = {1694-1708},
          doi = {10.1093/mnras/sts710},
archivePrefix = {arXiv},
       eprint = {1301.4922},
 primaryClass = {astro-ph.HE},
       adsurl = {https://ui.adsabs.harvard.edu/abs/2013MNRAS.430.1694D}
}

@ARTICLE{2018MNRAS.474.5351P,
       author = {{Pal}, Main and {Naik}, Sachindra},
        title = "{Correlated X-ray/UV/optical emission and short-term variability in a Seyfert 1 galaxy NGC 4593}",
      journal = {\mnras},
         year = 2018,
        month = mar,
       volume = {474},
       number = {4},
        pages = {5351-5362},
          doi = {10.1093/mnras/stx3103},
archivePrefix = {arXiv},
       eprint = {1711.11194},
 primaryClass = {astro-ph.HE},
       adsurl = {https://ui.adsabs.harvard.edu/abs/2018MNRAS.474.5351P}
}

@ARTICLE{2021PASA...38...42K,
       author = {{Kumari}, Neeraj and {Pal}, Main and {Naik}, Sachindra and {Jana}, Arghajit and {Jaisawal}, Gaurava K. and {Kushwaha}, Pankaj},
        title = "{Complex optical/UV and X-ray variability in Seyfert 1 galaxy Mrk 509}",
      journal = {\pasa},
         year = 2021,
        month = sep,
       volume = {38},
          eid = {e042},
        pages = {e042},
          doi = {10.1017/pasa.2021.41},
archivePrefix = {arXiv},
       eprint = {2107.11994},
 primaryClass = {astro-ph.HE},
       adsurl = {https://ui.adsabs.harvard.edu/abs/2021PASA...38...42K}
}

@ARTICLE{2025Natur.638..370M,
       author = {{Masterson}, Megan and {Kara}, Erin and {Panagiotou}, Christos and {Alston}, William N. and {Chakraborty}, Joheen and {Burdge}, Kevin and {Ricci}, Claudio and {Laha}, Sibasish and {Arcavi}, Iair and {Arcodia}, Riccardo and {Cenko}, S. Bradley and {Fabian}, Andrew C. and {Garc{\'\i}a}, Javier A. and {Giustini}, Margherita and {Ingram}, Adam and {Kosec}, Peter and {Loewenstein}, Michael and {Meyer}, Eileen T. and {Miniutti}, Giovanni and {Pinto}, Ciro and {Remillard}, Ronald A. and {Sadaula}, Dev R. and {Shuvo}, Onic I. and {Trakhtenbrot}, Benny and {Wang}, Jingyi},
        title = "{Millihertz oscillations near the innermost orbit of a supermassive black hole}",
      journal = {\nat},
         year = 2025,
        month = feb,
       volume = {638},
       number = {8050},
        pages = {370-375},
          doi = {10.1038/s41586-024-08385-x},
archivePrefix = {arXiv},
       eprint = {2501.01581},
 primaryClass = {astro-ph.HE},
       adsurl = {https://ui.adsabs.harvard.edu/abs/2025Natur.638..370M}
}

@ARTICLE{2022MNRAS.515..353X,
       author = {{Xiang}, X. and {Ballantyne}, D.~R. and {Bianchi}, S. and {De Rosa}, A. and {Matt}, G. and {Middei}, R. and {Petrucci}, P.-O. and {R{\'o}{\.z}a{\'n}ska}, A. and {Ursini}, F.},
        title = "{REXCOR: a model of the X-ray spectrum of active galactic nuclei that combines ionized reflection and a warm corona}",
      journal = {\mnras},
         year = 2022,
        month = sep,
       volume = {515},
       number = {1},
        pages = {353-368},
          doi = {10.1093/mnras/stac1646},
archivePrefix = {arXiv},
       eprint = {2206.06825},
 primaryClass = {astro-ph.HE},
       adsurl = {https://ui.adsabs.harvard.edu/abs/2022MNRAS.515..353X}
}

@ARTICLE{2021ApJ...913...13X,
       author = {{Xu}, Yerong and {Garc{\'\i}a}, Javier A. and {Walton}, Dominic J. and {Connors}, Riley M.~T. and {Madsen}, Kristin and {Harrison}, Fiona A.},
        title = "{The Nature of Soft Excess in ESO 362-G18 Revealed by XMM-Newton and NuSTAR Spectroscopy}",
      journal = {\apj},
         year = 2021,
        month = may,
       volume = {913},
       number = {1},
          eid = {13},
        pages = {13},
          doi = {10.3847/1538-4357/abf430},
archivePrefix = {arXiv},
       eprint = {2103.17002},
 primaryClass = {astro-ph.HE},
       adsurl = {https://ui.adsabs.harvard.edu/abs/2021ApJ...913...13X}
}

@ARTICLE{2019ApJ...871...88G,
       author = {{Garc{\'\i}a}, Javier A. and {Kara}, Erin and {Walton}, Dominic and {Beuchert}, Tobias and {Dauser}, Thomas and {Gatuzz}, Efrain and {Balokovic}, Mislav and {Steiner}, James F. and {Tombesi}, Francesco and {Connors}, Riley M.~T. and {Kallman}, Timothy R. and {Harrison}, Fiona A. and {Fabian}, Andrew and {Wilms}, J{\"o}rn and {Stern}, Daniel and {Lanz}, Lauranne and {Ricci}, Claudio and {Ballantyne}, David R.},
        title = "{Implications of the Warm Corona and Relativistic Reflection Models for the Soft Excess in Mrk 509}",
      journal = {\apj},
         year = 2019,
        month = jan,
       volume = {871},
       number = {1},
          eid = {88},
        pages = {88},
          doi = {10.3847/1538-4357/aaf739},
archivePrefix = {arXiv},
       eprint = {1812.03194},
 primaryClass = {astro-ph.HE},
       adsurl = {https://ui.adsabs.harvard.edu/abs/2019ApJ...871...88G}
}

@ARTICLE{2004MNRAS.349L...7G,
       author = {{Gierli{\'n}ski}, Marek and {Done}, Chris},
        title = "{Is the soft excess in active galactic nuclei real?}",
      journal = {\mnras},
         year = 2004,
        month = mar,
       volume = {349},
       number = {1},
        pages = {L7-L11},
          doi = {10.1111/j.1365-2966.2004.07687.x},
archivePrefix = {arXiv},
       eprint = {astro-ph/0312271},
 primaryClass = {astro-ph},
       adsurl = {https://ui.adsabs.harvard.edu/abs/2004MNRAS.349L...7G}
}

@article{waddell2024erosita,
  title={The eROSITA Final Equatorial Depth Survey (eFEDS): Complex absorption and soft excesses in hard X-ray--selected active galactic nuclei},
  author={Waddell, Sophia GH and Nandra, Kirpal and Buchner, Johannes and Wu, Qiaoya and Shen, Yue and Arcodia, Riccardo and Merloni, Andrea and Salvato, Mara and Dauser, Thomas and Boller, Th and others},
  journal={Astronomy \& Astrophysics},
  volume={690},
  pages={A132},
  year={2024},
  publisher={EDP Sciences}
}

@software{2017ascl.soft02002F,
       author = {{Foreman-Mackey}, Daniel},
        title = "{corner.py: Corner plots}",
 howpublished = {Astrophysics Source Code Library, record ascl:1702.002},
         year = 2017,
        month = feb,
          eid = {ascl:1702.002},
archivePrefix = {ascl},
       eprint = {1702.002},
       adsurl = {https://ui.adsabs.harvard.edu/abs/2017ascl.soft02002F}
}

@INPROCEEDINGS{2014xru..confE..25U,
       author = {{Uttley}, P.},
        title = "{X-ray time lags and reverberation from accreting black holes}",
    booktitle = {The X-ray Universe 2014},
         year = 2014,
       editor = {{Ness}, Jan-Uwe},
        month = jul,
          eid = {25},
        pages = {25},
       adsurl = {https://ui.adsabs.harvard.edu/abs/2014xru..confE..25U}
}

@ARTICLE{2010MNRAS.404..738C,
       author = {{Cabanac}, C. and {Henri}, G. and {Petrucci}, P.-O. and {Malzac}, J. and {Ferreira}, J. and {Belloni}, T.~M.},
        title = "{Variability of X-ray binaries from an oscillating hot corona}",
      journal = {\mnras},
         year = 2010,
        month = may,
       volume = {404},
       number = {2},
        pages = {738-748},
          doi = {10.1111/j.1365-2966.2010.16340.x},
archivePrefix = {arXiv},
       eprint = {1001.2116},
 primaryClass = {astro-ph.HE},
       adsurl = {https://ui.adsabs.harvard.edu/abs/2010MNRAS.404..738C}
}

@ARTICLE{Risaliti2002,
       author = {{Risaliti}, G. and {Elvis}, M. and {Nicastro}, F.},
        title = "{Ubiquitous Variability of X-Ray-absorbing Column Densities in Seyfert 2 Galaxies}",
      journal = {\apj},
         year = 2002,
        month = may,
       volume = {571},
       number = {1},
        pages = {234-246},
          doi = {10.1086/324146},
archivePrefix = {arXiv},
       eprint = {astro-ph/0107510},
 primaryClass = {astro-ph},
       adsurl = {https://ui.adsabs.harvard.edu/abs/2002ApJ...571..234R}
}

@ARTICLE{Risaliti2009,
       author = {{Risaliti}, G. and {Miniutti}, G. and {Elvis}, M. and {Fabbiano}, G. and {Salvati}, M. and {Baldi}, A. and {Braito}, V. and {Bianchi}, S. and {Matt}, G. and {Reeves}, J. and {Soria}, R. and {Zezas}, A.},
        title = "{Variable Partial Covering and A Relativistic Iron Line in NGC 1365}",
      journal = {\apj},
         year = 2009,
        month = may,
       volume = {696},
       number = {1},
        pages = {160-171},
          doi = {10.1088/0004-637X/696/1/160},
archivePrefix = {arXiv},
       eprint = {0901.4809},
 primaryClass = {astro-ph.CO},
       adsurl = {https://ui.adsabs.harvard.edu/abs/2009ApJ...696..160R}
}

@ARTICLE{Nenkova2008,
       author = {{Nenkova}, Maia and {Sirocky}, Matthew M. and {Nikutta}, Robert and {Ivezi{\'c}}, {\v{Z}}eljko and {Elitzur}, Moshe},
        title = "{AGN Dusty Tori. II. Observational Implications of Clumpiness}",
      journal = {\apj},
         year = 2008,
        month = sep,
       volume = {685},
       number = {1},
        pages = {160-180},
          doi = {10.1086/590483},
archivePrefix = {arXiv},
       eprint = {0806.0512},
 primaryClass = {astro-ph},
       adsurl = {https://ui.adsabs.harvard.edu/abs/2008ApJ...685..160N}
}

@ARTICLE{Markowitz2014,
       author = {{Markowitz}, A.~G. and {Krumpe}, M. and {Nikutta}, R.},
        title = "{First X-ray-based statistical tests for clumpy-torus models: eclipse events from 230 years of monitoring of Seyfert AGN}",
      journal = {\mnras},
         year = 2014,
        month = apr,
       volume = {439},
       number = {2},
        pages = {1403-1458},
          doi = {10.1093/mnras/stt2492},
archivePrefix = {arXiv},
       eprint = {1402.2779},
 primaryClass = {astro-ph.GA},
       adsurl = {https://ui.adsabs.harvard.edu/abs/2014MNRAS.439.1403M}
}

@ARTICLE{2025ApJ...994..216L,
       author = {{Layek}, Narendranath and {Nandi}, Prantik and {Naik}, Sachindra and {Chhotaray}, Birendra and {Jana}, Arghajit and {Dash}, Priyadarshee P. and {Kumari}, Neeraj and {Stalin}, C.~S. and {Bandari}, Srikanth and {Muneer}, S.},
        title = "{Discovery of Changing-look Behavior in AGN NGC 3822: A Long-term Multiwavelength Study}",
      journal = {\apj},
         year = 2025,
        month = dec,
       volume = {994},
       number = {2},
          eid = {216},
        pages = {216},
          doi = {10.3847/1538-4357/ae10ae},
archivePrefix = {arXiv},
       eprint = {2510.05599},
 primaryClass = {astro-ph.HE},
       adsurl = {https://ui.adsabs.harvard.edu/abs/2025ApJ...994..216L}
}

@ARTICLE{Nandi2026,
       author = {{Nandi}, Prantik and {Layek}, Narendranath and {Chakrabarti}, Sandip K. and {Naik}, Sachindra and {Dash}, Priyadarshee P.},
        title = "{Evolution of accretion properties in Mrk 1040 using long-term X-ray observations}",
      journal = {\mnras},
         year = 2026,
        month = mar,
       volume = {546},
       number = {3},
          eid = {stag072},
        pages = {stag072},
          doi = {10.1093/mnras/stag072},
archivePrefix = {arXiv},
       eprint = {2512.24141},
 primaryClass = {astro-ph.HE},
       adsurl = {https://ui.adsabs.harvard.edu/abs/2026MNRAS.546ag072N}
}




\appendix

\section{Some Additional Material}

\begin{figure*}
    \centering

    \begin{subfigure}[b]{0.24\textwidth}
        \includegraphics[width=\textwidth]{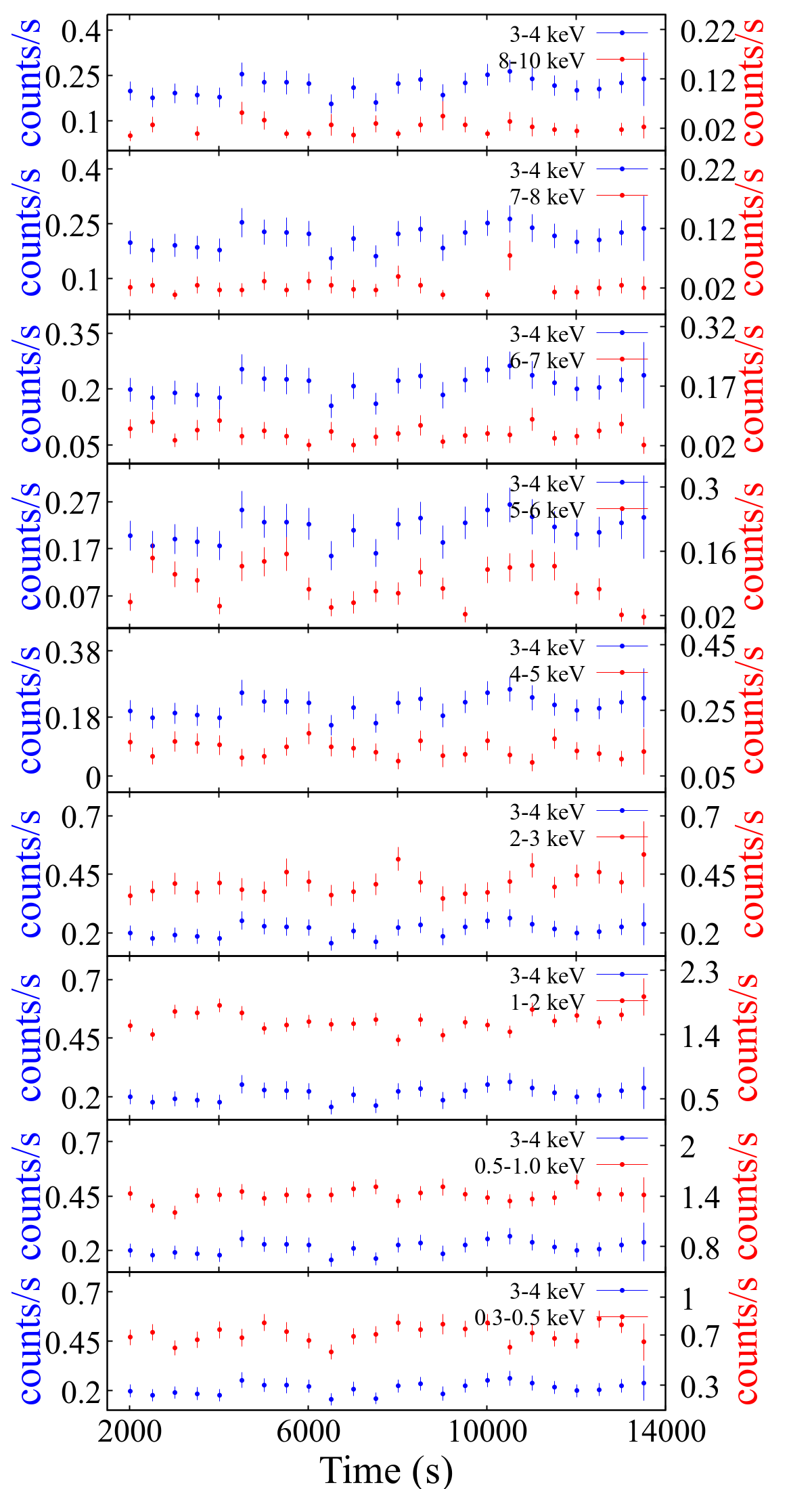}
    \end{subfigure}
    \begin{subfigure}[b]{0.24\textwidth}
        \includegraphics[width=\textwidth]{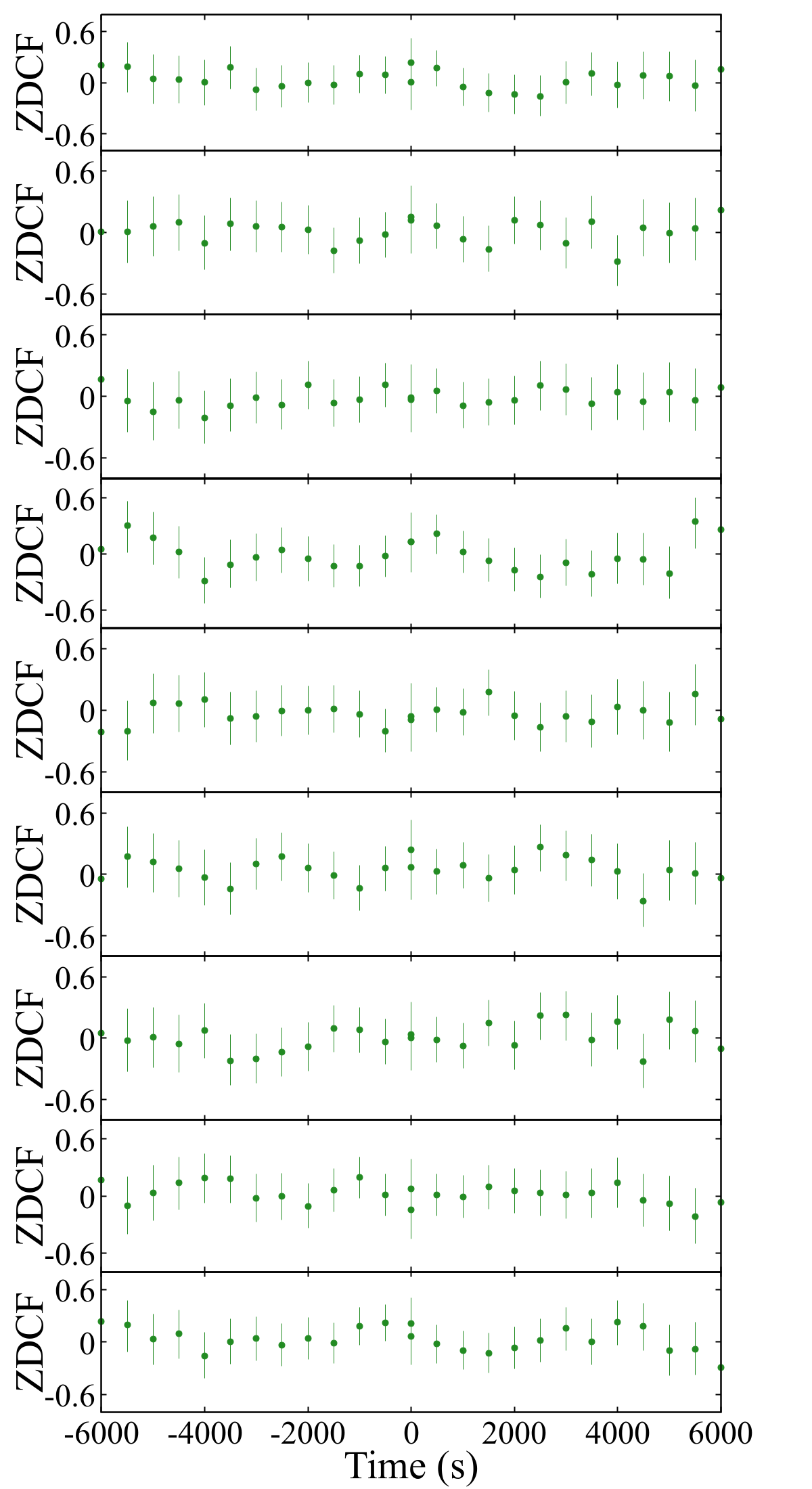}
    \end{subfigure}
    \begin{subfigure}[b]{0.24\textwidth}
        \includegraphics[width=\textwidth]{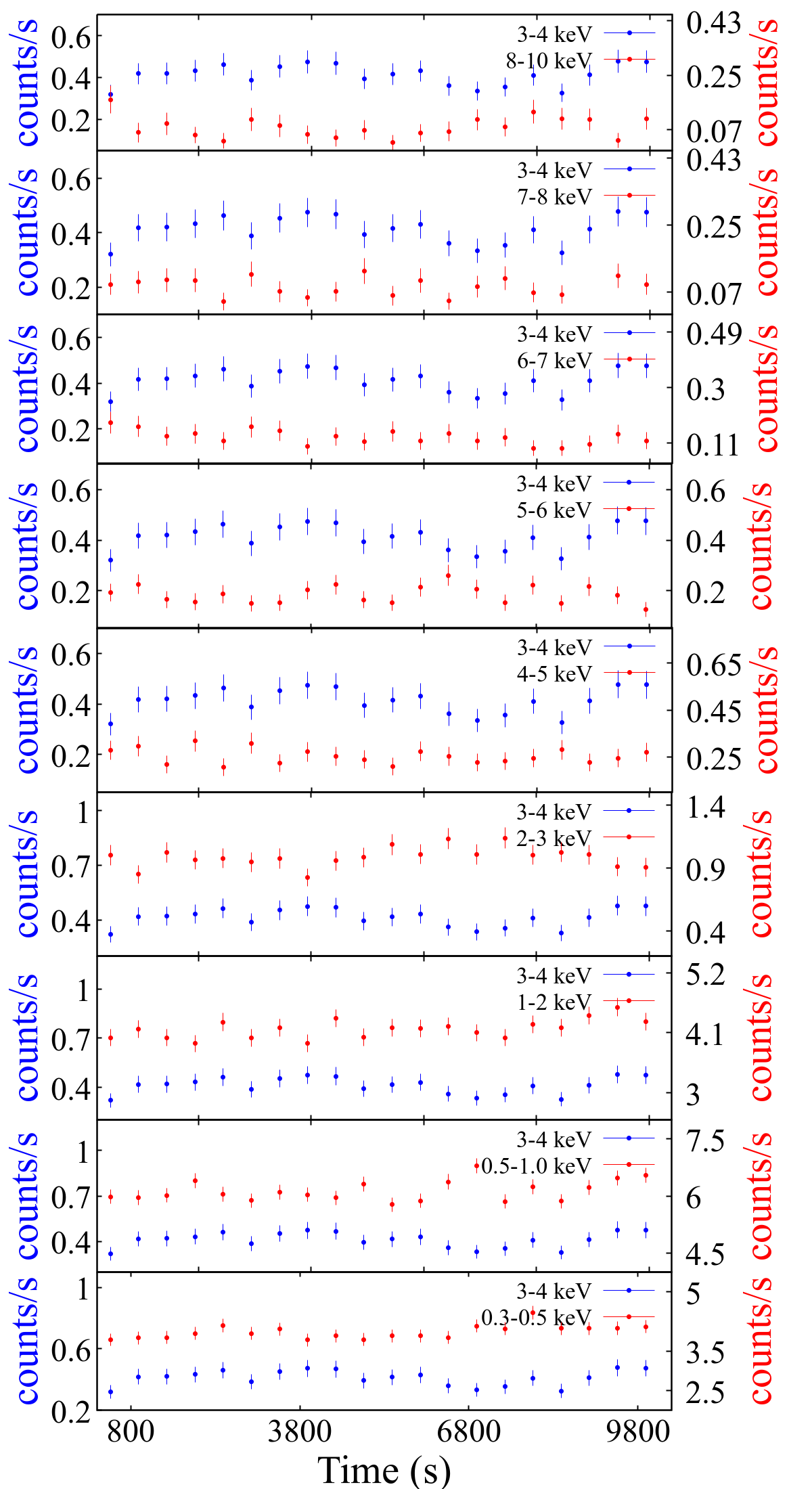}
    \end{subfigure}
    \begin{subfigure}[b]{0.24\textwidth}
        \includegraphics[width=\textwidth]{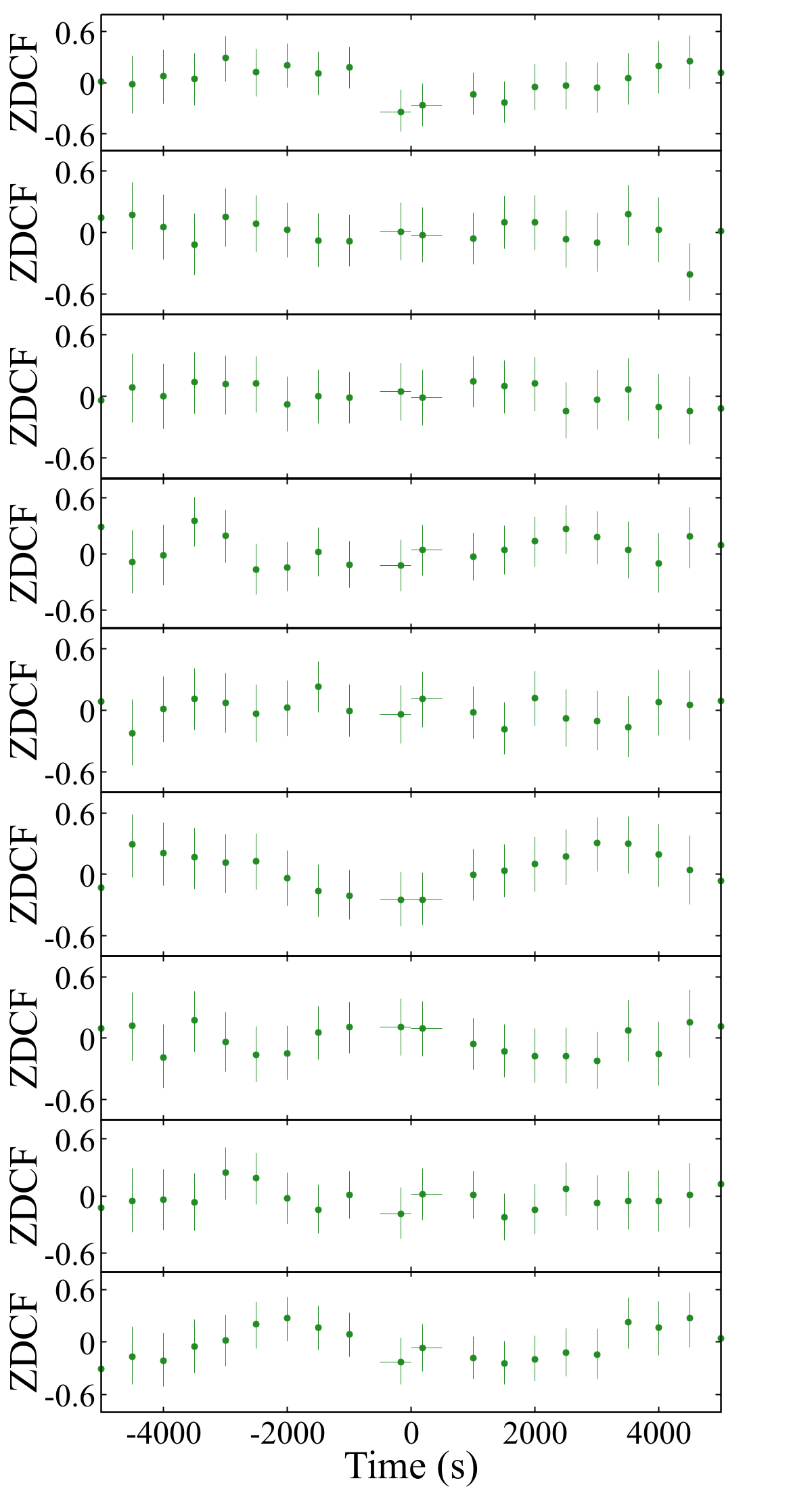}
    \end{subfigure}

    \vspace{0.5cm}
    \begin{subfigure}[b]{0.24\textwidth}
        \includegraphics[width=\textwidth]{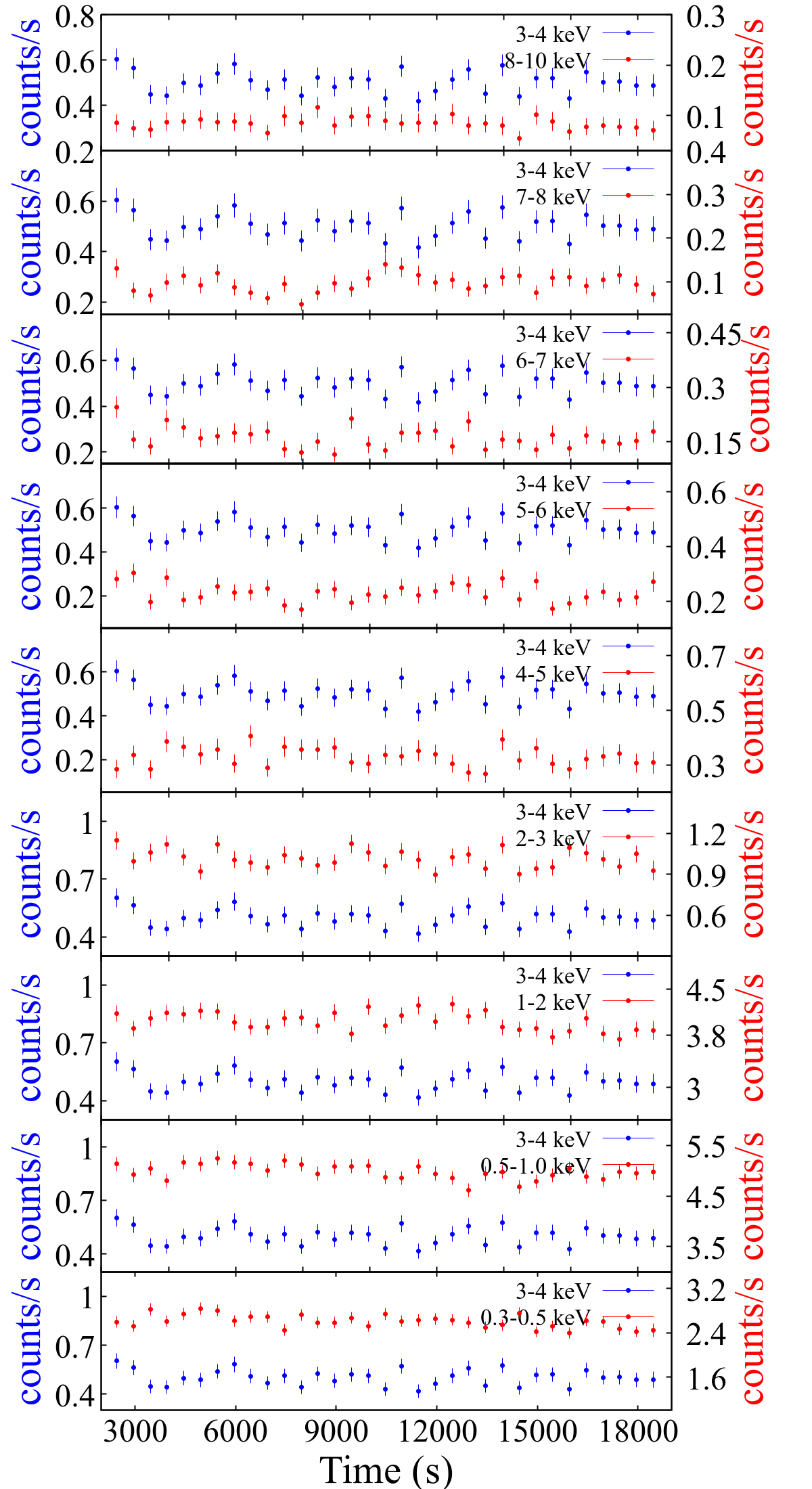}
    \end{subfigure}
    \hspace{0.2cm}
    \begin{subfigure}[b]{0.24\textwidth}
        \includegraphics[width=\textwidth]{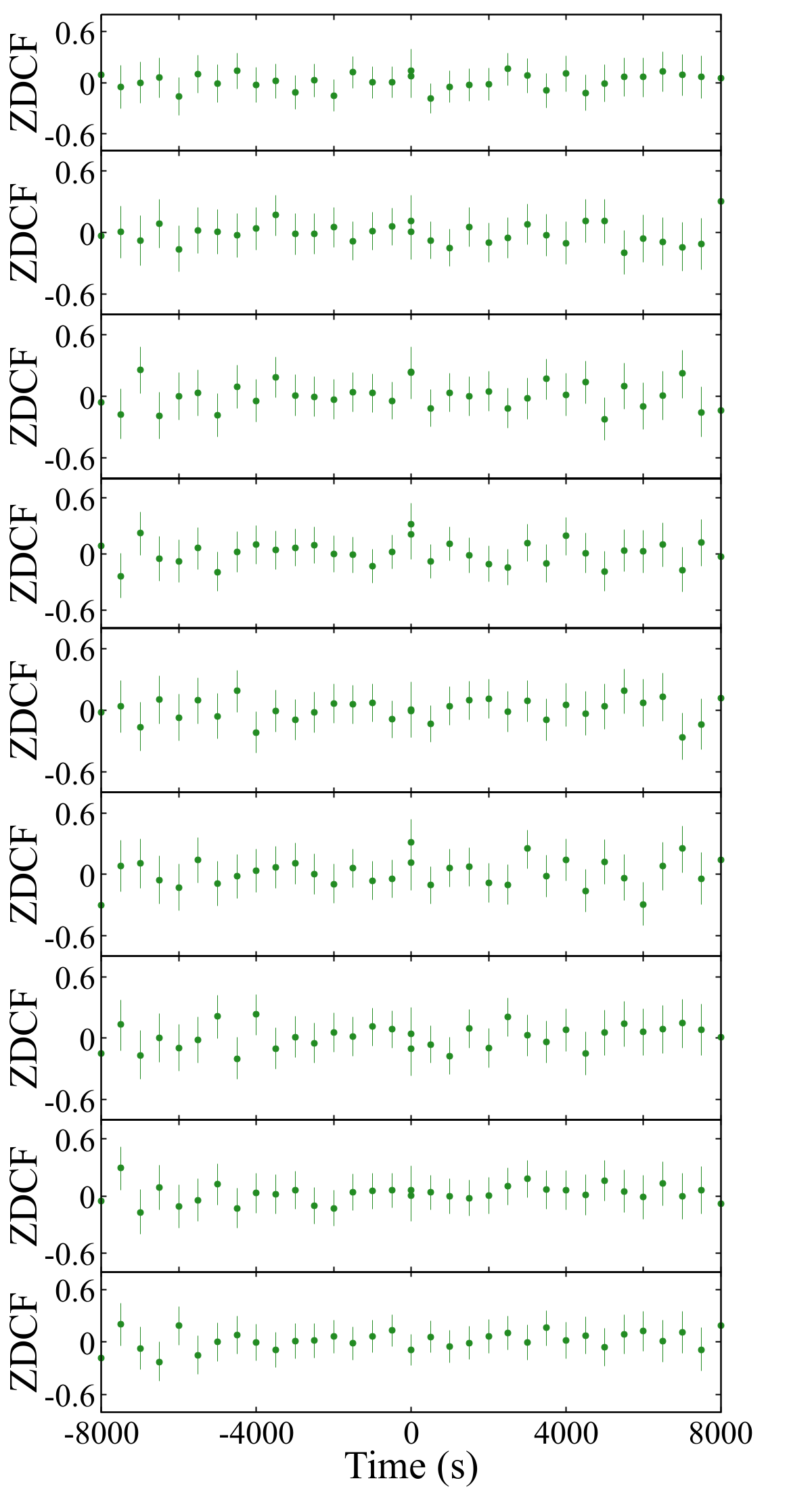}
    \end{subfigure}

    \caption{The lightcurves and the corresponding correlation functions (ZDCF) for the XMM1 (top left), XMM2 (top right) and XMM3 (bottom) observations in different energy bands.}
    \label{fig:lightcurves}
\end{figure*}

\begin{table*}
\caption{The best fit parameters for the baseline model \texttt{Tbabs $\times$ powerlaw} for all \textit{Swift}/XRT observations of Mrk 530 in the year 2018.}
    \begin{tabular}{c|c|c|c|c|c|c}
    \hline\\
         Date & $\Gamma$ & $Norm^{\dagger}$ & log $L_{total}$ & log $L_{hard}$ & log $L_{soft}$ & $C-stat/dof$  \\
         (MJD) & & ($10^{-3}$) & (erg $s{^-1}$) & (erg $s^{-1}$) & (erg $s^{-1}$) & \\
          \hline\\
          58120 & $2.24\pm0.07$ & $10.69^{+0.49}_{-0.48}$ & $43.96\pm0.02$ & $43.52\pm0.05$ & $43.77\pm0.02$ & 212.45/191 \\ 
          \vspace{0.1cm}
          58124 & $2.10\pm0.10$ & $8.43^{+0.58}_{-0.56}$ & $43.88\pm0.03$ & $43.50\pm0.06$ & $43.65\pm0.04$ & 81.05/94 \\\vspace{0.1cm}
          58128 & $2.12\pm0.09$ & $7.33^{+0.43}_{-0.42}$ & $43.82\pm0.03$ & $43.43\pm0.06$ & $43.59\pm0.03$ & 168.86/149 \\\vspace{0.1cm}
          58131 & $1.79\pm0.12$ & $7.45^{+0.67}_{-0.64}$ & $43.90\pm0.04$ & $43.64\pm0.07$ & $43.56\pm0.04$ & 60.60/70 \\\vspace{0.1cm}
          58133 & $1.96\pm0.12$ & $6.20^{+0.50}_{-0.48}$ & $43.78\pm0.04$ & $43.46\pm0.07$ & $43.50\pm0.04$ & 91.52/79 \\\vspace{0.1cm}
          58135 & $1.83\pm0.15$ & $6.02^{+0.64}_{-0.60}$ & $43.80\pm0.05$ & $43.53\pm0.09$ & $43.47\pm0.05$ & 76.97/52 \\\vspace{0.1cm}
          58137 & $1.83\pm0.13$ & $5.49^{+0.59}_{-0.55}$ &$43.78\pm0.05$ & $43.51\pm0.08$ & $43.45\pm0.05$ & 55.76/57 \\\vspace{0.1cm}
          58141 & $1.87\pm0.25$ & $7.34^{+1.26}_{-1.14}$ & $43.87\pm0.08$ & $43.59\pm0.16$ & $43.56\pm0.08$ & 15.37/20 \\\vspace{0.1cm}
          58236 & $1.86\pm0.18$ & $5.25^{+0.73}_{-0.69}$ & $43.73\pm0.06$ & $43.45\pm0.10$ & $43.41\pm0.07$ & 33.70/32 \\\vspace{0.1cm}
          58238 & $1.92\pm0.11$ & $6.73^{+0.53}_{-0.51}$ & $43.82\pm0.04$ & $43.51\pm0.07$ & $43.53\pm0.04$ & 100.07/82 \\\vspace{0.1cm}
          58240 & $1.81\pm0.10$ & $6.49^{+0.50}_{-0.48}$ & $43.84\pm0.04$ & $43.57\pm0.06$ & $43.50\pm0.04$ & 98.00/95 \\\vspace{0.1cm}
          58242 & $1.99\pm0.20$ & $6.09^{+0.88}_{-0.80}$ & $43.76\pm0.06$ & $43.43\pm0.12$ & $43.49\pm0.07$ & 28.08/27 \\\vspace{0.1cm}
          58244 & $1.98\pm0.10$ & $7.36^{+0.54}_{-0.52}$ & $43.85\pm0.03$ & $43.52\pm0.06$ & $43.57\pm0.04$ & 81.63/96 \\\vspace{0.1cm}
          58246 & $1.92\pm0.12$ & $7.86^{+0.66}_{-0.62}$ & $43.89\pm0.04$ & $43.56\pm0.07$ & $43.59\pm0.04$ & 92.01/78 \\\vspace{0.1cm}
          58250 & $1.84\pm0.10$ & $9.04^{+0.66}_{-0.63}$ & $43.97\pm0.03$ & $43.70\pm0.06$ & $43.65\pm0.04$ & 119.30/102 \\\vspace{0.1cm}
          58252 & $2.01\pm0.10$ & $8.69^{+0.61}_{-0.59}$ & $43.91\pm0.03$ & $43.57\pm0.06$ & $43.65\pm0.04$ & 107.91/100 \\\vspace{0.1cm}
          58254 & $2.00\pm0.12$ & $8.05^{+0.71}_{-0.67}$ & $43.88\pm0.04$ & $43.54\pm0.08$ & $43.61\pm0.04$ & 75.93/69 \\\vspace{0.1cm}
          58256 & $1.81\pm0.12$ & $5.93^{+0.54}_{-0.51}$ & $43.80\pm0.04$ & $43.53\pm0.07$ & $43.46\pm0.04$ & 77.04/73 \\\vspace{0.1cm}
          58258 & $2.03\pm0.19$ & $4.91^{+0.65}_{-0.60}$ & $43.66\pm0.06$ & $43.31\pm0.12$ & $43.40\pm0.07$ & 19.85/32 \\\vspace{0.1cm}
          58260 & $1.95\pm0.11$ & $5.02^{+0.40}_{-0.38}$ & $43.69\pm0.04$ & $43.37\pm0.07$ & $43.40\pm0.04$ & 110.01/83 \\\vspace{0.1cm}
          58262 & $2.07\pm0.17$ & $7.51^{+0.93}_{-0.86}$ & $43.83\pm0.05$ & $43.47\pm0.11$ & $43.59\pm0.06$ & 25.23/36 \\\vspace{0.1cm}
          58264 & $1.85\pm0.11$ & $5.41^{+0.42}_{-0.40}$ & $43.75\pm0.04$ & $43.47\pm0.06$ & $43.42\pm0.04$ & 86.74/89 \\\vspace{0.1cm}
          58265 & $1.81\pm0.10$ & $6.05^{+0.47}_{-0.45}$ & $43.81\pm0.04$ & $43.54\pm0.06$ & $43.47\pm0.04$ & 87.42/94 \\\vspace{0.1cm}
          58268 & $1.81\pm0.12$ & $4.84^{+0.42}_{-0.40}$ & $43.71\pm0.04$ & $43.45\pm0.07$ & $43.37\pm0.04$ & 80.22/75 \\\vspace{0.1cm}
          58270 & $1.58\pm0.14$ & $3.22^{+0.37}_{-0.34}$ & $43.61\pm0.05$ & $43.42\pm0.08$ & $43.18\pm0.05$ & 43.88/52 \\\vspace{0.1cm}
          58272 & $1.75\pm0.13$ & $3.35^{+0.34}_{-0.32}$ & $43.57\pm0.05$ & $43.33\pm0.08$ & $43.21\pm0.05$ & 64.82/61 \\\vspace{0.1cm}
          58278 & $1.78\pm0.15$ & $2.66^{+0.30}_{-0.28}$ & $43.46\pm0.05$ & $43.21\pm0.09$ & $43.11\pm0.05$ & 35.38/46 \\\vspace{0.1cm}
          58280 & $1.65\pm0.17$ & $1.90^{+0.24}_{-0.23}$ & $43.36\pm0.06$ & $43.15\pm0.10$ & $42.95\pm0.06$ & 56.62/40 \\\vspace{0.1cm}
          58282 & $1.69\pm0.21$ & $2.01^{+0.33}_{-0.30}$ & $43.37\pm0.08$ & $43.14\pm0.13$ & $42.98\pm0.08$ & 19.84/25 \\\vspace{0.1cm}
          58283 & $1.78\pm0.14$ & $3.28^{+0.35}_{-0.33}$ & $43.55\pm0.05$ & $43.30\pm0.09$ & $43.20\pm0.05$ & 58.06/52 \\\vspace{0.1cm}
          58286 & $1.70\pm0.15$ & $2.67^{+0.31}_{-0.28}$ & $43.49\pm0.05$ & $43.26\pm0.09$ & $43.10\pm0.05$ & 39.32/48 \\\vspace{0.1cm}
          58288 & $1.81\pm0.15$ & $2.88^{+0.33}_{-0.30}$ & $43.49\pm0.05$ & $43.22\pm0.09$ & $43.15\pm0.05$ & 46.00/47 \\\vspace{0.1cm}
          58292 & $1.59\pm0.21$ & $2.73^{+0.47}_{0.42}$ & $43.54\pm0.08$ & $43.34\pm0.12$ & $43.10\pm0.08$ & 31.29/24 \\\vspace{0.1cm}
          58294 & $1.73\pm0.11$ & $4.73^{+0.40}_{-0.38}$ & $43.73\pm0.04$ & $43.49\pm0.07$ & $43.35\pm0.04$ & 84.46/84 \\\vspace{0.1cm}
          58296 & $1.90\pm0.17$ & $5.31^{+0.63}_{-0.59}$ & $43.72\pm0.06$ & $43.42\pm0.11$ & $43.42\pm0.06$ & 44.14/40 \\\vspace{0.1cm}
          58298 & $1.98\pm0.13$ & $7.14^{+0.66}_{-0.62}$ & $43.83\pm0.04$ & $43.51\pm0.08$ & $43.56\pm0.05$ & 87.94/62 \\\vspace{0.1cm}
          58301 & $1.71\pm0.12$ & $3.87^{+0.35}_{-0.33}$ & $43.65\pm0.04$ & $43.41\pm0.07$ & $43.27\pm0.04$ & 69.65/71 \\\vspace{0.1cm}
          58306 & $1.84\pm0.12$ & $4.94^{+0.42}_{-0.40}$ & $43.71\pm0.04$ & $43.43\pm0.07$ & $43.38\pm0.04$ & 87.81/78\\\vspace{0.1cm}
          58308 & $1.83\pm0.43$ & $5.68^{+1.59}_{-1.35}$ & $43.77\pm0.15$ & $43.50\pm0.27$ & $43.44\pm0.14$ & 6.59/7 \\\vspace{0.1cm}
          58310 & $1.89\pm0.11$ & $6.40^{+0.53}_{-0.50}$ & $43.81\pm0.04$ & $43.52\pm0.07$ & $43.50\pm0.04$ & 71.09/78 \\\vspace{0.1cm}
          58312 & $1.52\pm0.30$ & $8.55^{+1.65}_{-1.48}$ & $44.07\pm0.12$ & $43.89\pm0.18$ & $43.60\pm0.09$ & 14.10/18 \\\vspace{0.1cm}
          58314 & $1.91\pm0.09$ & $5.50^{+0.36}_{-0.34}$ & $43.74\pm0.03$ & $43.44\pm0.06$ & $43.44\pm0.03$ & 112.32/120 \\\vspace{0.1cm}
          58316 & $1.79\pm0.12$ & $5.19^{+0.47}_{-0.44}$ & $43.75\pm0.04$ & $43.49\pm0.08$ & $43.40\pm0.04$ & 69.00/69 \\\vspace{0.1cm}
          58320 & $1.97\pm0.26$ & $3.69^{+0.59}_{-0.54}$ & $43.55\pm0.08$ & $43.23\pm0.16$ & $43.27\pm0.08$ & 30.92/21 \\\vspace{0.1cm}
          58324 & $1.88\pm0.23$ & $3.55^{+0.62}_{-0.56}$ & $43.55\pm0.08$ & $43.26\pm0.14$ & $43.25\pm0.09$ & 30.82/20 \\
          \hline
    \end{tabular}
    \label{tab:2018_pheno}
    \leftline{$\dagger$ in the units of $photons/keV/s/cm^2$}
\end{table*}
\begin{figure*}
    \centering

    \begin{minipage}[b]{0.48\linewidth}
        \centering
        \includegraphics[width=\linewidth]{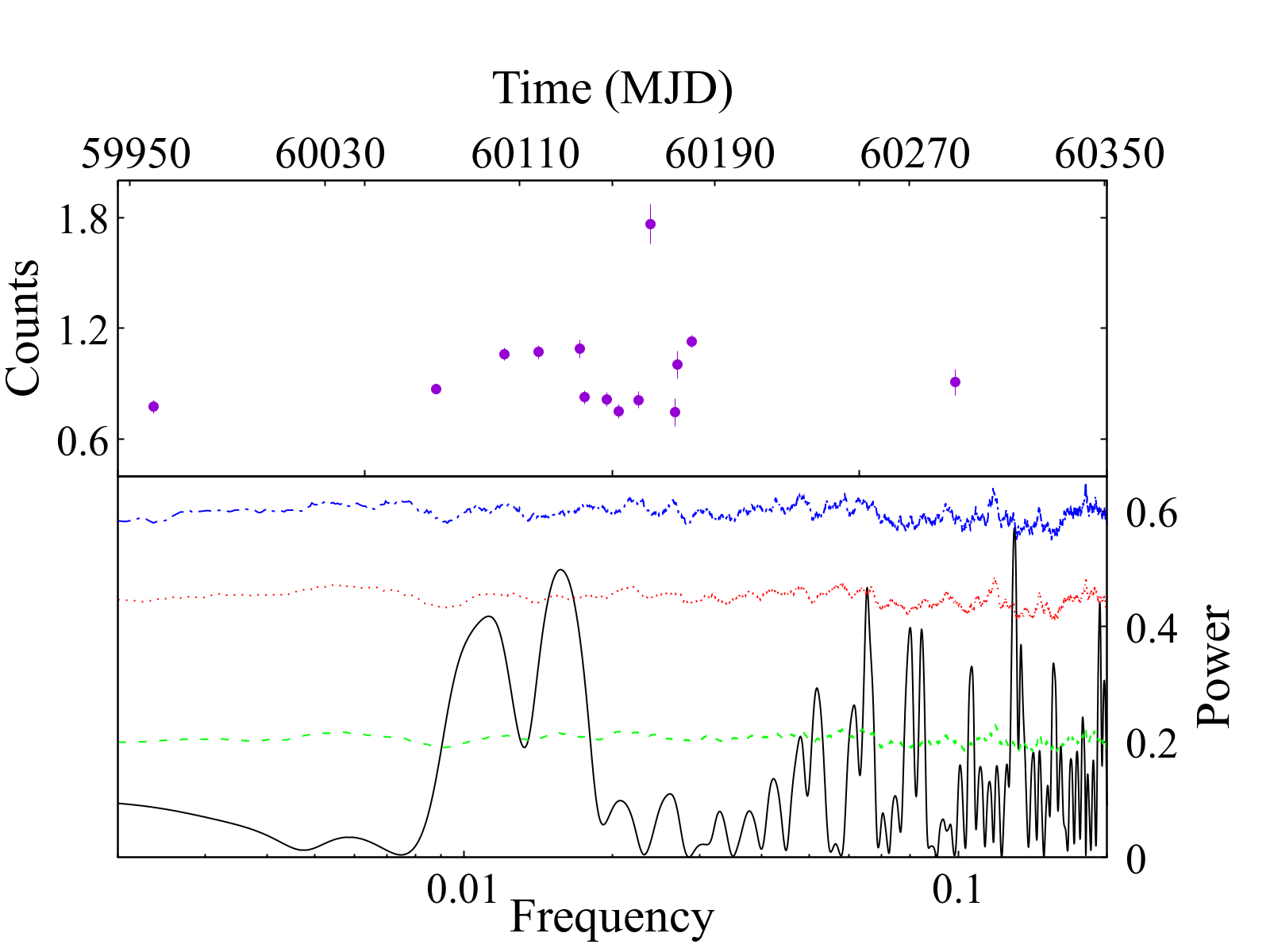}
    \end{minipage}
    \hspace{0.1cm}
    \begin{minipage}[b]{0.48\linewidth}
        \centering
        \includegraphics[width=\linewidth]{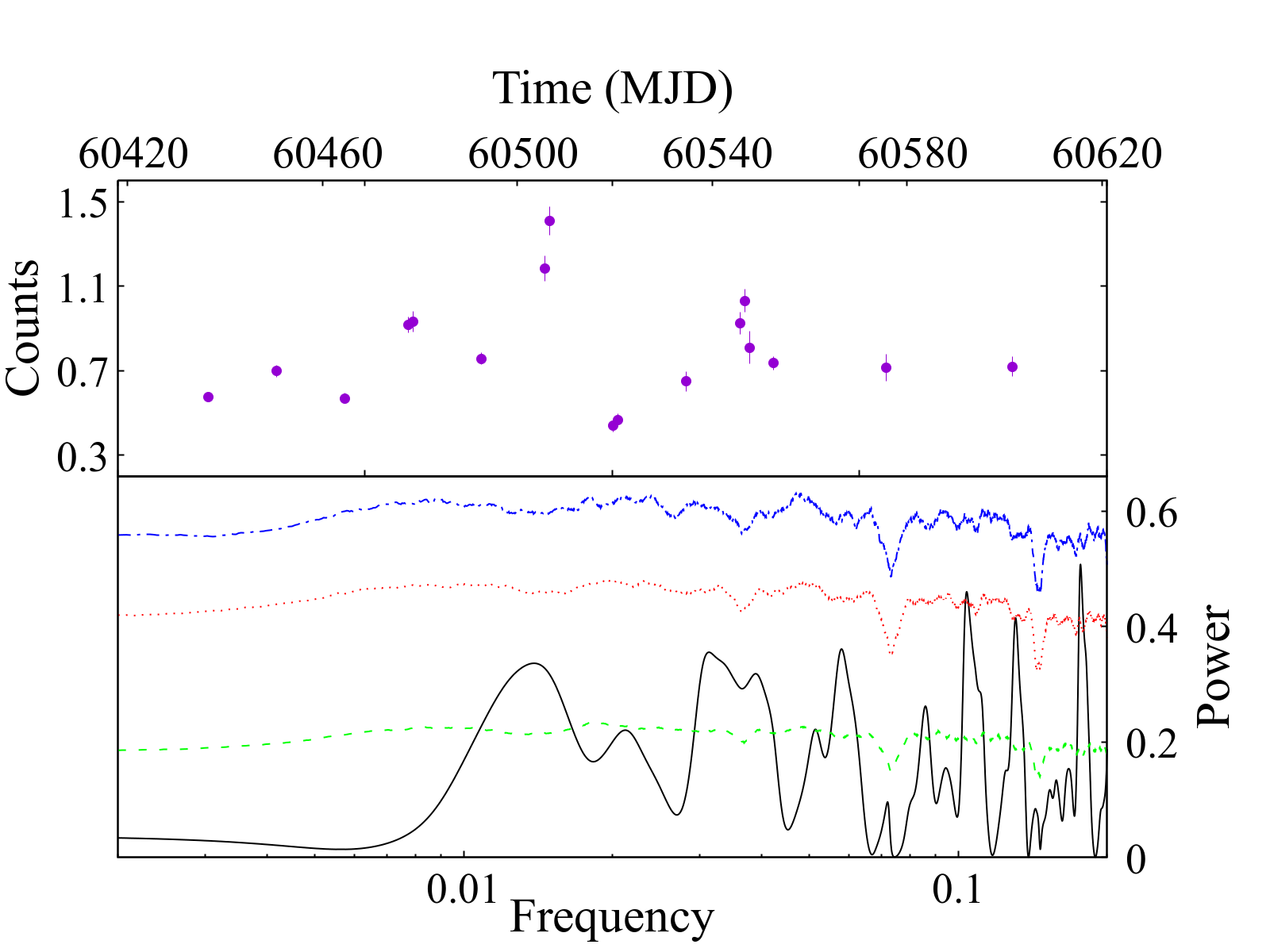}
    \end{minipage}
    \hspace{0.1cm}
    \caption{Lightcurves (top) and the corresponding periodograms (bottom) obtained from the {\it Swift}/XRT monitoring of Mrk 530 in the year 2023 (left) and 2024 (right). The 1$\sigma$, 2$\sigma$ and 3$\sigma$ confidence levels estimated from simulating surrogate light curves are overplotted in the periodogram.} 
    \label{fig:periodogram}
\end{figure*}

\begin{figure*}
    \centering
    \includegraphics[width=0.8\linewidth]{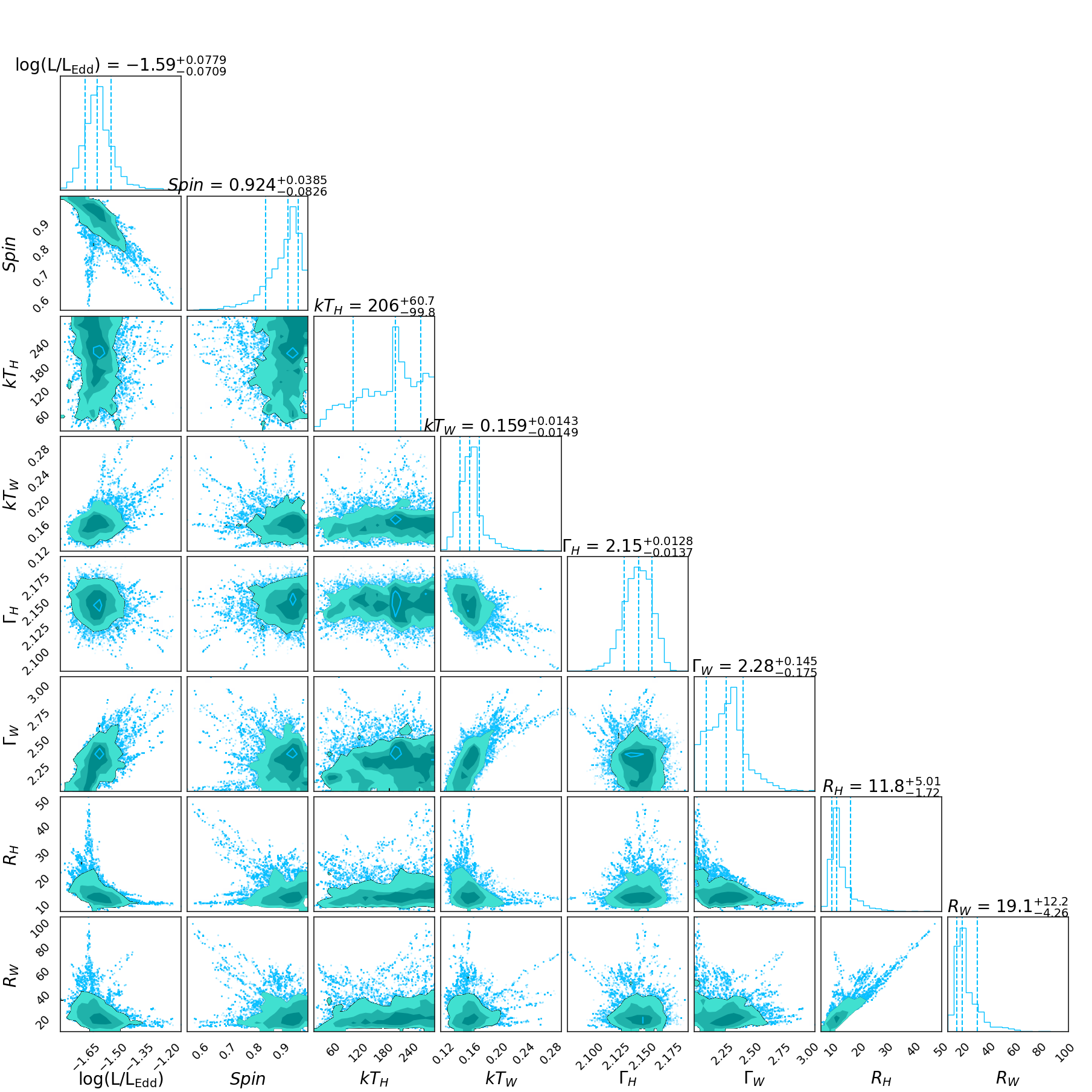}
    \caption{Corner plots of spectral parameters from MCMC analysis for the 2006 XMM3 observation using the {\tt AGNSED} model. The plots were created using the public code {\tt CORNER.PY} by \protect\cite{2017ascl.soft02002F}}
    \label{fig:mcmc}
\end{figure*}
\begin{figure*}
    \centering
    \includegraphics[width=0.6\linewidth]{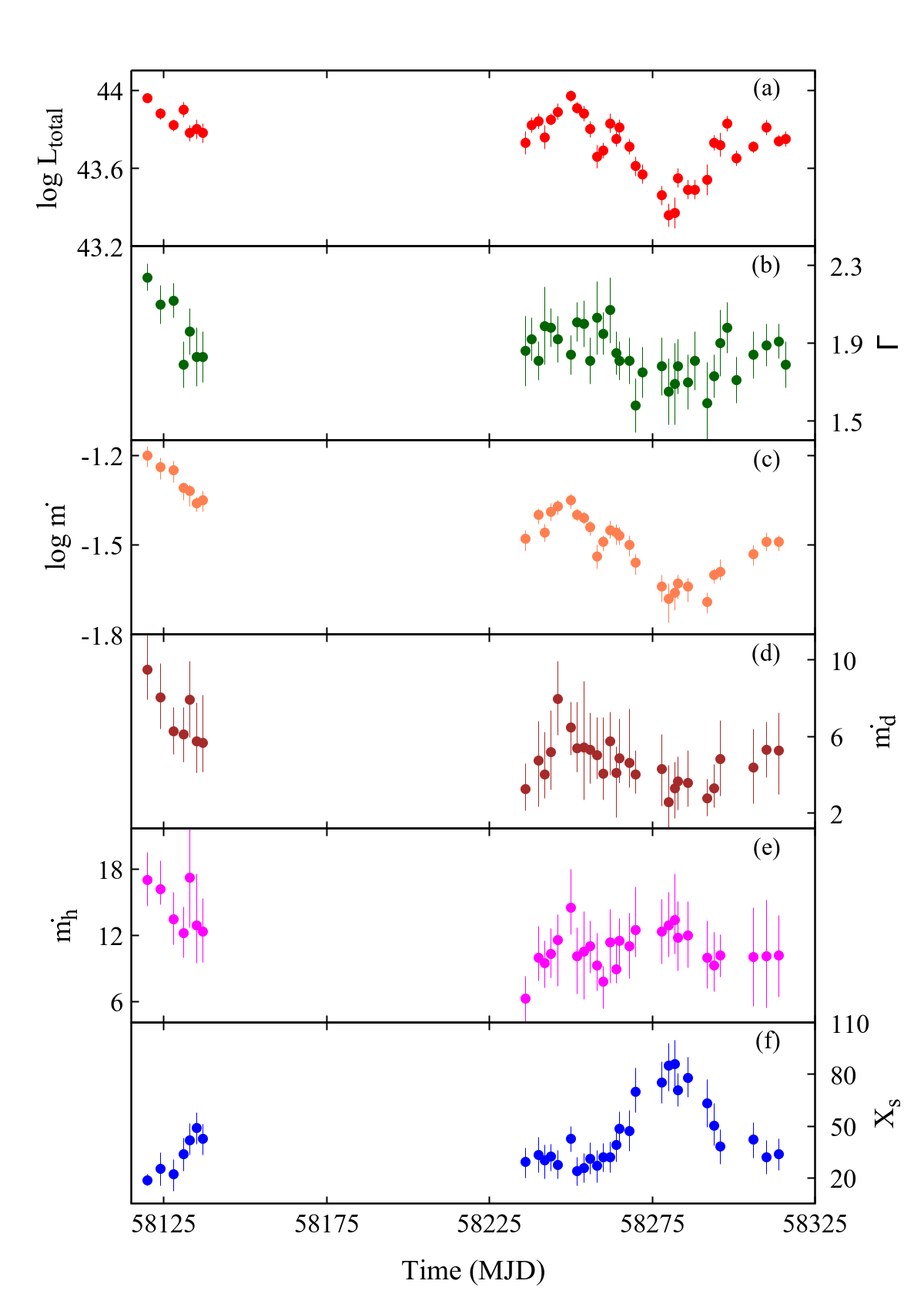}
    \caption{Variation of spectral parameters derived from the combined {\it Swift}/XRT and UVOT  observations in 2018. During the joint UV-X-ray spectral fitting exercise, the hot coronal temperature reached the upper bound of the model and was therefore fixed at $kT_{e,hot} = 100$ keV. Since no significant soft excess was detected in these observations, the warm corona parameters were fixed at values of  $kT_{e,warm} = 0.2$ keV, $\Gamma_{warm}=2.5$, and $R_{warm} = 6 ~Rg$. The inferred mass accretion rates obtained from the {\tt AGNSED} model (panel~c) follow a similar trend to that of the $0.3-10$ keV X-ray luminosity (panel~a) and photon index (panel~b). The disk accretion rates derived within the {\tt TCAF} framework (panel-d) also correlate with the accretion rates inferred from {\tt AGNSED}, while the halo accretion rate (panel~e) remains invariant within uncertainties. The shock location (panel~f) anti-correlates with both the X-ray luminosity and the accretion rates, supporting the physical scenario inferred from the X-ray-only spectral analysis of the 2018 {\it Swift} datasets.}
    \label{fig:2018_param_uv}
\end{figure*}

\begin{table*}
\caption{The best fit parameters for the physical models TCAF and AGNSED for the broadband UV-X-ray spectral analysis of the \textit{Swift}/XRT and UVOT observations of Mrk 530 in 2018.}
    \resizebox{\textwidth}{!}{\begin{tabular}{c|c|c|c|c|c|c|c|c|c|c|c}
        \hline 
        & & & TCAF & & & & & & AGNSED & & \\
        \hline \\
        Date & $\dot{m}_d$ & $\dot{m}_h$ & $X_s$ & $R$ & $Norm^\dagger$ & C-stat/dof & & log $\dot{m}$ & $\Gamma_{hot}$ & $R_{hot}$ & C-stat/dof \\ 
        & ($\times10^{-3}$) & ($\times10^{-3}$) & ($R_g$) & ($ \times10^{-4}$) & & & & & & ($R_g$) & \\ \hline \vspace{0.2cm}
    
        58120 & $9.48^{+1.85}_{-1.56}$ & $17.00^{+2.51}_{-2.34}$ & $19^{+3}_{-3}$ & $1.79^{+0.78}_{-0.56}$ & $10.5^{+3.4}_{-2.3}$ & 210.35/189 & & $-1.20^{+0.03}_{-0.04}$ & $2.41^{+0.01}_{-0.01}$ & $294^{+86}_{-99}$ & 227.34/191 \\ \vspace{0.2cm}
        
        58124 & $8.05^{+1.76}_{-1.66}$ & $16.21^{+2.54}_{-1.45}$ & $25^{10}_{-9}$ & $1.86^{+0.66}_{-0.56}$ & $10.0^{+4.4}_{-4.3}$ & 109.22/92 & & $-1.24^{+0.03}_{-0.04}$ & $2.43^{+0.01}_{-0.01}$ & $272^{+89}_{-97}$ & 111.01/94 \\ \vspace{0.2cm}
        
        58128 & $6.29^{+1.25}_{-1.22}$ & $13.47^{+2.43}_{-2.33}$ & $22^{+10}_{-8}$ & $1.89^{+0.63}_{-0.77}$ & $13.2^{+5.5}_{-4.5}$ & 165.20/147 & & $-1.25^{+0.03}_{-0.04}$ & $2.46^{+0.01}_{-0.01}$ & $306^{+69}_{-79}$ & 202.86/149 \\ \vspace{0.2cm}
        
        58131 & $6.10^{+1.42}_{-1.43}$ & $12.23^{+2.37}_{-2.25}$ & $34^{+10}_{-9}$ & $1.79^{+0.77}_{-0.65}$ & $10.2^{+4.4}_{-3.4}$ & 110.03/71 & & $-1.31^{+0.02}_{-0.04}$ & $2.40^{+0.01}_{-0.01}$ & $285^{+76}_{-92}$ & 127.76/70 \\ \vspace{0.2cm}
        
        58133 & $7.94^{+2.00}_{-1.98}$ & $17.25^{+4.43}_{-4.54}$ & $42^{+8}_{-10}$ & $1.90^{+0.92}_{-0.89}$ & $9.89^{+5.4}_{-5.2}$ & 109.56/77 & & $-1.32^{+0.02}_{-0.05}$ & $2.44^{+0.01}_{-0.01}$ & $307^{+107}_{-65}$ & 139.53/79 \\ \vspace{0.2cm}
        
        58135 & $5.75^{+2.03}_{-1.66}$ & $12.91^{+4.66}_{-3.45}$ & $49^{+9}_{-9}$ & $1.86^{+0.55}_{-0.81}$ & $10.3^{+5.6}_{-5.4}$ & 85.19/50 & & $-1.36^{+0.02}_{-0.03}$ & $2.44^{+0.01}_{-0.01}$ & $226^{+57}_{-68}$ & 95.14/52 \\ \vspace{0.2cm}
        
        58137 & $5.69^{+2.46}_{-1.56}$ & $12.32^{+3.01}_{-2.78}$ & $43^{+10}_{-9}$ & $1.80^{+0.99}_{-0.79}$ & $8.35^{+4.4}_{-3.3}$ & 85.17/55 & & $-1.35^{+0.03}_{-0.04}$ & $2.44^{+0.01}_{-0.01}$ & $260^{+80}_{-83}$ & 114.22/57 \\ \vspace{0.2cm}
        
        58236 & $3.24^{+1.34}_{-1.12}$ & $6.30^{+2.01}_{-2.21}$ & $29^{+9}_{-8}$ & $1.84^{+0.79}_{-0.81}$ & $9.89^{+3.3}_{-3.1}$ & 33.60/30 & & $-1.48^{+0.03}_{-0.04}$ & $2.40^{+0.02}_{-0.01}$ & $221^{+80}_{-96}$ & 56.80/32 \\ \vspace{0.2cm}
        
        58240 & $4.76^{+2.02}_{-2.44}$ & $9.97^{+2.87}_{-2.11}$ & $33^{+10}_{-10}$ & $1.86^{+0.88}_{-0.81}$ & $7.02^{+3.3}_{-3.2}$ & 125.45/93 & & $-1.40^{+0.02}_{-0.03}$ & $2.38^{+0.01}_{-0.01}$ & $251^{+45}_{-75}$ & 181.52/95 \\ \vspace{0.2cm}
        
        58242 & $4.01^{+2.22}_{-1.23}$ & $9.49^{+2.01}_{-2.26}$ & $30^{+11}_{-9}$ & $1.94^{+0.99}_{-0.92}$ & $15.9^{+5.5}_{-5.3}$ & 22.73/25 & & $-1.46^{+0.03}_{-0.03}$ & $2.40^{+0.02}_{-0.02}$ & $191^{+69}_{-76}$ & 39.28/27 \\ \vspace{0.2cm}
        
        58244 & $5.19^{+2.19}_{-1.97}$ & $10.32^{+2.29}_{-2.15}$ & $32^{+8}_{-7}$ & $1.83^{+0.99}_{-0.82}$ & $10.8^{+5.6}_{-4.4}$ & 90.11/94 & & $-1.39^{+0.03}_{-0.03}$ & $2.38^{+0.01}_{-0.01}$ & $239^{+75}_{-81}$ & 122.25/96 \\ \vspace{0.2cm}
        
        58246 & $7.97^{+1.95}_{-1.88}$ & $11.57^{+2.32}_{-4.21}$ & $28^{+8}_{-8}$ & $1.90^{+0.88}_{-0.87}$ & $28.0^{+6.4}_{-2.3}$ & 90.57/76 & & $-1.37^{+0.02}_{-0.03}$ & $2.37^{+0.01}_{-0.01}$ & $256^{+71}_{-84}$ & 131.67/78 \\ \vspace{0.2cm}
        
        58250 & $6.47^{+1.34}_{-1.45}$ & $14.47^{+3.52}_{-2.44}$ & $42^{+7}_{-7}$ & $1.90^{+0.88}_{-0.82}$ & $1.05^{+3.3}_{-3.4}$ & 119.63/100 & & $-1.35^{+0.02}_{-0.03}$ & $2.33^{+0.01}_{-0.01}$ & $285^{+65}_{-94}$ & 187.87/102 \\ \vspace{0.2cm}
        
        58252 & $5.38^{+2.43}_{-1.23}$ & $10.11^{+2.56}_{-3.43}$ & $24^{+8}_{-8}$ & $1.85^{+0.99}_{-0.82}$ & $7.76^{+2.3}_{-3.3}$ & 101.20/98 & & $-1.40^{+0.02}_{-0.02}$ & $2.35^{+0.01}_{-0.01}$ & $190^{+41}_{-48}$ & 138.35/100 \\ \vspace{0.2cm}
        
        58254 & $5.44^{+3.44}_{-2.74}$ & $10.50^{+3.67}_{-4.32}$ & $26^{+8}_{-9}$ & $1.89^{+0.98}_{-0.88}$ & $8.38^{+3.9}_{-2.6}$ & 77.51/67 & & $-1.41^{+0.02}_{-0.02}$ & $2.36^{+0.01}_{-0.01}$ & $184^{+42}_{-55}$ & 98.51/69 \\ \vspace{0.2cm}
        
        58256 & $5.31^{+1.93}_{-1.78}$ & $11.03^{+2.31}_{-2.45}$ & $31^{+9}_{-9}$ & $1.77^{+0.93}_{-0.77}$ & $4.02^{+3.1}_{-3.2}$ & 97.77/71 & & $-1.44^{+0.02}_{-0.03}$ & $2.39^{+0.01}_{-0.01}$ & $223^{+61}_{-74}$ & 141.93/73 \\ \vspace{0.2cm}
        
        58258 & $5.01^{+1.98}_{-1.21}$ & $9.29^{+2.92}_{-2.34}$ & $27^{+10}_{-9}$ & $1.80^{+0.93}_{-0.80}$ & $6.38^{+3.3}_{-2.3}$ & 33.52/30 & & $-1.54^{+0.04}_{-0.04}$ & $2.40^{+0.02}_{-0.01}$ & $208^{+87}_{-84}$ & 30.15/32 \\ \vspace{0.2cm}
        
        58260 & $4.08^{+2.92}_{-1.37}$ & $7.84^{+1.34}_{-2.45}$ & $32^{+8}_{-8}$ & $1.86^{+0.88}_{-0.86}$ & $11.4^{+2.6}_{-2.6}$ & 111.64/81 & & $-1.49^{+0.02}_{-0.03}$ & $2.42^{+0.01}_{-0.01}$ & $202^{+52}_{-64}$ & 159.29/83 \\ \vspace{0.2cm}
        
        58262 & $5.74^{+1.54}_{-1.76}$ & $11.35^{+2.98}_{-3.65}$ & $32^{+9}_{-9}$ & $1.93^{+0.94}_{-0.90}$ & $16.5^{+3.1}_{-2.4}$ & 22.70/34 & & $-1.45^{+0.03}_{-0.03}$ & $2.35^{+0.02}_{-0.01}$ & $210^{+84}_{-84}$ & 32.02/36 \\ \vspace{0.2cm}
        
        58264 & $4.11^{+1.35}_{-2.34}$ & $8.89^{+2.45}_{-1.23}$ & $39^{+10}_{-10}$ & $1.80^{+0.87}_{-0.78}$ & $8.97^{+2.6}_{-2.1}$ & 108.41/87 & & $-1.46^{+0.03}_{-0.04}$ & $2.39^{+0.01}_{-0.01}$ & $276^{+88}_{-95}$ & 159.66/89 \\ \vspace{0.2cm}
        
        58265 & $4.85^{+2.06}_{-1.27}$ & $11.49^{+2.02}_{-2.39}$ & $49^{+9}_{-10}$ & $1.85^{+1.23}_{-0.84}$ & $8.44^{+22.9}_{-2.7}$ & 115.48/91 & & $-1.47^{+0.03}_{-0.03}$ & $2.36^{+0.01}_{-0.01}$ & $265^{+86}_{-90}$ & 163.20/93 \\ \vspace{0.2cm}
        
        58268 & $4.63^{+2.81}_{-1.29}$ & $10.98^{+3.04}_{-2.91}$ & $47^{+12}_{-12}$ & $1.84^{+1.06}_{-0.84}$ & $6.89^{+2.9}_{-1.1}$ & 110.81/73 & & $-1.50^{+0.03}_{-0.04}$ & $2.40^{+0.01}_{-0.01}$ & $242^{+74}_{-81}$ & 146.45/75 \\ \vspace{0.2cm}
        
        58270 & $4.01^{+1.25}_{-1.01}$ & $12.51^{+3.91}_{-2.47}$ & $70^{+12}_{-14}$ & $1.85^{+0.88}_{-0.82}$ & $10.1^{+3.3}_{-3.2}$ & 95.21/50 & & $-1.56^{+0.03}_{-0.04}$ & $2.43^{+0.01}_{-0.01}$ & $242^{+72}_{-79}$ & 144.83/52 \\ \vspace{0.2cm}
        
        58278 & $4.29^{+1.83}_{-1.90}$ & $12.32^{+2.97}_{-2.94}$ & $75^{+12}_{-12}$ & $1.84^{+0.93}_{-0.89}$ & $13.4^{+3.7}_{-3.8}$ & 51.33/44 & & $-1.64^{+0.04}_{-0.05}$ & $2.46^{+0.01}_{-0.01}$ & $223^{+83}_{-84}$ & 91.61/46 \\ \vspace{0.2cm}
        
        58280 & $2.58^{+1.94}_{-1.38}$ & $12.90^{+3.01}_{-2.89}$ & $85^{+15}_{-13}$ & $1.85^{+1.03}_{-0.84}$ & $7.01^{+2.2}_{-2.5}$ & 81.39/38 & & $-1.68^{+0.05}_{-0.08}$ & $2.62^{+0.02}_{-0.02}$ & $230^{+90}_{-96}$ & 127.56/40 \\ \vspace{0.2cm}
        
        58282 & $3.30^{+1.37}_{-1.59}$ & $13.40^{+4.21}_{-3.09}$ & $86^{+13}_{-14}$ & $1.85^{+1.01}_{-0.84}$ & $8.18^{+2.2}_{-2.3}$ & 35.22/23 & & $-1.66^{+0.04}_{-0.06}$ & $2.50^{+0.02}_{-0.02}$ & $226^{+81}_{-91}$ & 58.54/25 \\ \vspace{0.2cm}
        
        58283 & $3.65^{+1.29}_{-1.47}$ & $11.79^{+3.26}_{-3.02}$ & $71^{+9}_{-10}$ & $1.85^{+0.87}_{-0.83}$ & $10.9^{+3.3}_{-3.1}$ & 68.22/50 & & $-1.63^{+0.03}_{-0.05}$ & $2.41^{+0.01}_{-0.01}$ & $254^{+84}_{-98}$ & 112.23/52 \\ \vspace{0.2cm}
        
        58286 & $3.60^{+1.65}_{-1.26}$ & $12.01^{+3.02}_{-2.96}$ & $78^{+11}_{-12}$ & $1.86^{+1.11}_{-0.86}$ & $7.54^{+4.0}_{-2.1}$ & 75.41/48 & & $-1.64^{+0.03}_{-0.05}$ & $2.44^{+0.01}_{-0.01}$ & $266^{+66}_{-102}$ & 108.26/48 \\ \vspace{0.2cm}
        
        58292 & $2.79^{+0.98}_{-0.93}$ & $10.00^{+3.30}_{-2.83}$ & $63^{+14}_{-14}$ & $1.84^{+0.81}_{-0.80}$ & $7.94^{+3.2}_{-2.4}$ & 35.06/22 & & $-1.69^{+0.03}_{-0.04}$ & $2.41^{+0.02}_{-0.02}$ & $189^{+73}_{-74}$ & 73.77/24 \\ \vspace{0.2cm}
        
        58294 & $3.30^{+1.24}_{-1.01}$ & $9.28^{+2.96}_{-2.36}$ & $50^{+11}_{-13}$ & $1.86^{+1.02}_{-0.81}$ & $9.13^{+2.1}_{-2.4}$ & 109.78/82 & & $-1.60^{+0.02}_{-0.03}$ & $2.33^{+0.01}_{-0.01}$ & $238^{+77}_{-84}$ & 169.60/84 \\ \vspace{0.2cm}
        
        58296 & $4.84^{+1.99}_{-1.93}$ & $10.17^{+1.90}_{-1.94}$ & $38^{+10}_{-10}$ & $1.79^{+1.12}_{-0.77}$ & $6.84^{+2.1}_{-2.8}$ & 51.09/38 & & $-1.59^{+0.04}_{-0.03}$ & $2.34^{+0.01}_{-0.01}$ & $213^{+83}_{-85}$ & 62.05/40 \\ \vspace{0.2cm}
        
        58306 & $4.40^{+1.99}_{-1.92}$ & $10.01^{+4.53}_{-4.45}$ & $42^{+11}_{-10}$ & $1.84^{+1.00}_{-0.84}$ & $7.31^{+3.0}_{-3.1}$ & 105.37/76 & & $-1.53^{+0.03}_{-0.04}$ & $2.37^{+0.01}_{-0.01}$ & $256^{+90}_{-94}$ & 146.67/78 \\ \vspace{0.2cm}
        
        58310 & $5.30^{+1.46}_{-1.45}$ & $10.13^{+5.10}_{-4.67}$ & $32^{+10}_{-10}$ & $1.80^{+1.06}_{-0.80}$ & $10.9^{+4.3}_{-4.5}$ & 85.70/76 & & $-1.49^{+0.03}_{-0.03}$ & $2.34^{+0.01}_{-0.01}$ & $252^{+91}_{-80}$ & 116.06/78 \\ \vspace{0.2cm}
        
        58314 & $5.28^{+1.97}_{-2.32}$ & $10.16^{+3.64}_{-3.75}$ & $34^{+10}_{-9}$ & $1.79^{+1.09}_{-0.76}$ & $10.9^{+1.2}_{-2.3}$ & 129.75/118 & & $-1.49^{+0.02}_{-0.03}$ & $2.38^{+0.01}_{-0.01}$ & $243^{+63}_{-71}$ & 188.13/120 \\
        \hline
    \end{tabular}}
    \label{tab:2018_physical_uv}
\end{table*}


\bsp	
\label{lastpage}
\end{document}